\documentclass[aps,notitlepage,nofootinbib,longbibliography,twocolumn,superscriptaddress,10pt]{revtex4-1}
\usepackage{amsmath}
\usepackage{amsthm, amssymb}
\usepackage{slashed}
\usepackage{graphicx}
\usepackage{xcolor}
\usepackage{tikz}
\usetikzlibrary{calc,fadings,decorations.pathreplacing,shapes,shapes.multipart,arrows,shapes.misc,intersections,positioning,patterns}
\usepackage{bm}
\usepackage[colorlinks=true,citecolor=blue,linkcolor=magenta]{hyperref}
\usepackage{dsfont}
\usepackage{changepage}
\usepackage{array}

\newcommand{\bit}{\begin{itemize}}
\newcommand{\eit}{\end{itemize}}

\newcommand{\f}{\frac}
\renewcommand{\>}{\right\rangle}
\newcommand{\<}{\left\langle}
\newcommand{\ba}{\begin{align}}
\newcommand{\ea}{\end{align}}
\newcommand{\be}{\begin{equation}}
\newcommand{\ee}{\end{equation}}
\newcommand{\bi}{\begin{itemize}}
\newcommand{\ei}{\end{itemize}}
\newcommand{\lf}{\left(}
\newcommand{\ri}{\right)}
\newcommand{\dd}{\mathrm{d}}
\newcommand{\DD}{\mathcal{D}}

\newcommand{\Tr}{\operatorname{Tr}}
\newcommand{\tr}{\operatorname{tr}}

\def\braket#1#2{\left\langle #1|#2\right\rangle}

\newcommand{\bra}[1]{\< #1 \right|}
\newcommand{\ket}[1]{\left| #1 \>}

\newcommand{\znk}{\overline{Z_n^k}}
\newcommand{\zk}{\overline{Z_2^k}}

\newcommand{\seq}{s_\text{eq}}
\newcommand{\I}{\mathbb{I}}
\newcommand{\lt}{\mathcal{E}}

\graphicspath{{}{./images/}}

\begin{document}
\title{Emergent statistical mechanics of entanglement in random unitary circuits}

\author{Tianci Zhou}
\email{tzhou13@illinois.edu}
\affiliation{University of Illinois, Department of Physics, 1110 W. Green St. Urbana, IL 61801 USA}
\author{Adam Nahum}
\email{adam.nahum@physics.ox.ac.uk}
\affiliation{Theoretical Physics, Oxford University, 1 Keble Road, Oxford OX1 3NP, United Kingdom}

\date{\today}

\newcommand{\ugatew}[2]{
  \draw[line width = 1pt] (#1,#2)--++(0,0.25);
  \draw[line width = 1pt] (#1,#2+1.25)--++(0,0.25);
  \draw[line width = 1pt] (#1+1,#2)--++(0,0.25);
  \draw[line width = 1pt] (#1+1,#2+1.25)--++(0,0.25);
  \draw[line width = 1pt] (#1-0.2,#2+0.25) rectangle (#1+1.2,#2+1.25);
}

\newcommand{\trilatticecode}{
  \foreach \x in {0,...,2}{
    \fill[green!20] (\x*1.732,0)--++(60:1.732)--++(-1.732,0)--cycle;
    \draw (\x*1.732,0)--++(60:1.732)--++(-1.732,0)--cycle;
  }
  \foreach \x in {0.5,1.5}{
    \fill[green!20] (\x*1.732,1.5)--++(60:1.732)--++(-1.732,0)--cycle;
    \draw (\x*1.732,1.5)--++(60:1.732)--++(-1.732,0)--cycle;
  }
  \foreach \x in {0,...,2}{
    \fill[green!20] (\x*1.732,3)--++(60:1.732)--++(-1.732,0)--cycle;
    \draw (\x*1.732,3)--++(60:1.732)--++(-1.732,0)--cycle;
  }
}

\newcommand{\spint}[3]{
    \begin{tikzpicture}[baseline=(current  bounding  box.center), scale = 0.5]
    \draw[dashed] (0,-0.5)--(2,-0.5)--(1,1.232)--cycle;
    \node[left] () at (0,-0.5) {$#2$};
    \node[right] () at (2,-0.5) {$#3$};
    \node[above] () at (1,1.232) {$#1$};
  \end{tikzpicture}
}

\newcommand{\dptri}[3]{
  \begin{tikzpicture}[baseline=(current  bounding  box.center), scale = 0.5]
  \draw (0,0)--(-1,1.732)--(1,1.732)--cycle;
  \node[below] () at (0,0) {$#1$};
  \node[left] () at (-1,1.732) {$#2$};
  \node[right] () at (1,1.732) {$#3$};
  \end{tikzpicture}
}

\newcommand{\vacbubb}{
  \begin{tikzpicture}[baseline=(current  bounding  box.center), scale = 0.5]
  \draw (0,0)--(-1,1.732)--(1,1.732)--cycle;
  \draw[line width = 1pt] (0,2/3*1.732) --++ (210:1) node[below] () {\footnotesize $\mu$};
  \draw[line width = 1pt] (0,2/3*1.732) --++ (-30:1) coordinate (A);
  \node () at ([shift=({0.2,-0.37})]A) {\footnotesize $\mu^{-1}$};
  \end{tikzpicture}
}

\newcommand{\sdwbubb}{
  \begin{tikzpicture}[baseline=(current  bounding  box.center), scale = 0.5]
    \draw (0,0)--(-1,1.732)--(1,1.732)--cycle;
    \draw (0,1.3) -- +(90:1.077);
    \draw (0,1.3) -- +(210:1.077);
    \draw[line width = 1pt] (0,1.1) -- +(210:1.02) node[below] () {\footnotesize $\mu$};
    \draw[line width = 1pt] (0,1.1) -- +(-30:1.02) coordinate (A);
    \node () at ([shift=({0.2,-0.37})]A) {\footnotesize $\mu^{-1}$};
    \node () at (-1.1,0) {};
    \node () at (1.1,0) {};
    \draw (0,0)--(-1,1.732)--(1,1.732)--cycle;
  \end{tikzpicture}
}

\newcommand{\stvert}[2]{\draw (#1,#2)--++(0,1);
\draw (#1,#2)--++(-30:0.5);
\draw (#1,#2)--++(210:0.5);
\draw (#1,#2+1)--++(30:0.5);
\draw (#1,#2+1)--++(150:0.5);
\fill (#1,#2) circle (0.05);
\fill (#1,#2+1) circle (0.05);}

\newcommand\fourU[2]{
  \fill (#1,#2) circle (0.08);
  \fill (#1,#2+1) circle (0.08);
  \fill (#1+1,#2) circle (0.08);
  \fill (#1+1,#2+1) circle (0.08);
  \fill[blue!50] (#1+0.3,#2+0.3) rectangle (#1+0.7,#2+0.7);
  \draw (#1+0.3,#2+0.3)--(#1+0,#2+0);
  \draw (#1+0.3,#2+0.7)--(#1+0,#2+1);
  \draw (#1+0.7,#2+0.3)--(#1+1,#2+0);
  \draw (#1+0.7,#2+0.7)--(#1+1,#2+1);
}

\newcommand{\sdwfull}[0]{
    \begin{tikzpicture}[baseline=(current  bounding  box.center), scale = 0.5]
      \draw (0,0)--(-1,1.732)--(1,1.732)--cycle;
      \draw (0,1.155) -- +(90:1.077);
      \draw (0,1.155) -- +(210:1.077);
      \node () at (-1.1,0) {};
      \node () at (1.1,0) {};
    \end{tikzpicture}
}

\newcommand{\sdwr}[0]{
    \begin{tikzpicture}[baseline=(current  bounding  box.center), scale = 0.5]
      \draw (0,0)--(-1,1.732)--(1,1.732)--cycle;
      \draw (0,1.155) -- +(90:1.077);
      \draw (0,1.155) -- +(-30:1.077);
      \node () at (-1.1,0) {};
      \node () at (1.1,0) {};
    \end{tikzpicture}
}

\newcommand{\sdwi}[0]{
    \begin{tikzpicture}[baseline=(current  bounding  box.center), scale = 0.5]
      \draw (0,0)--(-1,1.732)--(1,1.732)--cycle;
      \draw[dashed] (0,0)--++(0,1.732*2/3)--(0-1,0+1.732);
      \draw[dashed] (0,1.732*2/3)--(0+1,0+1.732);
      \draw (0,1.732+0.5)--(0,1.732)--(-0.5,0.5*1.732)--++(-0.25*1.732,-0.25);
      \node () at (-1.1,0) {};
      \node () at (1.1,0) {};
    \end{tikzpicture}
}

\newcommand{\sdwturn}[0]{
    \begin{tikzpicture}[baseline=(current  bounding  box.center), scale = 0.5]
      \draw (0,0)--(-1,1.732)--(1,1.732)--cycle;
      \draw[dashed] (0,0)--++(0,1.732*2/3)--(0-1,0+1.732);
      \draw[dashed] (0,1.732*2/3)--(0+1,0+1.732);
      \draw (0,1.732+0.5)--(0,1.732)--(+0.5,0.5*1.732);
      \draw[red] (0.5, 0.5*1.732)--(-0.5,0.5*1.732);
      \draw (-0.5, 0.5*1.732)--++(-0.25*1.732,-0.25);
      \node () at (-1.1,0) {};
      \node () at (1.1,0) {};
    \end{tikzpicture}
}

\newcommand{\sdwtri}[0]{
    \begin{tikzpicture}[baseline=(current  bounding  box.center), scale = 0.5]
      \draw (0,0)--(-1,1.732)--(1,1.732)--cycle;
      \draw[dashed] (0,0)--++(0,1.732*2/3)--(0-1,0+1.732);
      \draw[dashed] (0,1.732*2/3)--(0+1,0+1.732);
      \draw (0,1.732+0.5)--(0,1.732)--(-0.5,0.5*1.732)--++(-0.25*1.732,-0.25);
      \draw[red] (-0.4,0.5*1.732)--++(0.85,0)--++(120:0.85)--cycle;
      \node () at (-1.1,0) {};
      \node () at (1.1,0) {};
    \end{tikzpicture}
}

\newcommand{\sdwpfull}[0]{
    \begin{tikzpicture}[baseline=(current  bounding  box.center), scale = 0.5]
      \draw (0,0)--(-1,1.732)--(1,1.732)--cycle;
      \draw (0,1.2) -- +(90:1.077);
      \draw (0,1.2) -- +(210:1.077);
      \draw (0,1.1) -- +(-30:1.02);
      \draw (0,1.1) -- +(210:1.02);
      \node () at (-1.1,0) {};
      \node () at (1.1,0) {};
    \end{tikzpicture}
}

\newcommand{\sdwpi}[0]{
    \begin{tikzpicture}[baseline=(current  bounding  box.center), scale = 0.5]
      \draw (0,0)--(-1,1.732)--(1,1.732)--cycle;
      \draw[dashed] (0,0)--++(0,1.732*2/3)--(0-1,0+1.732);
      \draw[dashed] (0,1.732*2/3)--(0+1,0+1.732);
      \draw (0,1.732+0.5)--(0,1.732)--(-0.5,0.5*1.732)--++(-0.25*1.732,-0.25);
      \draw (-0.45,0.45*1.732)--++(60:0.9)--++(-60:0.9);
      \draw (-0.45,0.45*1.732)--++(-0.25*1.732,-0.25);
      \draw (0.45,0.45*1.732)--++(0.25*1.732,-0.25);
      \node () at (-1.1,0) {};
      \node () at (1.1,0) {};
    \end{tikzpicture}
}

\newcommand{\sdwpw}[0]{
    \begin{tikzpicture}[baseline=(current  bounding  box.center), scale = 0.5]
      \draw (0,0)--(-1,1.732)--(1,1.732)--cycle;
      \draw[dashed] (0,0)--++(0,1.732*2/3)--(0-1,0+1.732);
      \draw[dashed] (0,1.732*2/3)--(0+1,0+1.732);
      \draw (0,1.732+0.5)--(0,1.732)--(-0.5,0.5*1.732)--++(-0.25*1.732,-0.25);
      \draw (-0.45,0.45*1.732)--++(-0.25*1.732,-0.25);
      \draw (0.45,0.45*1.732)--++(0.25*1.732,-0.25);
      \draw[red] (-0.45,0.45*1.732)--(0.45,0.45*1.732);
      \node () at (-1.1,0) {};
      \node () at (1.1,0) {};
    \end{tikzpicture}
}

\newcommand{\sdwpiturn}[0]{
    \begin{tikzpicture}[baseline=(current  bounding  box.center), scale = 0.5]
      \draw (0,0)--(-1,1.732)--(1,1.732)--cycle;
      \draw[dashed] (0,0)--++(0,1.732*2/3)--(0-1,0+1.732);
      \draw[dashed] (0,1.732*2/3)--(0+1,0+1.732);
      \draw (0,1.732+0.5)--(0,1.732)--(+0.5,0.5*1.732);
      \draw[red] (0.5, 0.5*1.732)--(-0.5,0.5*1.732);
      \draw (-0.5, 0.5*1.732)--++(-0.25*1.732,-0.25);
      \draw (-0.45,0.45*1.732)--++(60:0.9)--++(-60:0.9);
      \draw (-0.45,0.45*1.732)--++(-0.25*1.732,-0.25);
      \draw (0.45,0.45*1.732)--++(0.25*1.732,-0.25);
      \node () at (-1.1,0) {};
      \node () at (1.1,0) {};
    \end{tikzpicture}
}

\newcommand{\sdwpwturn}[0]{
    \begin{tikzpicture}[baseline=(current  bounding  box.center), scale = 0.5]
      \draw (0,0)--(-1,1.732)--(1,1.732)--cycle;
      \draw[dashed] (0,0)--++(0,1.732*2/3)--(0-1,0+1.732);
      \draw[dashed] (0,1.732*2/3)--(0+1,0+1.732);
      \draw (0,1.732+0.5)--(0,1.732)--(+0.5,0.5*1.732);
      \draw[red] (0.5, 0.5*1.732)--(-0.5,0.5*1.732);
      \draw (-0.5, 0.5*1.732)--++(-0.25*1.732,-0.25);
      \draw (-0.45,0.45*1.732)--++(-0.25*1.732,-0.25);
      \draw (0.45,0.45*1.732)--++(0.25*1.732,-0.25);
      \draw[red] (-0.45,0.45*1.732)--(0.45,0.45*1.732);
      \node () at (-1.1,0) {};
      \node () at (1.1,0) {};
    \end{tikzpicture}
}

\newcommand{\ddwfull}[0]{
    \begin{tikzpicture}[baseline=(current  bounding  box.center), scale = 0.5]
      \draw (0,0)--(-1,1.732)--(1,1.732)--cycle;
      \draw (0,1.155) -- +(90:1.077);
      \draw (0,1.155) -- +(210:1.077);
      \draw (-0.1,1.23) -- +(90:1.008);
      \draw (-0.1,1.23) -- +(210:1.02);
      \node () at (-1.1,0) {};
      \node () at (1.1,0) {};
    \end{tikzpicture}
}

\newcommand{\ddwlr}[0]{
    \begin{tikzpicture}[baseline=(current  bounding  box.center), scale = 0.5]
      \draw (0,0)--(-1,1.732)--(1,1.732)--cycle;
      \draw (0.06,1.155) -- +(90:1.077);
      \draw (0.06,1.155) -- +(-30:1.077);
      \draw (-0.06,1.155) -- +(90:1.077);
      \draw (-0.06,1.155) -- +(210:1.077);
      \node () at (-1.1,0) {};
      \node () at (1.1,0) {};
    \end{tikzpicture}
}

\newcommand{\ddwlri}[0]{
    \begin{tikzpicture}[baseline=(current  bounding  box.center), scale = 0.5]
      \draw (0,0)--(-1,1.732)--(1,1.732)--cycle;
      \draw[dashed] (0,0)--++(0,1.732*2/3)--(0-1,0+1.732);
      \draw[dashed] (0,1.732*2/3)--(0+1,0+1.732);
      \draw (-0.05,1.732+0.5)--++(0,-0.5)--(-0.55,0.55*1.732)--++(-0.25*1.732,-0.25);
      \draw (0.05,1.732+0.5)--++(0,-0.5)--(0.55,0.55*1.732)--++(0.25*1.732,-0.25);
      \node () at (-1.1,0) {};
      \node () at (1.1,0) {};
    \end{tikzpicture}
}

\newcommand{\ddwlriturn}[0]{
    \begin{tikzpicture}[baseline=(current  bounding  box.center), scale = 0.5]
      \draw (0,0)--(-1,1.732)--(1,1.732)--cycle;
      \draw[dashed] (0,0)--++(0,1.732*2/3)--(0-1,0+1.732);
      \draw[dashed] (0,1.732*2/3)--(0+1,0+1.732);
      \draw (-0.05,1.732+0.5)--++(0,-0.5)--(-0.55,0.55*1.732)--++(-0.25*1.732,-0.25);
      \draw (0.05,1.732+0.5)--++(0,-0.5)--(-0.5,0.5*1.732);
      \draw[red] (-0.5,0.5*1.732)--(0.5,0.5*1.732);
      \draw (0.5,0.5*1.732)--++(0.25*1.732,-0.25);
      \node () at (-1.1,0) {};
      \node () at (1.1,0) {};
    \end{tikzpicture}
}

\newcommand{\ddwlritt}[0]{
    \begin{tikzpicture}[baseline=(current  bounding  box.center), scale = 0.5]
      \draw (0,0)--(-1,1.732)--(1,1.732)--cycle;
      \draw[dashed] (0,0)--++(0,1.732*2/3)--(0-1,0+1.732);
      \draw[dashed] (0,1.732*2/3)--(0+1,0+1.732);
      \draw (-0.05,1.732+0.5)--++(0,-0.5)--(0.55,0.55*1.732);
      \draw[red] (0.55,0.55*1.732)--(-0.55,0.55*1.732);
      \draw (-0.55,0.55*1.732)--++(-0.25*1.732,-0.25);
      \draw (0.05,1.732+0.5)--++(0,-0.5)--(-0.5,0.5*1.732);
      \draw[red] (-0.5,0.5*1.732)--(0.5,0.5*1.732);
      \draw (0.5,0.5*1.732)--++(0.25*1.732,-0.25);
      \node () at (-1.1,0) {};
      \node () at (1.1,0) {};
    \end{tikzpicture}
}

\newcommand{\ddwlrtri}[0]{
    \begin{tikzpicture}[baseline=(current  bounding  box.center), scale = 0.5]
      \draw (0,0)--(-1,1.732)--(1,1.732)--cycle;
      \draw[dashed] (0,0)--++(0,1.732*2/3)--(0-1,0+1.732);
      \draw[dashed] (0,1.732*2/3)--(0+1,0+1.732);
      \draw (-0.05,1.732+0.5)--++(0,-0.5)--(-0.55,0.55*1.732)--++(-0.25*1.732,-0.25);
      \draw (0.05,1.732+0.5)--++(0,-0.5)--(0.55,0.55*1.732)--++(0.25*1.732,-0.25);
      \draw[red] (-0.45,0.5*1.732)--++(0.9,0)--++(120:0.9)--cycle;
      \node () at (-1.1,0) {};
      \node () at (1.1,0) {};
    \end{tikzpicture}
}

\newcommand{\ddwi}[0]{
    \begin{tikzpicture}[baseline=(current  bounding  box.center), scale = 0.5]
      \draw (0,0)--(-1,1.732)--(1,1.732)--cycle;
      \draw[dashed] (0,0)--++(0,1.732*2/3)--(0-1,0+1.732);
      \draw[dashed] (0,1.732*2/3)--(0+1,0+1.732);
      \draw (0,1.732+0.5)--(0,1.732)--(-0.5,0.5*1.732)--++(-0.25*1.732,-0.25);
      \draw (-0.1,1.732+0.5)--(-0.1,1.732)--(-0.55,0.55*1.732)--++(-0.25*1.732,-0.25);
      \node () at (-1.1,0) {};
      \node () at (1.1,0) {};
    \end{tikzpicture}
}

\newcommand{\ddwiturn}[0]{
    \begin{tikzpicture}[baseline=(current  bounding  box.center), scale = 0.5]
      \draw (0,0)--(-1,1.732)--(1,1.732)--cycle;
      \draw[dashed] (0,0)--++(0,1.732*2/3)--(0-1,0+1.732);
      \draw[dashed] (0,1.732*2/3)--(0+1,0+1.732);
      \draw (0,1.732+0.5)--(0,1.732)--(+0.5,0.5*1.732);
      \draw[red] (0.5, 0.5*1.732)--(-0.5,0.5*1.732);
      \draw (-0.1,1.732+0.5)--(-0.1,1.732)--(-0.55,0.55*1.732)--++(-0.25*1.732,-0.25);
      \draw (-0.5, 0.5*1.732)--++(-0.25*1.732,-0.25);
      \node () at (-1.1,0) {};
      \node () at (1.1,0) {};
    \end{tikzpicture}
}

\newcommand{\ddwtt}[0]{
    \begin{tikzpicture}[baseline=(current  bounding  box.center), scale = 0.5]
      \draw (0,0)--(-1,1.732)--(1,1.732)--cycle;
      \draw[dashed] (0,0)--++(0,1.732*2/3)--(0-1,0+1.732);
      \draw[dashed] (0,1.732*2/3)--(0+1,0+1.732);
      \draw (0,1.732+0.5)--(0,1.732)--(+0.5,0.5*1.732);
      \draw[red] (0.5, 0.5*1.732)--(-0.5,0.5*1.732);
      \draw (-0.5, 0.5*1.732)--++(-0.25*1.732,-0.25);
      \draw (-0.1,1.732+0.5)--(-0.1,1.732)--(0.35,0.55*1.732);
      \draw[red] (0.35,0.55*1.732)--(-0.55,0.55*1.732);
      \draw (-0.55,0.55*1.732)--++(-0.25*1.732,-0.25);
      \node () at (-1.1,0) {};
      \node () at (1.1,0) {};
    \end{tikzpicture}
}

\newcommand{\ddwtri}[0]{
    \begin{tikzpicture}[baseline=(current  bounding  box.center), scale = 0.5]
      \draw (0,0)--(-1,1.732)--(1,1.732)--cycle;
      \draw[dashed] (0,0)--++(0,1.732*2/3)--(0-1,0+1.732);
      \draw[dashed] (0,1.732*2/3)--(0+1,0+1.732);
      \draw (0,1.732+0.5)--(0,1.732)--(-0.5,0.5*1.732)--++(-0.25*1.732,-0.25);
      \draw[red] (-0.4,0.5*1.732)--++(0.85,0)--++(120:0.85)--cycle;
      \draw (-0.1,1.732+0.5)--(-0.1,1.732)--(-0.55,0.55*1.732)--++(-0.25*1.732,-0.25);
      \node () at (-1.1,0) {};
      \node () at (1.1,0) {};
    \end{tikzpicture}
}

\newcommand{\ddwpfull}[0]{
    \begin{tikzpicture}[baseline=(current  bounding  box.center), scale = 0.5]
      \draw (0,0)--(-1,1.732)--(1,1.732)--cycle;
      \draw (0,1.04) -- +(-30:1.02);
      \draw (0,1.04) -- +(210:1.02);
      \draw (-0.1,1.732+0.5)--++(0,-0.45-1/3*1.732)--++(210:1.0);
      \draw (-0.0,1.732+0.5)--++(0,-0.5-1/3*1.732)--++(210:1.077);
      \node () at (-1.1,0) {};
      \node () at (1.1,0) {};
    \end{tikzpicture}
}

\newcommand{\ddwpi}[0]{
    \begin{tikzpicture}[baseline=(current  bounding  box.center), scale = 0.5]
      \draw (0,0)--(-1,1.732)--(1,1.732)--cycle;
      \draw[dashed] (0,0)--++(0,1.732*2/3)--(0-1,0+1.732);
      \draw[dashed] (0,1.732*2/3)--(0+1,0+1.732);
      \draw (0,1.732+0.5)--(0,1.732)--(-0.5,0.5*1.732)--++(-0.25*1.732,-0.25);
      \draw (-0.45,0.45*1.732)--++(60:0.9)--++(-60:0.9);
      \draw (-0.45,0.45*1.732)--++(-0.25*1.732,-0.25);
      \draw (0.45,0.45*1.732)--++(0.25*1.732,-0.25);
      \draw (-0.1,1.732+0.5)--(-0.1,1.732)--(-0.55,0.55*1.732)--++(-0.25*1.732,-0.25);
      \node () at (-1.1,0) {};
      \node () at (1.1,0) {};
    \end{tikzpicture}
}

\newcommand{\ddwpw}[0]{
    \begin{tikzpicture}[baseline=(current  bounding  box.center), scale = 0.5]
      \draw (0,0)--(-1,1.732)--(1,1.732)--cycle;
      \draw[dashed] (0,0)--++(0,1.732*2/3)--(0-1,0+1.732);
      \draw[dashed] (0,1.732*2/3)--(0+1,0+1.732);
      \draw (0,1.732+0.5)--(0,1.732)--(-0.5,0.5*1.732)--++(-0.25*1.732,-0.25);
      \draw (-0.45,0.45*1.732)--++(-0.25*1.732,-0.25);
      \draw (0.45,0.45*1.732)--++(0.25*1.732,-0.25);
      \draw[red] (-0.45,0.45*1.732)--(0.45,0.45*1.732);
      \draw (-0.1,1.732+0.5)--(-0.1,1.732)--(-0.55,0.55*1.732)--++(-0.25*1.732,-0.25);
      \node () at (-1.1,0) {};
      \node () at (1.1,0) {};
    \end{tikzpicture}
}

\newcommand{\ddwpiturn}[0]{
    \begin{tikzpicture}[baseline=(current  bounding  box.center), scale = 0.5]
      \draw (0,0)--(-1,1.732)--(1,1.732)--cycle;
      \draw[dashed] (0,0)--++(0,1.732*2/3)--(0-1,0+1.732);
      \draw[dashed] (0,1.732*2/3)--(0+1,0+1.732);
      \draw (0,1.732+0.5)--(0,1.732)--(+0.5,0.5*1.732);
      \draw[red] (0.5, 0.5*1.732)--(-0.5,0.5*1.732);
      \draw (-0.5, 0.5*1.732)--++(-0.25*1.732,-0.25);
      \draw (-0.45,0.45*1.732)--++(60:0.9)--++(-60:0.9);
      \draw (-0.45,0.45*1.732)--++(-0.25*1.732,-0.25);
      \draw (0.45,0.45*1.732)--++(0.25*1.732,-0.25);
      \draw (-0.1,1.732+0.5)--(-0.1,1.732)--(-0.55,0.55*1.732)--++(-0.25*1.732,-0.25);
      \node () at (-1.1,0) {};
      \node () at (1.1,0) {};
    \end{tikzpicture}
}

\newcommand{\ddwpwturn}[0]{
    \begin{tikzpicture}[baseline=(current  bounding  box.center), scale = 0.5]
      \draw (0,0)--(-1,1.732)--(1,1.732)--cycle;
      \draw[dashed] (0,0)--++(0,1.732*2/3)--(0-1,0+1.732);
      \draw[dashed] (0,1.732*2/3)--(0+1,0+1.732);
      \draw (0,1.732+0.5)--(0,1.732)--(+0.5,0.5*1.732);
      \draw[red] (0.5, 0.5*1.732)--(-0.5,0.5*1.732);
      \draw (-0.5, 0.5*1.732)--++(-0.25*1.732,-0.25);
      \draw (-0.45,0.45*1.732)--++(-0.25*1.732,-0.25);
      \draw (0.45,0.45*1.732)--++(0.25*1.732,-0.25);
      \draw[red] (-0.45,0.45*1.732)--(0.45,0.45*1.732);
      \draw (-0.1,1.732+0.5)--(-0.1,1.732)--(-0.55,0.55*1.732)--++(-0.25*1.732,-0.25);
      \node () at (-1.1,0) {};
      \node () at (1.1,0) {};
    \end{tikzpicture}
}

\newcommand{\ddwbubble}[0]{
  \begin{tikzpicture}[baseline=(current  bounding  box.center), scale = 0.5]
    \draw (0,0)--(-1,1.732)--(1,1.732)--cycle;
    \draw (0,0)--(-1,-1.732)--(1,-1.732)--cycle;
    \draw (0.05,1.732+0.5)--++(0,-0.5-1/3*1.732)--++(1,-1/3*1.732)--++(0,-2/3*1.732)--++(-1,-1/3*1.732)--++(0,-0.5-1/3*1.732);
    \draw (-0.05,1.732+0.5)--++(0,-0.5-1/3*1.732)--++(-1,-1/3*1.732)--++(0,-2/3*1.732)--++(1,-1/3*1.732)--++(0,-0.5-1/3*1.732);
    \draw (0,0.616*1.732)--++(210:2/3*1.65)--++(0,-2/3*1.55)--(0,-0.616*1.732);
    \draw (0,0.616*1.732)--++(-30:2/3*1.65)--++(0,-2/3*1.55)--(0,-0.616*1.732);
    \node () at (-1.1,0) {};
    \node () at (1.1,0) {};
    \fill (0,2/3*1.732-0.03) circle (0.1);
    \fill (0,-2/3*1.732+0.03) circle (0.1);
  \end{tikzpicture}
}

\newcommand{\ddwbubbone}[0]{
  \begin{tikzpicture}[baseline=(current  bounding  box.center), scale = 0.5]
    \draw (0,0)--(-1,1.732)--(1,1.732)--cycle;
    \draw (0,0)--(-1,-1.732)--(1,-1.732)--cycle;
    \draw (0.05,1.732+0.5)--++(0,-0.5-1/3*1.732)--++(1,-1/3*1.732)--++(0,-2/3*1.732)--++(-1,-1/3*1.732)--++(0,-0.5-1/3*1.732);
    \draw (-0.05,1.732+0.5)--++(0,-0.5-1/3*1.732)--++(-1,-1/3*1.732)--++(0,-2/3*1.732)--++(1,-1/3*1.732)--++(0,-0.5-1/3*1.732);
    \draw (0,0.616*1.732)--++(210:2/3*1.65)--++(0,-2/3*1.55)--(0,-0.616*1.732);
    \draw (0,0.616*1.732)--++(-30:2/3*1.65)--++(0,-2/3*1.55)--(0,-0.616*1.732);
    \fill (0,2/3*1.732-0.03) circle (0.1);
    \fill (0,-2/3*1.732+0.03) circle (0.1);
    \node[above] () at (0,1.732+0.5) {\footnotesize $(12)(34)$};
    \node[below] () at (0,-1.732-0.5) {\footnotesize $(12)(34)$};
    \node[left] () at (-1,0) {\footnotesize $(14)(23)$};
    \node[right] () at (1,0) {\footnotesize $(24)(13)$};
  \end{tikzpicture}
}

\newcommand{\ddwbubbtwo}[0]{
  \begin{tikzpicture}[baseline=(current  bounding  box.center), scale = 0.5]
    \draw (0,0)--(-1,1.732)--(1,1.732)--cycle;
    \draw (0,0)--(-1,-1.732)--(1,-1.732)--cycle;
    \draw (0.05,1.732+0.5)--++(0,-0.5-1/3*1.732)--++(1,-1/3*1.732)--++(0,-2/3*1.732)--++(-1,-1/3*1.732)--++(0,-0.5-1/3*1.732);
    \draw (-0.05,1.732+0.5)--++(0,-0.5-1/3*1.732)--++(-1,-1/3*1.732)--++(0,-2/3*1.732)--++(1,-1/3*1.732)--++(0,-0.5-1/3*1.732);
    \draw (0,0.616*1.732)--++(210:2/3*1.65)--++(0,-2/3*1.55)--(0,-0.616*1.732);
    \draw (0,0.616*1.732)--++(-30:2/3*1.65)--++(0,-2/3*1.55)--(0,-0.616*1.732);
    \node () at (-1.1,0) {};
    \node () at (1.1,0) {};
    \fill (0,2/3*1.732-0.03) circle (0.1);
    \fill (0,-2/3*1.732+0.03) circle (0.1);
    \node[above] () at (0,1.732+0.5) {\footnotesize $(12)(34)$};
    \node[below] () at (0,-1.732-0.5) {\footnotesize $(12)(34)$};
    \node[left] () at (-1,0) {\footnotesize $(24)(13)$};
    \node[right] () at (1,0) {\footnotesize $(14)(23)$};
  \end{tikzpicture}
}

\newcommand{\ddwtwostep}[0]{
  \begin{tikzpicture}[baseline=(current  bounding  box.center), scale = 0.5]
    \draw (0,0)--(-1,1.732)--(1,1.732)--cycle;
    \draw (0,0)--(-1,-1.732)--(1,-1.732)--cycle;
    \draw (0.05,1.732+0.5)--++(0,-0.5-1/3*1.732);
    \draw (-0.05,1.732+0.5)--++(0,-0.5-1/3*1.732);
    \draw (0.05,-1.732-0.5)--++(0,0.5+1/3*1.732);
    \draw (-0.05,-1.732-0.5)--++(0,0.5+1/3*1.732);
    \node () at (0,2/3*1.732-0.2) {$\vdots$};
    \node () at (0,-2/3*1.732+0.5) {$\vdots$};
    \node () at (-1.1,0) {};
    \node () at (1.1,0) {};
  \end{tikzpicture}
}

\newcommand{\ddwtwostepcode}[0]{
  \draw (0,0)--(-1,1.732)--(1,1.732)--cycle;
  \draw (0,0)--(-1,-1.732)--(1,-1.732)--cycle;
  \node () at (-1.1,0) {};
  \node () at (1.1,0) {};
}

\newcommand{\ddwtwostepl}[0]{
  \begin{tikzpicture}[baseline=(current  bounding  box.center), scale = 0.5]
    \ddwtwostepcode
    \draw (-0.05,1.732+0.5)--++(0,-0.5-1/3*1.732)--++(-1,-1/3*1.732)--++(0,-2/3*1.732)--++(1,-1/3*1.732)--++(0,-0.5-1/3*1.732);
  \end{tikzpicture}
}

\newcommand{\ddwtwostepr}[0]{
  \begin{tikzpicture}[baseline=(current  bounding  box.center), scale = 0.5]
    \ddwtwostepcode
    \draw (0.05,1.732+0.5)--++(0,-0.5-1/3*1.732)--++(1,-1/3*1.732)--++(0,-2/3*1.732)--++(-1,-1/3*1.732)--++(0,-0.5-1/3*1.732);
  \end{tikzpicture}
}

\newcommand{\ddwtwostepll}[0]{
  \begin{tikzpicture}[baseline=(current  bounding  box.center), scale = 0.5]
    \ddwtwostepcode
    \draw (-0.05,1.732+0.5)--++(0,-0.5-1/3*1.732)--++(-1,-1/3*1.732)--++(0,-2/3*1.732)--++(1,-1/3*1.732)--++(0,-0.5-1/3*1.732) coordinate (A);
    \draw (0.05,1.732+0.5)--++(0,-0.5-0.05-1/3*1.732)--(0,0.616*1.732);
    \draw (0,0.616*1.732)--++(210:2/3*1.65)--++(0,-2/3*1.55)--(0,-0.616*1.732)--++(0.05,-0.05/1.732)--([shift=({0.1,0})]A);
  \end{tikzpicture}
}

\newcommand{\ddwtwosteplr}[0]{
  \begin{tikzpicture}[baseline=(current  bounding  box.center), scale = 0.5]
    \ddwtwostepcode
    \draw (-0.05,1.732+0.5)--++(0,-0.5-1/3*1.732)--++(-1,-1/3*1.732)--++(0,-2/3*1.732)--++(1,-1/3*1.732)--++(0,-0.5-1/3*1.732);
    \draw (0.05,1.732+0.5)--++(0,-0.5-1/3*1.732)--++(1,-1/3*1.732)--++(0,-2/3*1.732)--++(-1,-1/3*1.732)--++(0,-0.5-1/3*1.732);
  \end{tikzpicture}
}

\newcommand{\ddwtwosteprl}[0]{
  \begin{tikzpicture}[baseline=(current  bounding  box.center), scale = 0.5]
    \ddwtwostepcode
    \draw (-0.05,1.732+0.5)--++(0,-0.5-1/3*1.732)--++(1,-1/3*1.732)--++(0,-2/3*1.732)--++(-1,-1/3*1.732)--++(0,-0.5-1/3*1.732);
    \draw (0.05,1.732+0.5)--++(0,-0.5-1/3*1.732)--++(-1,-1/3*1.732)--++(0,-2/3*1.732)--++(1,-1/3*1.732)--++(0,-0.5-1/3*1.732);
  \end{tikzpicture}
}

\newcommand{\ddwtwosteprr}[0]{
  \begin{tikzpicture}[baseline=(current  bounding  box.center), scale = 0.5]
    \ddwtwostepcode
    \draw (0.05,1.732+0.5)--++(0,-0.5-1/3*1.732)--++(1,-1/3*1.732)--++(0,-2/3*1.732)--++(-1,-1/3*1.732)--++(0,-0.5-1/3*1.732) coordinate (A);
    \draw (-0.05,1.732+0.5)--++(0,-0.5-0.05-1/3*1.732)--(0,0.616*1.732);
    \draw (0,0.616*1.732)--++(-30:2/3*1.65)--++(0,-2/3*1.55)--(0,-0.616*1.732)--++(-0.05,-0.05/1.732)--([shift=({-0.1,0})]A);
  \end{tikzpicture}
}

\newcommand{\tdwlll}[0]{
    \begin{tikzpicture}[baseline=(current  bounding  box.center), scale = 0.5]
      \draw (0,0)--(-1,1.732)--(1,1.732)--cycle;
      \draw (-0.1,1.732+0.5)--++(0,-0.45-1/3*1.732)--++(210:1.0);
      \draw (0,1.732+0.5)--++(0,-0.5-1/3*1.732)--++(210:1.077);
      \draw (0.1,1.732+0.5)--++(0,-0.55-1/3*1.732)--++(210:1.154);
      \node () at (-1.1,0) {};
      \node () at (1.1,0) {};
    \end{tikzpicture}
}

\newcommand{\tdwllr}[0]{
    \begin{tikzpicture}[baseline=(current  bounding  box.center), scale = 0.5]
      \draw (0,0)--(-1,1.732)--(1,1.732)--cycle;
      \draw (-0.1,1.732+0.5)--++(0,-0.45-1/3*1.732)--++(210:1.0);
      \draw (0,1.732+0.5)--++(0,-0.5-1/3*1.732)--++(210:1.077);
      \draw (0.1,1.732+0.5)--++(0,-0.45-1/3*1.732)--++(-30:1.0);
      \node () at (-1.1,0) {};
      \node () at (1.1,0) {};
    \end{tikzpicture}
}

\newcommand{\tdwlrr}[0]{
    \begin{tikzpicture}[baseline=(current  bounding  box.center), scale = 0.5]
      \draw (0,0)--(-1,1.732)--(1,1.732)--cycle;
      \draw (-0.1,1.732+0.5)--++(0,-0.45-1/3*1.732)--++(210:1.0);
      \draw (0,1.732+0.5)--++(0,-0.5-1/3*1.732)--++(-30:1.077);
      \draw (0.1,1.732+0.5)--++(0,-0.45-1/3*1.732)--++(-30:1.0);
      \node () at (-1.1,0) {};
      \node () at (1.1,0) {};
    \end{tikzpicture}
}

\newcommand{\tdwrrr}[0]{
    \begin{tikzpicture}[baseline=(current  bounding  box.center), scale = 0.5]
      \draw (0,0)--(-1,1.732)--(1,1.732)--cycle;
      \draw (-0.1,1.732+0.5)--++(0,-0.55-1/3*1.732)--++(-30:1.154);
      \draw (0,1.732+0.5)--++(0,-0.5-1/3*1.732)--++(-30:1.077);
      \draw (0.1,1.732+0.5)--++(0,-0.45-1/3*1.732)--++(-30:1.0);
      \node () at (-1.1,0) {};
      \node () at (1.1,0) {};
    \end{tikzpicture}
}

\newcommand{\regvert}[0]{
    \begin{tikzpicture}[baseline=(current  bounding  box.center), scale = 0.5]
      \draw (0,0)--(-1,1.732)--(1,1.732)--cycle;
      \draw (0.05,1.732+0.5)--++(0,-0.5-1/3*1.732)--++(1,-1/3*1.732);
      \draw (-0.05,1.732+0.5)--++(0,-0.5-1/3*1.732)--++(-1,-1/3*1.732);
      \draw (0,0.616*1.732)--++(210:2/3*1.65) coordinate (A);
      \node[left] () at (A) {\footnotesize $(12)$};
      \draw (0,0.616*1.732)--++(-30:2/3*1.65) coordinate (B);
      \node[right] () at (B) {\footnotesize$(34)$};
      \node[above] () at (0,1.732+0.5) {\footnotesize $(12)(34)$};
      \node () at (-1-0.3,-1/1.732) {\footnotesize $\alpha$};
      \node () at (1+0.65,-1/1.732+0.1) {\footnotesize $\alpha^{-1}$};
      \draw[dashed] (A)--++(0,-2/3*1.55)--(0,-0.616*1.732);
      \draw[dashed] (B)--++(0,-2/3*1.55)--(0,-0.616*1.732);
      \node () at (-1.1,0) {};
      \node () at (1.1,0) {};
    \end{tikzpicture}
}

\newcommand{\ugate}[2]{
  \draw (#1,#2)--++(0,1.5);
  \draw (#1+1,#2)--++(0,1.5);
  \fill[blue!50] (#1,#2+0.25) rectangle (#1+1,#2+1.25);
}

\newcommand{\oneP}[0]{
  \begin{tikzpicture}[baseline=(current  bounding  box.center), scale = 0.5]
    \stvert{0}{0};
    \node () at (-.1,0) {};
    \node () at (0.5,0) {};
  \end{tikzpicture}
}

\newcommand{\stackP}[0]{
  \begin{tikzpicture}[baseline=(current  bounding  box.center), scale = 0.5]
    \stvert{0}{0};
    \stvert{0}{1.5};
    \node () at (-.1,0) {};
    \node () at (0.5,0) {};
  \end{tikzpicture}
}

\begin{abstract}

We map the dynamics of entanglement in random unitary circuits, with finite on-site Hilbert space dimension $q$, to an effective classical statistical mechanics, 
 and develop general diagrammatic tools for calculations in random unitary circuits.
We demonstrate explicitly the emergence of a `minimal membrane' governing entanglement growth, which in 1+1D is a directed random walk in spacetime (or a variant thereof).
Using the  replica trick to handle the logarithm in the definition of the $n$th R\'enyi entropy $S_n$, we map the calculation of the entanglement after a quench to a problem of interacting random walks.  A key role is played by effective classical spins (taking values in a permutation group) which distinguish between different ways of pairing spacetime histories in the replicated system.  For the second R\'enyi entropy, $S_2$, we are able to take the replica limit explicitly. This gives a mapping between entanglement growth and a directed polymer in a random medium at finite temperature (confirming Kardar-Parisi-Zhang (KPZ) scaling  for entanglement growth in generic noisy systems).
We find that the entanglement growth rate (`speed') $v_n$ depends on the R\'enyi index $n$, and we calculate $v_2$ and $v_3$ in an expansion in the inverse local Hilbert space dimension, $1/q$.
These rates are determined by the free energy of a random walk, and of a bound state of two random walks, respectively, and include contributions of `energetic' and `entropic' origin.
We give a combinatorial interpretation of the Page-like subleading corrections to the entanglement at late times and discuss the dynamics of the entanglement close to and after saturation.
We briefly discuss the application of these insights to time-independent Hamiltonian dynamics.

 \end{abstract}

\maketitle

\section{Introduction}

To understand nonequilibrium dynamics in generic quantum many-body systems, we need models that are analytically tractable but which are {not} integrable. 
Randomness is a key tool for constructing such models, even if our aim is ultimately to learn about systems that are not random. This philosophy is familiar from random matrix theory 
\cite{mehta2004random, haake2013quantum,tao_topics_2012}, as well as more recent examples like the Sachdev-Ye-Kitaev model \cite{sachdev_gapless_1993,maldacena_comments_2016}.
Random unitary circuits \cite{oliveira2007generic,vznidarivc2008exact,hamma2012quantum,brandao2016local,hosur_chaos_2016,nahum_quantum_2017,nahum_operator_2017,von_keyserlingk_operator_2017,banchi2017driven} are minimal models for chaotic quantum evolution.
They retain two fundamental features of  realistic systems,
namely unitarity and spatial locality, but dispense with any other structure:
 The interactions (between spins or qubits) are taken to be random in both space and time. 
Randomizing the interactions yields models that are analytically tractable 
to a large extent despite being nonintegrable.
These models offer the hope of revealing universal `hydrodynamic' structures 
that are shared by a broad class of many-body systems.

The entanglement entropy is a fundamental quantity whose dynamics remains nontrivial even in the absence of any conventional hydrodynamic modes \cite{calabrese2005evolution,kim2013ballistic,liu2014entanglement,kaufman2016quantum,asplund2015entanglement,casini2016spread,hosur_chaos_2016,ho2017entanglement,nahum_quantum_2017,mezei2017entanglement,mezei2017entanglement2,gu_spread_2017,jonay_coarse-grained_2018,mezei2018membrane}.
Spatially local random circuits have  led to long-wavelength dynamical equations for entanglement production \cite{nahum_quantum_2017,jonay_coarse-grained_2018}
and also for operator spreading \cite{nahum_operator_2017,von_keyserlingk_operator_2017}, i.e. for the `quantum butterfly effect' \cite{lieb_robinson,kitaev2014hidden,shenker_black_2014, shenker2014multiple,maldacena_bound_2015, roberts_two-dimensional_2015,roberts2015localized,roberts_lieb-robinson_2016,aleiner2016microscopic,patel2017quantum,chowdhury2017onset,bohrdt2017scrambling,li2017measuring,garttner2017measuring}
 in spatially local systems.
They have also elucidated effects of conserved quantities \cite{rakovszky_diffusive_2017,khemani_operator_2017} and quenched disorder \cite{nahum_dynamics_2017} on the spreading of quantum information. 
Very recently exact results have also been obtained for the dynamics of random Floquet circuits and related models \cite{chan_solution_2017,kos2017many,chan2018spectral}.
Here we will be interested in universality associated with the propagation of information through space, so we consider spatially local circuits, but we note that interesting lessons have been learned even from `zero-dimensional' random circuits in which any qubit can couple with any other \cite{hayden2007black,sekino2008fast,harrow2009random,dankert2009exact,lashkari2013towards,shenker2015stringy}.

In this paper we establish general tools for calculating dynamical observables in random unitary circuits and apply these tools to the R\'enyi entanglement entropies after a quench. 
We construct mappings between dynamical observables and a hierarchy (labelled by a replica-like index) of effective classical models involving permutations as effective spins (generalizing the mapping of \cite{nahum_operator_2017}). 
In particular we show  how the `entanglement membrane'   \cite{nahum_quantum_2017,jonay_coarse-grained_2018} emerges from these classical models under coarse-graining. 
These mappings go beyond previous simplifying limits (in particular the limit of infinite local Hilbert space dimension, $q=\infty$). They reveal new universal phenomena, such as a phase transition in the entanglement membrane, and new  combinatorial structures.

\begin{figure}[t]
\centering
\includegraphics[width=.92\linewidth]{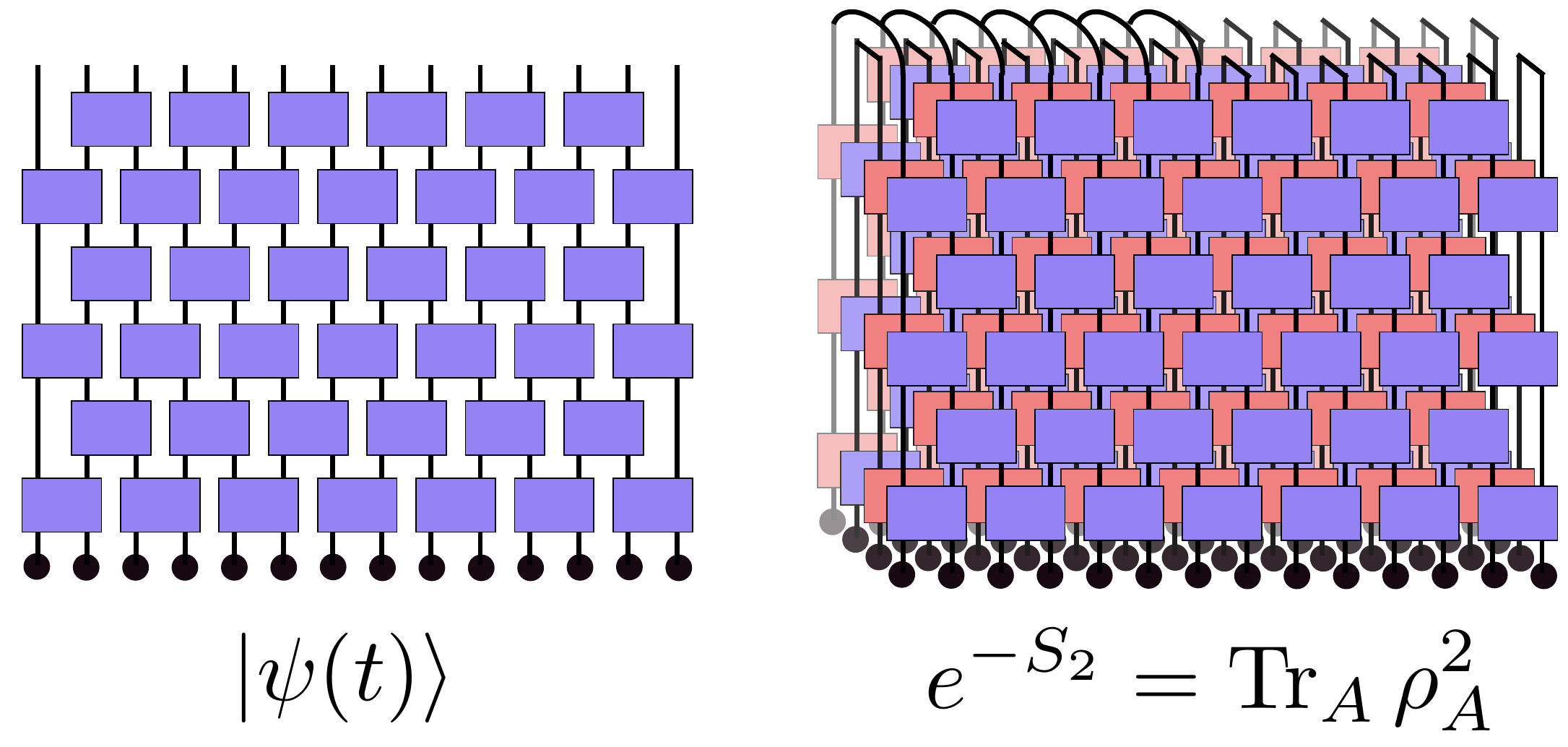}
\caption{Left: the time-evolved wavefunction represented as a unitary circuit (schematic).
Time runs vertically.
The circles at the bottom represent an initial product state.
The legs on the top carry the wavefunction's spin indices.
Right: the quantities we study (e.g. the second R\'enyi entropy, shown) are contractions of powers of the circuit (blue) and its conjugate (red). The replica trick requires $k$ copies of this `stack' (not shown).}
\label{fig:unitarystackfigure1}
\end{figure}

The coarse-grained picture for the growth of entanglement conjectured in \cite{nahum_quantum_2017,jonay_coarse-grained_2018} involves a `minimal membrane' in spacetime. (This picture has now been shown to apply in holographic conformal field theories \cite{mezei2018membrane}.)
 In 1+1D, the `membrane' is a one-dimensional path in two-dimensional spacetime that is characterized by a velocity $v(t)=\dd x/ \dd t$  (its slope in spacetime). It has an `entanglement line tension' $\mathcal{E}(v)$ that depends on this velocity \cite{jonay_coarse-grained_2018}. Leading-order entanglement calculations reduce to a classical optimization of the line tension for this path, which can be thought of as minimizing the free energy of a `polymer'.
 
 In the simplifying limit $q\rightarrow\infty$ for the local Hilbert space dimension $q$
 (and in a certain model)
 the polymer can be identified \cite{nahum_quantum_2017} with a coarse-grained `minimal cut' \cite{swingle2012entanglement, pastawski2015holographic, casini2016spread, hayden2016holographic} 
 through the unitary circuit generating the dynamics.
 For finite $q$, the computation of the circuit--averaged purity, $\overline{e^{-S_2}}$, leads to a related directed walk problem \cite{nahum_operator_2017}. 
 But for finite $q$ the minimal cut formula is no longer accurate, and the calculation of $\overline{S_n}$ is complicated by the need for a replica limit to handle the logarithm in the definition of the entropy (since e.g. ${\overline{S_2} \neq - \ln \overline{e^{-S_2}}}$).
It is important to tackle this in order to derive the membrane picture from explicit microscopic calculations at finite local Hilbert space dimension $q$.

Concretely, we take the time evolution operator (quantum circuit) to be a regular array of Haar-random two-site unitaries as shown in Fig.~\ref{fig:unitarystackfigure1} (Left).
This is quantum evolution with no conserved quantities, which equilibrates locally to the infinite temperature state.

\begin{figure}[t]
\centering
\includegraphics[width=.97\linewidth]{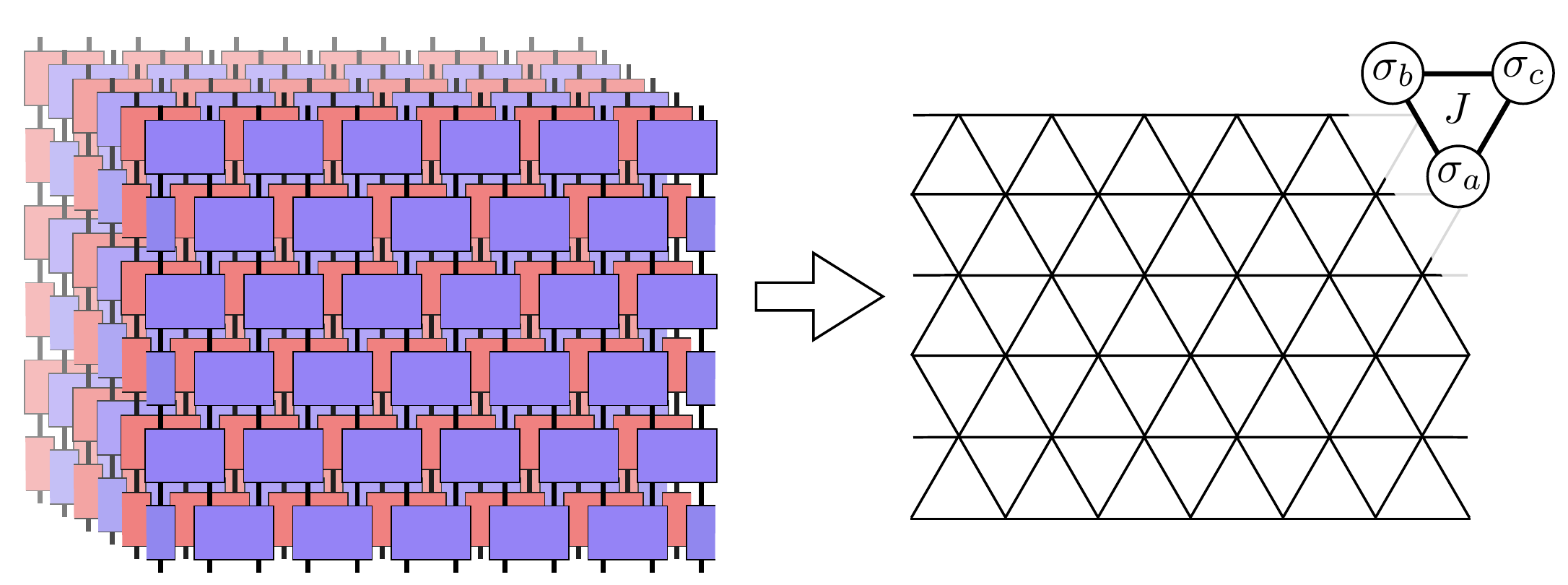}
\caption{By averaging a `stack' with $N$ layers of $U(t)$ and $N$ layers of $U(t)^*$ (Left) we obtain an effective classical model of interacting spins (Right) with three-body interactions. These spins take values in the permutation group $S_N$. The important configurations involve domain walls between different spin values, see e.g. Fig.~\ref{fig:bd_state}.}
\label{fig:unitarystack}
\end{figure}

In our mappings the minimal membrane arises from domain walls between two kinds of permutations.
These permutations appear in the average over the random unitaries in the circuit:  similar permutational degrees of freedom appear for random  tensor networks that are not made of unitaries {\cite{hayden_holographic_2016,qi_space-time_2018}}.
Mathematically the permutations represent different patterns of index contractions. 
For unitary dynamics these permutations can be understood more physically as distinct ways of pairing spacetime trajectories in the `path integral' for (say) the R\'enyi entropy, which involves multiple copies of the system (Fig.~\ref{fig:unitarystack}) \cite{nahum_operator_2017}. We expect this idea to be more generally applicable.

The domain walls between permutations can be viewed as a collection of interacting, directed random walks, with interactions of several kinds. The entanglement is related to the free energy of these walks (in the language of the classical problem) and has both energetic and entropic contributions. 
We obtain the time-dependence of the entropies in systematic expansions in $1/q$ 
(accounting for both the mean behaviour and fluctuations)
and we show how the late-time saturation can be understood in the minimal curve picture. 
We also briefly discuss the operator entanglement of the time evolution operator itself \cite{zanardi_entangling_2000,  zhou_operator_2017, prosen_operator_2007, dubail_entanglement_2016}.

\begin{figure}[b]
\centering
\includegraphics[width=.98\linewidth]{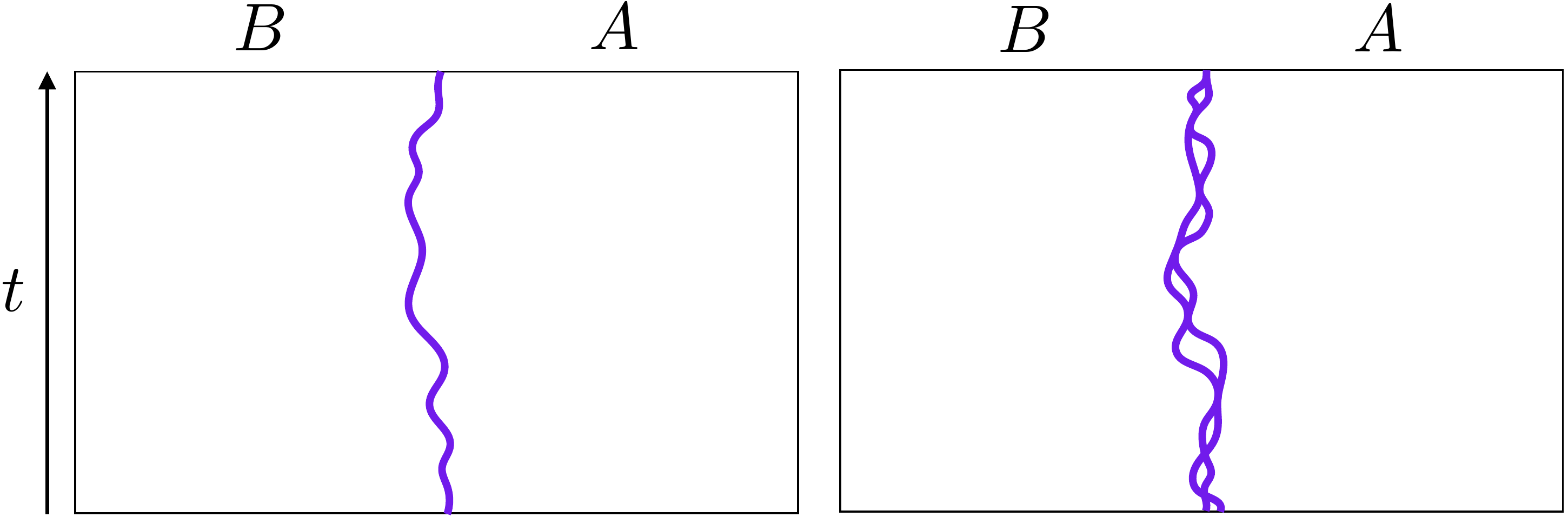}
\caption{In the calculation of $S_2$ for $n=2$ (cartoon on Left) each replica contains a single elementary walk (domain wall).
For $n>2$, e.g. $n=3$ (Right),  each replica contains multiple walks, which can form a `bound state' with a finite typical width.
To calculate averages involving $S_n$ we must use $k$ replicas (which multiplies the number of elementary walks by $k$) and take $k\rightarrow 0$ at the end of the calculation.}
\label{fig:bd_state}
\end{figure}

In more detail: For computing $\overline{e^{-S_2}}$ it is sufficient to consider a single walk which represents an `elementary' domain wall (Fig.~\ref{fig:bd_state}, Left). Collections of multiple interacting walks appear for two reasons which we describe below. In this paper we focus largely on the R\'enyi entropies with $n\geq 2$.

First, if we consider a higher R\'enyi entropy, the relevant domain wall is in fact a composite of $(n-1)$ `elementary' domain walls, i.e. $(n-1)$  walks (see also a similar picture in a Floquet circuit \cite{chan_solution_2017}). These attract each other strongly through a combinatorial mechanism and can form a `bound state' (Fig.~\ref{fig:bd_state}, Right).
In the continuum this bound state forms the minimal membrane.
 There is also an \textit{unbinding} phase transition for these walks as a function of their velocity: this unbinding is important in allowing general constraints conjectured in \cite{jonay_coarse-grained_2018}, relating entanglement growth to the butterfly velocity $v_B$, to be satisfied.

Second, to compute, say, $\overline{S_n}$ or the fluctuations in $S_n$, we must employ the replica trick.
We consider a $k$-fold replicated system and take the limit $k \rightarrow 0$ at the end of the calculation. There are then $k$ sets of domain walls, one for each replica. Distinct sets interact with a weak interaction that we compute by expanding in $1/q$.

This replica treatment allows us to pin down universal fluctuations in the entanglement that are due to randomness in the circuit. 
It was argued that for dynamics that is random in time these fluctuations are governed by the Kardar--Parisi--Zhang (KPZ) universality class \cite{kardar_dynamic_1986,HuseHenleyFisherRespond},
and have a magnitude that grows in time as $t^{1/3}$  \cite{nahum_quantum_2017}.
(These fluctuations are therefore subleading at large time compared to the leading order deterministic growth.)
We confirm these universal properties by an explicit mapping between the dynamics of the R\'enyi entropies and a problem that is equivalent to KPZ, namely the problem of a directed polymer in a random medium (at finite temperature \cite{huse_pinning_1985,KardarRoughening,HuseHenleyFisherRespond,kardar_dynamic_1986,kardar_replica_1987}).

Strikingly, the replica limit needed to handle the logarithm in the definition of the entanglement entropy  
is transmuted by this mapping into the replica limit associated with the disorder in the {classical} polymer problem.
As a result, the mapping to the directed polymer in a random medium can be carried through exactly on the lattice when $q$ is large but finite.
This polymer can be coarse-grained to give the exact leading $q$ dependence of the constants in the \textit{continuum} KPZ equation describing the entanglement growth.
At large $q$ there are several early-time crossovers in the entanglement growth. In fact  the timescale required to see KPZ fluctuations is numerically large even at $q=2$: we suggest that this is why quantum simulations of this model at short times did not show signatures of KPZ \cite{von_keyserlingk_operator_2017}, resolving an apparent paradox.

\section{Overview of results}
\label{sec:overview}

An appealing feature of random circuits is the possibility of useful quantum--classical mappings for \textit{real time} (as opposed to imaginary time) dynamics.  The R\'enyi entropies illustrate these mappings. 

\begin{figure}[t]
\centering

\includegraphics[width=0.8\columnwidth]{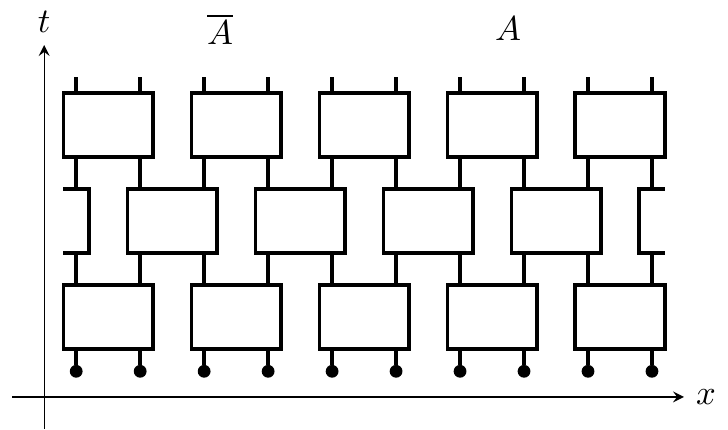}

 \caption{The structure of the random circuit.
Each two-site random gate is shown as a four-leg block. $x > 0$ corresponds to region A. Time evolution is going upward.}
\label{fig:circuit}
\end{figure}

Let us discuss the generation of entanglement after a `quench' from an initial product state (we will discuss some other setups later on).
For simplicity, take the chain to be infinite. The dynamics is generated by a random circuit. 
The time evolution operator $U(t)$ is made up $t$ `layers' of two-site random unitaries, each independently Haar-random, with unitaries applied to even bonds in even layers and odd bonds in odd layers: see Fig.~\ref{fig:circuit}.

Let $A$ denote the half-chain with $x>0$, and let $S_n(t)$ be the $n$th R\'enyi entropy of this region at time $t$. We define
\begin{equation}
Z_n  \equiv \Tr \rho_A^n = e^{-(n-1) S_n},
\end{equation}
where the $t$ dependence is implicit. 

The physical quantities of interest to us are averages such as $\overline{S}_n$ --- where the average is over the random unitaries in the circuit --- and also fluctuations around these averages. To obtain $\overline{S}_n$ we must average the logarithm of $Z_n$, since in general $\overline{\exp(- \alpha S_n)}\neq \exp(-\alpha \overline{S_n})$.  For this we will use the replica trick, studying the average of the $k$th power of $Z_n$ for an arbitrary integer number of `replicas' $k$, and then taking the formal limit $k\rightarrow 0$. 

The average entanglement is given by
\begin{equation}
\overline{S}_n = - \frac{1}{n-1} \left. \frac{\partial  \znk }{\partial k}\, \right|_{k = 0}
\end{equation}
and higher terms in the expansion about $k=0$ yield higher cumulants which quantify the fluctuations in the entanglement,
\begin{equation}
\label{eq:higher-moments}
\ln \overline{Z_n^{\,k}} = - k (n-1) \,\overline{S}_n + \f{k^2(n-1)^2}{2}  \, \overline{\lf S_n-\overline{S}_n\ri^2} + \ldots.
\end{equation} 

We will give a brief overview of the general features of this replica calculation of the entropies in Sec.~\ref{subsec:generalfeatures}. Then in the remainder of this section we summarize our basic results for the entanglement. We divide these into two classes. 
First the  \textit{leading order dynamics} of the entanglement entropy at large times (Sec.~\ref{subsec:overview_leading}). This leading order dynamics is deterministic, despite the randomness in the circuit.
Second, \textit{subleading fluctuations} arising from randomness in the circuit (Sec.~\ref{subsec:overview_kpz}). Although these fluctuations are subleading at large time, they have interesting universal structure. 

To clarify the distinction, consider the above example of $S_n(t)$ for an initial product state. When $t\gg 1$ the leading order behaviour is deterministic growth at a rate set by an `entanglement speed' $v_n$; we find this rate to be $n$-dependent. Randomness consists in subleading fluctuations, which obey KPZ scaling \cite{nahum_quantum_2017}, and are on the parametrically smaller scale $t^{1/3}$.

We may write $S_n(t)$ as
\begin{equation}
\label{eq:scaling_form}
S_n(t) \simeq \seq \,  \big[\,
v_n t 
+ 
B_n t^{1/3} \chi(t)\,
\big].
\end{equation}
The first factor is the equilibrium entropy density $\seq$: since the models we study have no conservation laws, they equilibrate locally to the infinite temperature state, and $\seq$ is set by the local Hilbert space dimension,
\begin{equation}
\seq = \ln q.
\end{equation} 
The first term inside the brackets in (\ref{eq:scaling_form}) is the deterministic leading order growth (the deterministic growth will have nontrivial time-dependence for e.g. a more general initial state, or for the entanglement of a finite region).
 The second term includes the KPZ fluctuations of size $t^{1/3}$. $B_n$ is the nonuniversal constant governing their strength. $\chi(t)$ is a random variable whose magnitude is of order 1 at late times, whose probability distribution is universal and given by the Tracy-Widom distribution $F_1$ \cite{CalabreseLeDoussalRosso2010,Dotsenko2010,SasamotoSpohn2010A,SasamotoSpohn2010B,SasamotoSpohn2010C,GideonCorwinQuastel2011,CalabreseLeDoussal2011,ProlhacSpohn2011,LeDoussalCalabrese2012,ImamuraSasamoto2012,ImamuraSasamoto2013,KriecherbauerKrug,CorwinReview,Halpin-Healy2015}.

We will obtain Eq.~\ref{eq:scaling_form} by explicit calculations at large but finite $q$ (therefore, modulo the fact that the replica calculation is nonrigorous, we confirm the conjectures of Ref.~\cite{nahum_quantum_2017} for the universal properties, also fixing various nonuniversal constants) and we will discuss various extensions.

 \subsection{General features of the mappings}
\label{subsec:generalfeatures}

We map $\znk$  to an effective classical statistical mechanics problem for interacting random walks which has some striking features, including nontrivial ``Feynman rules'' for allowed fusions of paths.
This significantly extends the mapping of Ref.~\cite{nahum_operator_2017} for the case $\overline{Z_2}$. We will use an expansion in powers of $1/q$ to obtain analytical control on the interaction constants in the effective classical problem, but we expect the resulting \textit{universal} results to hold for all $q$, including the minimal value $q=2$. In fact even our large $q$ results for various \textit{nonuniversal} constants should be reasonably close for small $q$, because of the numerical smallness of various constants.

The mapping involves several steps.
We first map $\znk$ to a partition function for a `classical magnet'. The `spins' in this magnet take values in a permutation group, as in work on random tensor networks \cite{hayden_holographic_2016}. The group relevant to us is $S_N$ for $N=n\times k$, as a result of the replica limit. Eventually we must consider the limit $k\rightarrow 0$. 

The interactions in the classical magnet are initially rather complicated but permit simplifications. In Ref.~\cite{nahum_operator_2017} it was shown that in the special case $n=2$, $k=0$ the partition function could be radically simplified by integrating out half of the spins.
We extend this idea to general $n$ and $k$.  This allows a much richer set of configurations, and the Boltzmann weight for a general configuration in the effective classical model remains complicated. However these Boltzmann weights obey crucial simplifying constraints, due to the \textit{unitarity} of the underlying quantum dynamics, which imply that many spin configurations do not contribute to the partition function.
We exploit these constraints, together with a large $q$ expansion, to reduce the partition function to one for multiple directed paths with interactions of various kinds. 

These paths arise as domain walls in the classical magnet. They may be viewed as living on a rotated square lattice, and they are directed in the time direction.
Each domain wall carries a label, analogous to a particle type.
This label is an element of $S_N$, which in the simplest case (an `elementary' domain wall) is a transposition such as $(12)$. We explain this structure in Sec.~\ref{sec:lattice_magnet}.

\begin{figure}[t]
\centering
\includegraphics[width=0.8\columnwidth]{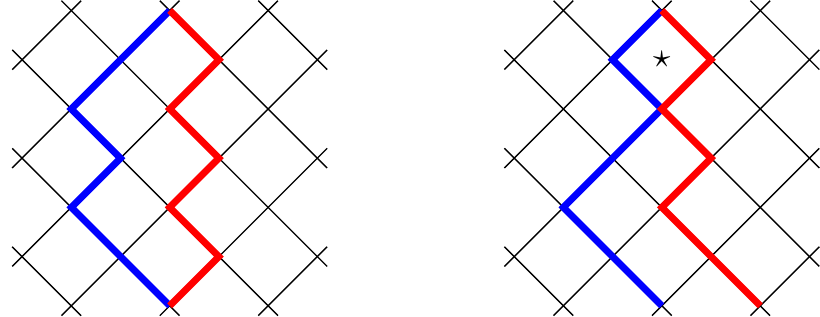}
\caption{ 
{
$\overline{Z_2^k}$ maps to $k$ directed walks on the tilted square lattice, with attractive interactions of order $1/q^4$ (vertical direction is physical time). The figure shows $k = 2$. Left: This configuration of the walks has no interaction contribution at this order. Right: In this configuration there is an interaction contribution from the square with a star, see Sec.~\ref{sec:kpz}.
}
}
\label{fig:walk_on_lattice}
\end{figure}

For $\overline{Z_2^k}$, which yields the second R\'enyi entropy, we obtain a partition function for $k$ directed paths, one for each replica: see Fig.~\ref{fig:walk_on_lattice}. There is an effective local attractive interaction between different replicas (different paths). This attraction is parametrically small when $q$ is large (of order $q^{-4}$). 

The problem of $k$ directed paths or `polymers' with attractive interactions, in the replica limit $k\rightarrow 0$, is a well-known one \cite{kardar_replica_1987,kardar_directed_1994,dotsenko2010bethe,calabrese2010free,calabrese_exact_2011}. It is the replica description of a single directed polymer in a random potential \cite{huse_pinning_1985,KardarRoughening,HuseHenleyFisherRespond}, a model which can be mapped to KPZ.
Therefore this sequence of mappings relates the universal properties of the entanglement to those of the directed polymer in a random potential. At large but finite $q$ it is even possible to make an explicit \textit{microscopic} correspondence with a specific lattice model for directed polymer in a random medium.

The entanglement $S_2(t)$ is the free energy of the polymer: the growth rate of the entanglement has both an `energetic' and an `entropic' contribution.
In addition to addressing the universal properties, we calculate some nonuniversal growth rates associated with $S_2$ by applying exact Bethe ansatz results for directed polymers in the continuum \cite{kardar_replica_1987,calabrese_exact_2011}.

The statistical mechanics problem becomes more intricate when $n>2$. Each replica now contributes $n-1$ `polymers'. Within each replica there are interactions which are \textit{not} small at large $q$. These interactions have an appealing combinatorial origin. We give some exact results for $n=3$ and a schematic picture for general $n$. 
For $n>2$, the interactions lead to the formation of a `bound state' of multiple walks, see Fig.~
\ref{fig:bd_state} for a cartoon.

Above we have focused on the mapping for the entanglement of quenched state. However many other quantities can be studied using our lattice magnet mapping. We  briefly discuss the operator entanglement entropy \cite{zanardi_entanglement_2001, bandyopadhyay_entangling_2005, prosen_operator_2007,pizorn_operator_2009,prosen_chaos_2007,zhou_operator_2017, dubail_entanglement_2016} of the  time-evolution operator (i.e. of the whole unitary random tensor network). This is obtained from the  same lattice magnet partition function, just with slightly different boundary conditions. This quantity may also be used to obtain the entanglement line tension Sec.~\ref{subsec:line_tension}. 

 \subsection{Entanglement: leading scaling at large $t$}
\label{subsec:overview_leading}

The leading order dynamics of $S_n(t)$ at large $t$ is deterministic, regardless of the value of $q$ --- this is why random circuits are reasonable minimal models for entanglement dynamics in realistic many-body systems with non-random Hamiltonians.
This deterministic dynamics extends beyond the simple linear-in-$t$ growth in Eq.~\ref{eq:scaling_form}: for example we also expect universal deterministic scaling forms for the saturation of the entanglement of a large finite region \cite{nahum_quantum_2017}, and for entanglement growth starting from a state with a nontrivial entanglement pattern \cite{jonay_coarse-grained_2018}.

The entanglement speeds $v_n$ may be calculated in a large $q$ expansion for $n=2$ and $n=3$. For $v_2$ we find
\begin{equation}
\begin{aligned}
\label{eq:v2}
  v_2 &=  \frac{1}{\ln q } \lf \ln \frac{q^2 + 1}{2q}  \ri +\f{1}{\ln q} \lf \frac{1}{384 q^8 }  + \ldots
   \ri\\
&= \, 1 + \frac{1}{\ln q}  \Bigg( - \ln 2  + \frac{1}{q^2} - \frac{1}{2q^4} \\
&\qquad\quad\quad\quad\,\,\,\,+ \frac{1}{3 q^6} - \frac{95}{384q^8}  + \mathcal{O}\lf \frac{1}{q^{10}} \ri \Bigg). 
\end{aligned}
\end{equation}
For $v_3$ we are only able to go to a lower order:
\begin{align}
\label{eq:v3}
v_3 & = 1 - \frac{\ln \lf 2 + \f{3}{\sqrt{2}} \ri}{2 \ln q} +\frac{3\sqrt{2}}{4} \frac{1}{q^2 \ln q} + \mathcal{O}\lf  \frac{1}{q^4\ln q } \ri.
\end{align}
This shows that the entanglement speed depends in general on $n$, which resolves a question left open in Ref.~\cite{nahum_quantum_2017}; see also Ref.~\cite{chan_solution_2017}.

The above entanglement speeds may be compared with the growth rates defined by averaging $Z_n$ instead of its logarithm, which we denote $\widetilde v_n$:
\begin{equation}
- \f{1}{n-1} \ln \overline{Z_n}\, \sim \,   \seq \widetilde v_n t.
\end{equation}
$\widetilde v_n$ does not include effects due to interactions between replicas. These interactions are suppressed at large $q$, so the difference between $\widetilde v$ and $v$ is parametrically small at large $q$, and for $v_2$ is numerically quite small even when $q=2$.
The average of the purity $\overline{Z_2}$ gives the `purity speed' $\widetilde v_2$ (called $v_P$ in Ref.~\cite{nahum_operator_2017}) which has been obtained previously in a variety of ways \cite{hamma_2012_quantum, nahum_quantum_2017, nahum_operator_2017, von_keyserlingk_operator_2017} 
\begin{equation}
\label{eq:purityspeed}
  \widetilde v_2  =  \frac{1}{\ln q} \left[ \ln  \lf \frac{q^2+1}{q} \ri -  \ln 2 \right] . 
\end{equation}
In the random walk picture,
$\overline{Z_2}$ is the partition function for a single random walk, and
 $\seq \widetilde v_2$ is its free energy per unit `length' in the time direction \cite{nahum_operator_2017}: in the above expression the first term is the energetic contribution to this free energy and the second term is the entropic one. The difference between $\widetilde v_2$ and $v_2$ is 
\begin{equation}
v_2 - \widetilde v_2 \simeq { \frac{1}{384 q^8 \ln q} + \mathcal{O}(\frac{1}{q^{10} \ln q}) } 
\end{equation}
at large $q$, and arises from interactions between replicas that are of order $q^{-4}$.

When we consider $S_3$, each replica contributes two walks, and these walks have effective interactions that are different from the interactions between replicas.
At the leading nontrivial order in $1/q$, the interaction arises for combinatorial reasons. In the simplest case, $\overline{Z_3}$, the walks are labelled by transpositions in the permutation group $S_3$:  either $(12)$, $(23)$, or $(13)$. 
Further, the labels on the two walks must multiply to give the 3-cycle $(123)$. This leaves three possibilities for how the walks are labelled, corresponding to the three ways to decompose the three-cycle:
\begin{equation}
\label{eq:123_split}
(123)= (12)\times (23) = (23)\times(13) = (13)\times(12).
\end{equation}
\begin{figure}[h]
\centering
\includegraphics[width=\columnwidth]{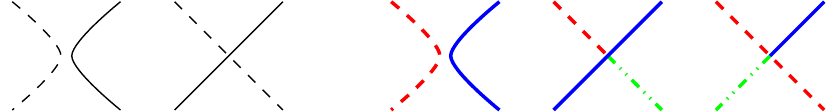}
\caption{Outcome of a splitting event. Left: Two commutative domain walls have two ways to split: exchange or passing through. Right: Two noncommutative domain walls have three ways to split. Red (dashed) represents (12), blue (solid) represents (23), green (dot-dashed) represents (13).}
\label{fig:splitting}
\end{figure}

 Each time the two walks meet, the labelling can change, as in the cartoon in Fig.~\ref{fig:splitting}. For a given spatiotemporal configuration of the domain walls we must therefore sum over all the consistent labellings. The resulting factor in the partition sum may be reinterpreted as a local attractive interaction between the two walks. This causes them to form a bound state: see Fig.~\ref{fig:bd_state} for a cartoon. 
(This phenomenon is similar to one appearing in the replica treatment of directed paths with random-sign weights \cite{kardar_directed_1994,medina_interference_1989}.)
 The attraction means that the constant $\ln 2$ appearing in Eq.~\ref{eq:purityspeed} is replaced with the constant ${(1/2) \ln ( 2  +  {3}/{\sqrt 2})}$ in Eq.~\ref{eq:v3}. As a result, the growth rate of $S_3$ is not the same as that of $S_2$. This combinatorial factor has also been obtained in the Floquet model of Ref.~\cite{chan_solution_2017}.

In the next order, $\mathcal{O}(\frac{1}{q^2})$, there is a weaker interaction arising from the non-commutativity of the constituents of the composite walk $(123)$. Taking it into consideration gives Eq.~\ref{eq:v3} for the growth rate $v_3$.

The entanglement speeds $v_n$ are in fact special cases of the more general quantities $\lt_n(v)$  which determine the entanglement dynamics for more general initial states, in which the entanglement is different across spatial cuts at different positions $x$ \cite{jonay_coarse-grained_2018}. For a detailed discussion of $\lt(v)$ see Ref.~\cite{jonay_coarse-grained_2018}, where it is argued that for typical initial states with a given initial entanglement profile $S(x,0)$, the leading order dynamics is
\begin{equation}
S_n(x,t) = \min_v \lf S_n\lf x-vt, 0\ri + \seq t \times \lt_n (v) \ri.
\end{equation}
The quantity  $\lt_n(v)$ has a transparent meaning in the present approach. $\seq \lt_2(v)$ is the coarse-grained free energy (per unit length) of a walk when its coarse-grained `speed' is equal to $v$ (in the replica limit).
For higher $n$, $(n-1)\times \lt_n(v)$ is the analogous free energy for a `bound state' of $n-1$ walks.
We refer to $\lt_n$ as the slope-dependent line tension. 

The line tension of a walk which is \textit{vertical} in the coarse-grained sense is $\lt_n(0)$, which is simply $v_n$. As noted in Ref.~\cite{jonay_coarse-grained_2018}, the random walk picture gives an explicit form for $\lt_2$ at large $q$. We quantify the size of the corrections to this large $q$ result, and  we also give explicit formulae for $\lt_3(v)$. These forms are consistent with the general constraints \cite{jonay_coarse-grained_2018} $\lt_n(v_B) = v_B$, $\lt'(v_B) = 1$. 

We also find an interesting phenomenon for the R\'enyi entropies $\overline{S_n}$ with $n>2$; there is a nonanalyticity in the line tension $\lt_n(v)$ at the value $v=v_B$. This is associated with an `unbinding' phase transition for the $(n-1)$ walks that appear in the calculation of $S_n$. This phase transition is crucial in allowing the constraint $\lt_n(v_B) = v_B$ to be satisfied.

So far we have discussed the entanglement of an infinite subsystem, which grows indefinitely. We also consider the saturation of the entanglement for a finite subsystem (Sec.~\ref{sec:saturation}). 
At asymptotically late times the entanglement saturates to a value given by Page's formula and its generalizations\cite{page_average_1993,nadal_statistical_2011}. We show that the universal constants appearing in this formula have an appealing combinatorial interpretation in terms of domain walls. We also confirm conjectured scaling forms for entanglement saturation \cite{nahum_quantum_2017}, and we show that at the moment of saturation there are subleading corrections to these scaling forms with a similar combinatorial origin to the constant in the Page formula.

 \subsection{Entanglement: KPZ fluctuations}
\label{subsec:overview_kpz}

The statistical mechanics problem becomes more interesting when interactions between replicas are considered. These interactions encode the fluctuations in the entanglement due to circuit randomness.\footnote{In the absence of interactions between replicas $\overline{Z_n^k}$ would be equal to  $(\overline{Z_n})^k$, so that the generating function $\overline{\exp \lf -k(n-1)S_n \ri}$ would be trivial and equal to ${\exp \lf -k(n-1) \overline{S}_n \ri}$.} They also determine subleading (in $q$)  corrections to $v_n$ and other constants.

Above we reviewed the basic features of KPZ scaling of fluctuations (Eq.~\ref{eq:scaling_form}) with their characteristic $t^{1/3}$ growth.
It was conjectured, on the basis of analytic results in particular limits\footnote{Haar-random circuits with a random geometrical structure at $q=\infty$, and Clifford circuits at $q = 2$.} and numerical results on some more generic circuits, that KPZ scaling should hold in any generic random circuit \cite{nahum_quantum_2017}. 
However until now there has not been an explicit analytic derivation of KPZ scaling in a generic circuit that does not have the simplifying feature of either $q=\infty$ or Clifford structure. 
 
The need for such a demonstration is pressing in the light of the recent numerics for the second R\'enyi entropy in regular Haar random circuits.
Numerical results for ${q=2}$, for times up to $t=20$ in Ref.~\cite{von_keyserlingk_operator_2017} showed no obvious sign of fluctuations growing with time, and it was conjectured there that KPZ fluctuations were absent. 
Here we find that KPZ fluctuations are indeed present, and the reason for their apparent absence at small times is that they are (numerically) surprisingly small. 
 
We focus on $S_2$, where a quantitative calculation is possible for large but finite $q$. We argue that a similar logic implies KPZ fluctuations also for the higher R\'enyi entropies, but with an additional coarse-graining step (we comment briefly on $S_1$).
This gives for the first time an analytic demonstration of KPZ scaling in a random circuit that is not Clifford and which has finite (albeit large) $q$.

\begin{figure}[t]
\centering
\includegraphics[width=0.66\linewidth]{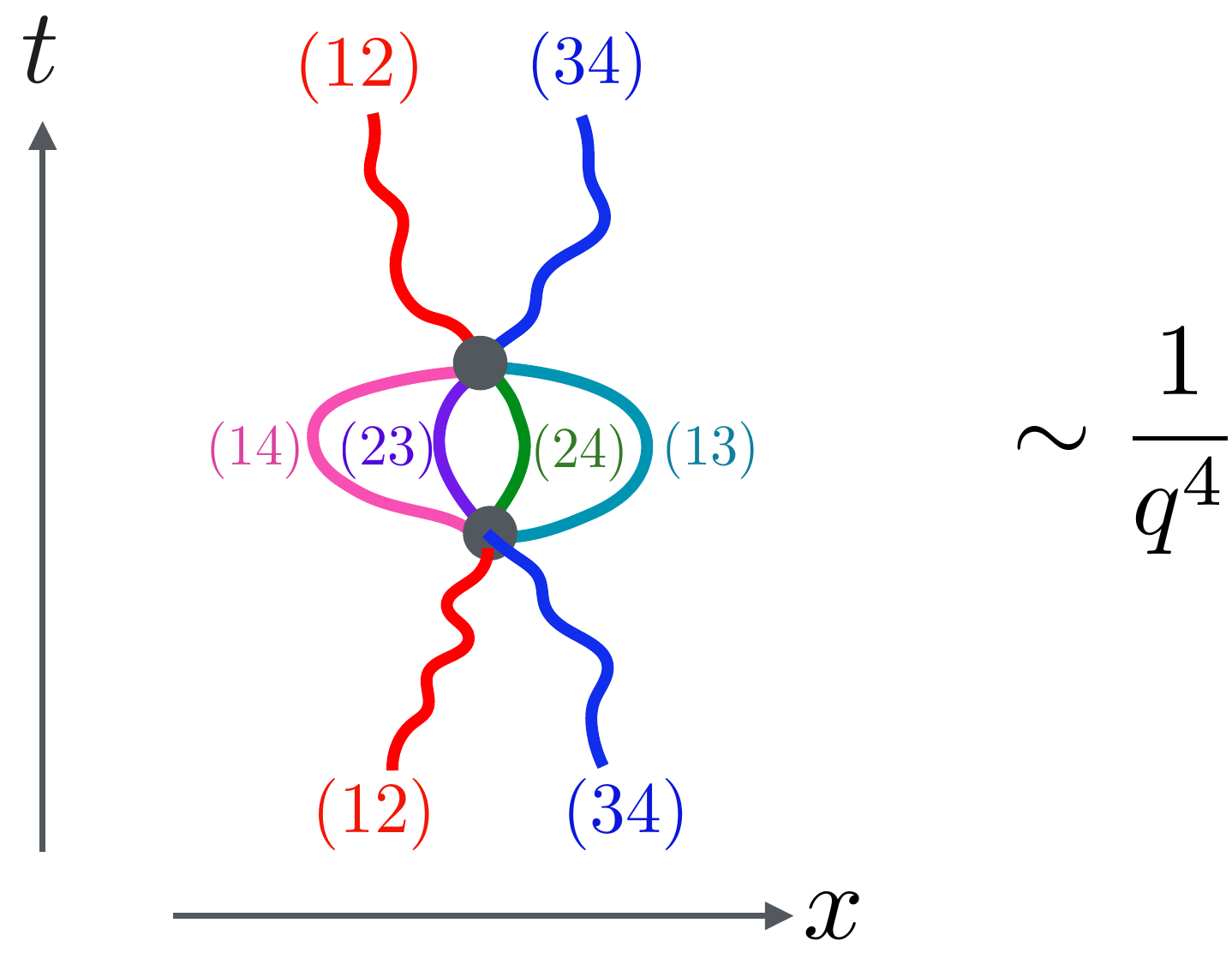}
\caption{At leading order in $q$, interactions between replicas arise from a particular real-space ``Feynman diagram'' in which the worldline's particle types have the combinatorial structure shown. (This continuum figure is only a cartoon: in our calculation, the ``Feynman diagram'' is a particular local configurations of paths on the lattice).}
\label{fig:feynmandiagram}
\end{figure}

For $S_2$, we find that the prefactor $B_2$ in Eq.~\ref{eq:scaling_form} governing the strength of fluctuations at asymptotically late times is (at large $q$)
\begin{equation}
 \label{eq:Bfmla}
B_2 \simeq - \f{1}{4\times 2^{1/3}\times q^{8/3}}.
\end{equation}
We obtain this using the directed polymer mapping together with the fact that at large $q$ the weakness of the interactions between replicas can be used to justify a continuum treatment. In this continuum treatment each polymer is interpreted as the worldline of a boson, so we have a problem of $k\rightarrow 0$ interacting bosons\cite{kardar_replica_1987}. The constants in their Hamiltonian are
\begin{equation}
  { \mathcal{H} =  k \ln \frac{q^2 + 1}{2 q} - \frac{1}{2} \sum_{ \alpha = 1}^k \frac{\partial^2}{\partial x_{\alpha}^2 } - \frac{1}{4 q^4} \sum_{ \alpha < \beta} \delta( x_\alpha - x_\beta ), }
\end{equation}
where the interaction arises from certain domain wall configurations that are absent in the $q=\infty$ limit. 
These configurations involve a ``Feynman diagram'' (in spacetime, not momentum space) whose combinatorial structure is shown schematically in Fig.~\ref{fig:feynmandiagram}.

We use numerical simulations for small $t$ to check various diagramatic calculations that go into the directed polymer mapping. While Eq.~\ref{eq:Bfmla} is valid at asymptotically late times, these small $t$ simulations are consistent with the weakness of interactions between replicas being the reason why KPZ growth of fluctuations cannot be seen, even for $q=2$, on timescales accessible using MPS techniques.

So far we have discussed only the Haar random quantum circuit with a fixed regular geometry. Perhaps the simplest modification of this circuit is to draw each local unitary from a modified probability distribution, which returns a Haar random 2-site unitary with probability $1-p$, and the 2-site identity with probability $p$. That is, we punch a density $p$ of holes in the circuit. In the limit of small $1-p$ this gives (after an appropriate rescaling of time) the model in which unitaries are applied in continuous time in a Poissonian fashion \cite{nahum_quantum_2017}. The strength of attraction between replicas --- the strength of disorder in the directed polymer language --- varies with $p$.  If $p$ is nonzero, there are nontrivial KPZ fluctuations  even in the strict $q=\infty$ limit \cite{nahum_quantum_2017}.

\section{Mapping to a `lattice magnet'}
\label{sec:lattice_magnet}

In this section we map the average of multiple copies of the circuit (and its conjugate) to a 'lattice magnet'. 
We will focus on the quantity
\begin{equation}\label{eq:startingpoint}
Z_n^k = \lf \Tr \rho_A^n \ri^k,
\end{equation}
where $\rho_A=\rho_A(t)$ is the reduced density matrix for a region $A$ in a chain that is globally in a pure state.
However the mappings below can  be applied to many other dynamical observables (for example various types of correlation functions) simply by modifying the boundary conditions.

\begin{figure}[h]
\centering

\includegraphics[width=0.8\columnwidth]{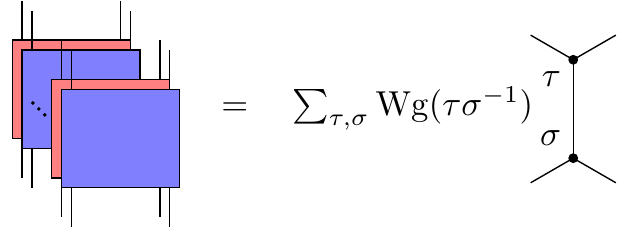}

 \caption{Graphical representation of the unitary averages in Eq.~\eqref{eq:fourU}. Each blue square with four legs is the two site local unitary gate and each red square is its complex conjugate. The ellipsis represents a total of $N =nk$ copies of each. On the right, the two top legs are $|\tau \rangle_i |\tau \rangle_{i+1} $ and the two bottom legs are ${}_i\langle \sigma | {}_{i+1} \langle \sigma | $. We associate `spins' $\sigma$ and $\tau$ with the vertices.}
\label{fig:fourU}
\end{figure}

Writing the RHS in terms of the circuit, we see that each local unitary $U$ as well as its complex conjugate $U^*$ appear $N$ times, with 
\begin{equation}
N\equiv nk.
\end{equation}
Specifically, each local unitary $U$ gives rise to the tensor product ${U\otimes U^*\otimes  \ldots \otimes U^*}$. This tensor product is shown graphically in Fig.~\ref{fig:fourU}, Left (there is one such `block' for each unitary in the circuit).
Next we perform the Haar average over the unitaries to obtain 
\begin{equation}
\overline{Z_n^k} = \overline{\lf \Tr \rho_A^n \ri^k}.
\end{equation}
where each unitary is averaged independently. Taking the replica limit $k\rightarrow 0$ in this quantity gives averages of the $n$th R\'enyi entropy, as described in Sec.~\ref{sec:overview}. In this section we set up the necessary machinery and in  the following sections we use it to calculate the entropies in various regimes.

\begin{figure}[h]
\centering

\includegraphics[width=\columnwidth]{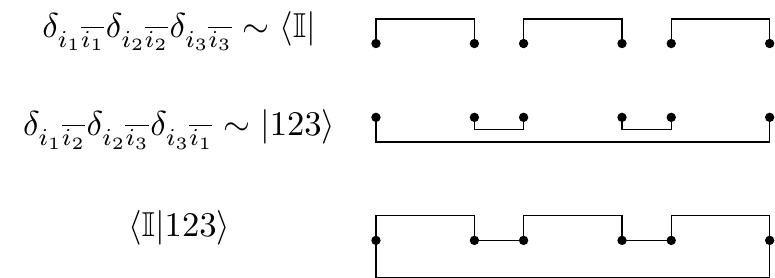}

 \caption{Contraction of the permutation states by counting the number of cycles (loops).}
\label{fig:Sn_inner_prod}
\end{figure}

A standard result gives this single-unitary average in terms of a sum over two elements, $\sigma$ and $\tau$, of the permutation group on $N$ elements \cite{weingarten1978asymptotic, collins_integration_2006}:
\begin{equation}
\label{eq:fourU}
\overline{U\otimes U^*\otimes  \ldots \otimes U^*} 
= 
\sum_{\sigma,\tau\in S_N} \mathrm{Wg} (\tau \sigma^{-1}) \ket{\tau\tau} \bra{\sigma\sigma}.
\end{equation}
Because we have $N\equiv nk$ copies of the circuit and $N=nk$ copies of its conjugate, at each physical site we now have the tensor product of $N$ factors of the physical Hilbert space and $N$ factors of its dual. The state $\ket{\sigma\sigma}=\ket{\sigma\sigma}_{i,i+1}$ is a product of identical states $\ket{\sigma}_i$ and $\ket{\sigma}_{i+1}$ on each of the two sites that the unitary acts on. The state $\ket{\sigma}$ is labelled by the permutation $\sigma \in S_N$.
In the natural basis, its components  are
\begin{equation}
\braket{a_1,\bar a_1, \ldots, a_N,\bar a_N}{\sigma} = \prod_j \delta(a_j, \overline{a}_{\sigma(j)}).
\end{equation}
Two examples of such states and their inner products are shown in Fig.~\ref{fig:Sn_inner_prod}.

Finally, Eq.~\eqref{eq:fourU} contains the Weingarten function, $\mathrm{Wg}( \tau\sigma^{-1})$. At this point, we only need to know that it is a function of the cycle structure of the permutation $\tau\sigma^{-1}$. We reserve App.~\ref{app:wein} to discuss its properties and perturbative expansion in  $1/q$. 

Graphically, Eq.~\ref{eq:fourU} can be represented as in Fig.~\ref{fig:fourU}. We will refer to $\sigma$ and $\tau$ as `spins'. Each unitary gives rise to an independent $\sigma$ spin and an independent $\tau$ spin living on the vertices connecting the vertical link. 

The full expression for $\overline{Z_n^k}$ is obtained by contracting the tensors (`blocks') defined in Eq.~\ref{eq:fourU} in accordance with the spatiotemporal structure of the circuit. Each non-vertical link connecting two blocks yields a power of $q$:
\begin{equation}
\label{eq:non_vertical_link}
\braket{\sigma}{\tau} = q^{N - |\tau\sigma^{-1}|}.
\end{equation}
The exponent ${N - |\tau\sigma^{-1}|}$ is simply the number of cycles in the permutation $\tau\sigma^{-1}$, which is at most $N$. The term $|\tau\sigma^{-1}|$ is the distance between $\sigma$ and $\tau$, which is minimized when $\sigma=\tau$. It is given by the minimal number of transpositions required to construct $\tau\sigma^{-1}$.

\begin{figure}[h]
\centering

\includegraphics[width=0.8\columnwidth]{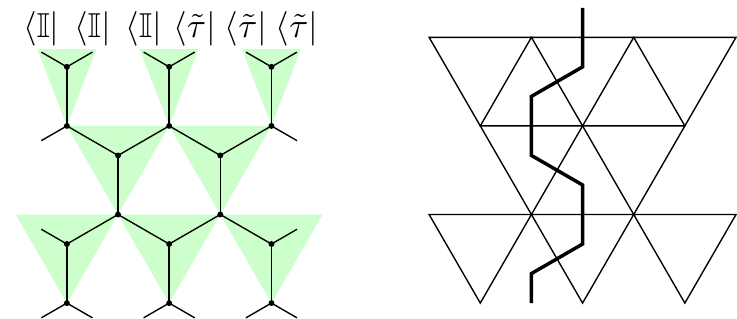}

 \caption{Left: $\overline{\tr( \rho_A^n ) }$ represented as a lattice magnet. The upper boundary is contracted with the boundary state $\langle \tilde{\tau} | = \langle \tau_{n,k}|$ for region $A$ and $\langle \I|$ for region $B$, the bottom boundary is identical to the top for the operator entanglement and free for the state entanglement. Each 4-leg block is the tensor in Eq.~\eqref{eq:fourU}. Right: The domain wall representation on the triangular lattice after integrating out the $\tau$ spins in each center of the down-pointing triangle (green).  }
\label{fig:tr_rho_n}
\end{figure}

After including these inner products, $\znk$ becomes a partition function for the $\sigma$ and $\tau$ degrees of freedom, with one $\sigma$ and one $\tau$ associated with each unitary in the circuit. This structure is shown in Fig.~\ref{fig:tr_rho_n}(left). 

At the  time $0$ (bottom) boundary, we obtain contractions with the initial density matrix. If the initial physical state is taken to be  a product state $\prod_i \ket{e}_i$, then at the time $0$ boundary we have $N$ copies of this state and $N$ copies of its dual, giving at each site $\ket{e\otimes \bar e\otimes \ldots \otimes \bar e}_i$. This is then contracted with a state $\ket{\sigma}$ associated with one of the unitaries in the lowest layer. This gives 
\begin{equation}
\braket{\sigma}{e\otimes \bar e \otimes \ldots \otimes \bar e}_i =
|{\braket{e}{e}_i}|^{2N} = 1.
\end{equation}
At the final time (top) boundary there are contractions which come from the traces in Eq.~\ref{eq:startingpoint}. This gives a weight which depends on the $\tau$s for the top row of unitaries. For each link, this is the inner product between that $\tau$ and another permutation which is determined by the structure of the trace. Outside region $A$ we contract row and column indices of $\rho$, which corresponds to contracting with the state $\bra{\mathbb{I}}$. Inside region $A$ we first take the product of $n$ copies of $\rho$ before taking the trace. This corresponds to contracting with the state
\begin{equation}
\bra{\tau_{n,k}} \equiv \bra{(1,2,3,\ldots,n)\,(n+1,\ldots, 2n)\,\ldots\,},
\end{equation}
given by a product of $n$-cycles, one for each power of $Z_n$ in $\znk$. 

Thus we have converted $\overline{Z_n^k}$ to a partition function for $\tau$ and $\sigma$ spins on the honeycomb lattice. Each \textit{non-vertical link} has weight specified by the inner product in Eq.~\eqref{eq:non_vertical_link}, and each \textit{vertical} link has a weight given by the Weingarten function. The boundary conditions are illustrated in Fig.~\ref{fig:tr_rho_n}(left). We will discuss the Boltzmann weight in terms of the domain wall picture in the following subsection. 

 \subsection{Domain walls on triangular lattice}

Now let us discuss the weight for a given spin configuration. At this point the weight is complicated because the Weingarten function in Eq.~\ref{eq:fourU} leads to a profusion of nonzero and also negative weights. (For example, if $\delta$ is an elementary transposition of two elements, $\mathrm{Wg}(\delta)$ is negative, while if $\delta=\mathbb{I}$ it is positive.) 

Remarkably, the partition function simplifies if we sum over the $\tau$ degrees of freedom \cite{nahum_operator_2017} associated with every unitary, giving a partition function for the $\sigma$s only. Each $\tau$ couples to three $\sigma$s, as can be seen in the green down-pointing triangles in Fig.~\ref{fig:tr_rho_n}, so integrating it out gives a three-spin interaction. These $\sigma$s form down-pointing triangles. We denote the weight for this triangle by
\begin{equation}
\begin{aligned}
\dptri{\sigma_a}{\sigma_b}{\sigma_c} = J(\sigma_b, \sigma_c;\sigma_a) \quad (\text{with }\, \sigma_a,\sigma_b,\sigma_c \in S_N).
\end{aligned}
\end{equation}
We will specify $J$ below for the cases of interest. For many values of $\{\sigma_a, \sigma_b, \sigma_c\}$ the weight $J(\sigma_b, \sigma_c;\sigma_a)$ vanishes, and this leads to considerable simplifications. Formally, we have
\begin{equation}
\label{eq:Jabc}
J(\sigma_b, \sigma_c;\sigma_a) = \sum_\tau \mathrm{Wg} (\sigma_a \tau^{-1}) q^{2N - | \sigma^{-1}_b \tau|  - |\tau^{-1} \sigma_c| }.
\end{equation}
From Eq.~\eqref{eq:Jabc}, we can explicitly verify that $J$ is invariant under left and right multiplication on all the spins,
\begin{equation}
J(\sigma_b, \sigma_c;\sigma_a) 
= J( \sigma \sigma_b, \sigma \sigma_c; \sigma \sigma_a) 
= J( \sigma_b \sigma', \sigma_c \sigma'; \sigma_a \sigma' ),
\end{equation}
as required from symmetries of the multi-layer circuit under permutations of the $U$ layers and of the $U^*$ layers.
Unitarity constraint can reduce the independent weights further, see Sec.~\ref{subsec:weight_down_tri}. 
This weight defines a partition function for spins on the vertices of the triangular lattice. 
At the top boundary we have triangles whose upper spins are fixed to be  $\tau_{n,k}$ inside region $A$ and  $\mathbb{I}$ outside region $A$ (the slightly slimmer triangles at the top of Fig.~\ref{fig:tr_rho_n}, Left); the equation above  also applies for their weights. At the lower boundary the spins are free.

\begin{figure}[h]
\centering

\includegraphics[width=0.8\columnwidth]{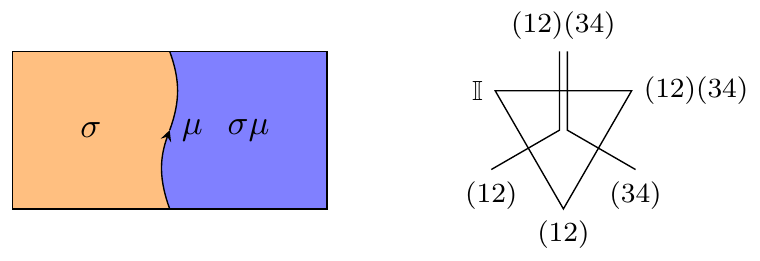}
 \caption{Definition of domain walls. Left: Domain wall labeling convention. Right:  $\sigma$ spins on the triangular lattice and domain walls on the dual lattice. The figure shows the splitting of two commutative elementary domain walls.}
\label{fig:dw_def}
\end{figure}

It is easiest to visualize the weights in terms of the domain walls. Each domain wall is itself labeled by a permutation $\mu$, as in Fig.~\ref{fig:dw_def}. To fix the labelling we must assign a direction to the domain wall, either upgoing or downgoing. This choice is arbitrary: a downgoing domain wall with label $\mu$ is equivalent to an upgoing domain wall with label $\mu^{-1}$. In our figures we take domain walls to be upgoing. 
Our convention is that if an upgoing domain wall labelled by $\mu$ has a domain of type $\sigma$ to the left, then the domain to the right is $\sigma\mu$,
 see Fig.~\ref{fig:dw_def}.

If a domain wall corresponds to a single transposition, for example $(12)$, we refer to it as an elementary domain wall.  When $|\mu|=m$, meaning that $\mu$ can be written  minimally as the product of $m$ transpositions, we will refer to a $\mu$ domain wall as a composite of $m$ elementary domain walls. However we must be careful to distinguish between e.g. $\mu=(123)$ and $\mu=(12)(34)$. Both of these have $|\mu|=2$ but they are not equivalent as they have different cycle structure. 

For simplicity let us take $A$ to be the region $x>0$ in a finite or infinite chain so that there is a single entanglement cut.
Then for $\znk$ the top boundary has a single domain wall of type ${\tau_{n,k}}$ which enters the system at the link of the entanglement cut.

We will show that $\znk$ can be regarded as a partition function for $k$ sets of $(n-1)$ elementary domain walls in ${\tau_{n,k}}$, with nontrivial interactions both within sets and between sets. These domain walls start at the top of the system at the position of the entanglement cut and undergo random walks downwards towards the bottom, where the boundary condition on the spins is free.

 \subsection{Triangle weights}
\label{subsec:weight_down_tri}

In Eq.~\eqref{eq:ex_weight} we give the {\it exact} results for the weights of the simplest configurations of a triangle, which involve at most 1 `incoming' elementary domain wall at the top of the triangle. 
\begin{equation}
\begin{aligned}
\dptri{}{}{} = 1, \qquad &\sdwfull = \sdwr = \frac{q}{q^2 + 1} \\
\vacbubb = 0 \qquad &\sdwbubb  = 0 
\qquad \quad \,\,\,\, (\mu \neq \mathbb{I}).
\end{aligned}
\label{eq:ex_weight}
\end{equation}
For a given triangle, we describe a domain wall at the top of the triangle as `incoming', and  domain walls at the bottom left and right as `outgoing'. 

The formula \eqref{eq:Jabc} involves a sum over $N!$ elements of the permutation group, with nontrivial weights. Remarkably, the final results for the weights above are independent of $N$. The non-vanishing diagrams are the ones that conserve the number (either 0 or 1) of incoming elementary domain walls. For example, it is not possible for the incoming elementary domain wall $(12)$ to split into  $(12) \mu$ and $\mu^{-1}$  with $\mu \neq \mathbb{I}$, despite the fact that this  splitting is consistent with the domain wall multiplication rule. Similarly, if the number of incoming domain walls is zero, there are no outgoing  domain walls: generation of  domain wall pairs out of the ``vacuum'' is forbidden.

We can summarize these rules algebraically as

\begin{equation}
\label{eq:ex_res}
\begin{aligned}
J(\sigma_b, \sigma_b; \sigma_a ) &= \delta_{ \sigma_a,  \sigma_b }\\
J(\sigma_b,(12) \sigma_b; \sigma_a ) &= \frac{q}{q^2 +1} \left( \delta_{ \sigma_a,  \sigma_b }  + \delta_{ \sigma_a, \sigma_b (12) } \right)
\end{aligned}
\end{equation}
We also give exact weights for the case $N=3$ in Appendix~\ref{app:ex_s3}. 

For a general configuration at large $N$ it is hard to evaluate the exact weights of the diagrams. 
However, we conjecture that, as in the example above,
$J$ does not depend directly on $N$: i.e. on the number of `additional' unused elements in permutations $\sigma_a$, $\sigma_b$, $\sigma_c$.
For example, we may evaluate 
$J(\mathbb{I}, (123); \mathbb{I} )$ 
for any $N\geq 3$, and we conjecture that the result is independent of $N$.
We have checked the conjectured $N$-independence of weights by explicitly evaluating all $J$s  for $N$ up to $4$.
However, for most of our purposes it will be sufficient to evaluate triangle weights in a large $q$ expansion, where we can obtain coefficients for all $N$.

\begin{figure}[h]
\centering

\includegraphics[width=0.6\columnwidth]{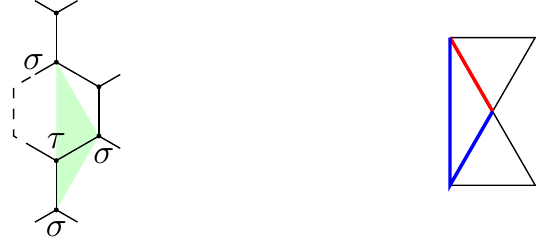}

 \caption{Spins on the spatial boundary. Left: The left-most legs of the unitary gates act on the same boundary site, so they are effectively connected, as shown by the dashed line. The $\tau$ spin on the boundary still connects to 3 $\sigma$ spins, and form a tilted triangle. Right: The boundary triangle on the triangular lattice. The red line is the top link, the blue lines are the bottom left and right links of the down-pointing triangle. }
\label{fig:bd_tri}
\end{figure}

Finally we specify the weights in the presence of spatial boundaries. These involve identical three-spin weights, but now the corresponding triangles are tilted: see Fig.~\ref{fig:bd_tri}.

\subsection{The $q=\infty$ limit}
\label{subsec:q_infty}

The partition function $\znk$ simplifies in the limit $q=\infty$, and this limit is a useful starting point for thinking about finite $q$.
When $q\rightarrow \infty$, the terms that survive in the partition function $\znk$ are those with the \textit{minimal total length} of elementary domain walls.
This means that domain walls cannot `split':  for each down-pointing triangle, the number of elementary domain walls entering from the top is equal to the number leaving from the bottom. 

At leading order in $q$, the weight of a triangle with $m$ elementary domain walls passing through it is just ${q^{-m}}$: for example (using doubled/tripled lines to represent composite domain walls)
\begin{equation}
\label{eq:tri_inf_q}
\tdwlll \simeq \tdwllr \simeq \tdwlrr \simeq \tdwrrr \simeq  \frac{1}{q^3}.
\end{equation}

This simplification of the weights means that at $q=\infty$ distinct replicas decouple: 
\begin{equation}
\label{eq:znk_fact}
\ln \znk \sim k \ln \overline{Z}_n \quad \text{at leading order in $q$}.
\end{equation}
This means that in this limit the fluctuations in $S_n$ are negligible (i.e. we must go to higher order to see them) and also that
\begin{equation}
\overline{S}_n \sim - \frac{1}{n-1} \ln \overline{Z_n}
\quad \text{at leading order in $q$}.
\end{equation}
Therefore leading order results in $q$ can be obtained by studying the partition function for a single replica, $\overline{Z_n}$.  In fact, this is sufficient to obtain not just the first term but the first few nontrivial terms in a large $q$ expansion for various quantities such as the growth rate of entanglement after a `quench'. We do this in the next section.

\section{Entanglement production rates}
\label{subsec:map_leading_in_q}

In this section we consider the partition functions $\overline{Z_2}$ and $\overline{Z_{n>2}}$ for a single replica. These suffice to obtain the first few orders in a large $q$ expansion of the R\'enyi entropy growth rates $v_2$ and $v_3$, as well as the `entanglement line tension' that generalizes these growth rates when the initial state is not a product state (Sec.~\ref{subsec:line_tension}). Later on we will address the effects of interactions between replicas. 

Let us first consider only the leading order contributions to the partition function at large $q$.
$\overline{Z_n}$ is then the partition function for ${n-1}$ elementary domain walls making up the permutation $(12\cdots n )$, and the entanglement entropy $\overline{S}_n$ is proportional to the free energy for this random walk problem. 

Since the number of domain walls is conserved at each time step, each layer in the triangular lattice contributes a factor of $q^{-(n-1)}$ to $\overline{Z_n}$. The logarithm of these factors gives the (negative of the) 'energy' of each configuration. There is also an \textit{entropy} term, coming from counting the number $\Omega_n(t)$ of distinct configurations:
\begin{align}
\overline{Z}_n & \sim \Omega_n(t) q^{-(n-1)t}, &
\overline{S}_n & \sim t \ln q - \frac{\ln \Omega_n(t)}{n-1}.
\end{align}
To go beyond this leading order result we use the more detailed weights in Eq.~\ref{eq:Jabc}.
Let us now consider various cases. 

\subsection{Second R\'enyi entropy}

The case $n=2$ has been treated in Ref.~\cite{nahum_quantum_2017}. There is only a single domain wall $(12)$ starting from the entanglement cut at the final time. In this case the mapping of $\overline{Z_2}$ to the partition function for a single simple walk is exact for any $q$ if we replace the approximate energy $\ln q$ with the logarithm  $-\ln K$ of the exact weight for a single triangle in Fig.~\ref{eq:ex_weight},
\begin{equation}
K = \frac{q}{q^2 +1}.
\end{equation}
The number of configurations is $2^t$. Therefore
\begin{align}
\overline{S}_2 
\sim & \seq\times  v_2 t, & v_2 &\simeq \widetilde v_2
\end{align}
with  $\seq=\ln q$ and the `purity speed'    \cite{zanardi_entangling_2000,hamma_2012_quantum, nahum_quantum_2017, nahum_operator_2017, von_keyserlingk_operator_2017}
\begin{equation}
\widetilde{v}_2  = \frac{1}{\ln q} \ln  \lf \frac{q^2 +1}{2q} \ri.
\end{equation}
Once interactions between replicas are taken into account this growth rate is corrected at the relatively high order $1/q^8 \ln q$ as we discuss in Sec.~\ref{sec:kpz}.

\subsection{Higher R\'enyi entropies}

For general $S_n$ we must consider the composite domain wall ${(12\ldots n)}$.
We may write this as a product of ${(n-1)}$ elementary domain walls labelled by transpositions. These transitions are non-commuting, which  gives rise to nontrivial combinatorial `interactions'. 

To see this, consider the  case $n=3$.
There are  3 ways to split a domain wall labelled $(123)$ into a product of two elementary domain walls, one on the left and one  on the right:
\begin{equation}
\label{eq:123_split}
(123)= (12)\times (23) = (23)\times(13) = (13)\times(12).
\end{equation}
This may be contrasted with the 2 ways to split a product of commutative transpositions:
\begin{equation}
\label{eq:1234_split}
(12)(34)= (12)\times (34) = (34) \times (12).
\end{equation}

The partition function for $(123)$ involves a nontrivial sum over how the elementary walks are labelled. 
Each time the walks meet, the labelling can change from one of the possibilities in Eq.~\ref{eq:123_split} to another.
 At first sight we must now keep track of the label on each domain wall, but in fact we can absorb the combinatorial factors associated with the labelling into a simple effective interaction. 

If two independent random walks $A$ and $B$ (for example two `commuting' elementary domain walls in the present problem) meet at a given time step, there are two possibilities for the configuration subsequently: either $A$ is on the left or $B$ is on the left (corresponding to the two terms on the RHS of Eq.~\ref{eq:1234_split}). For noncommuting domain walls such as $(12)$ and $(23)$, there are instead \textit{three} possibilities for the subsequent configuration. These are listed above in (\ref{eq:123_split}). This relative factor of $3/2$ means that $\overline{Z_3}$ maps to a partition function for a pair of (distinguishable) directed random walks with an attractive interaction. In a given configuration, let the number of `splitting events' be the number of times the walks meet and split. (When they meet, they may either split again immediately, or they may form a section of composite domain wall which extends for a finite period of time). Then
\begin{equation}
\label{eq:a3t-def}
\Omega_3(t) = \sum_{\substack{\text{configs of} \\ \text{2 walks} }} \lf \f{3}{2} \ri^\text{$\#$ splitting events}.
\end{equation}
The attraction means that the free energy is smaller than that of a pair of independent random walks. This means that $\overline{Z_3}=\overline{e^{-2S_3}}$ is larger than $\overline{Z_2^2} = \overline{e^{-2S_2}}$, so that the entanglement velocity $v_3$ is \textit{smaller} than $v_2$.

Interestingly, an effective combinatorial interaction between paths also arises in the replica treatment of directed polymers with Boltzmann weights of random signs, by a different mechanism \cite{medina_interference_1989, kardar_directed_1994}.

By a combinatorial computation in App.~\ref{app:z3t}, we obtain the exact asymptotic expression for $\Omega_3(t)$:
\begin{equation}
\Omega_3(t) \sim \lf 2 + \frac{{3}}{\sqrt 2} \ri^t.
\end{equation}
This constant was also obtained independently in a related Floquet model in Ref.~\cite{chan_solution_2017}, where it arises from essentially the same combinatorial mechanism.

The constant $\Omega_3$ gives the first nontrivial term in the expansion of $v_3$ at large $q$. We can go to one higher order by taking into account subleading \textit{repulsive} interactions of strength $\mathcal{O}(1/q^2)$ between the walks which appear when we go beyond the leading order expression for the triangle weights (exact results are in App.~\ref{app:ex_s3}, App.~\ref{app:pert_calc}). For example, if the composite domain wall on the LHS below is $(123)$,
\begin{equation}
\ddwfull \simeq \sdwfull \times \sdwfull \times \left[ 1 - \frac{1}{q^2}  \right],
\end{equation}
corresponding to a reduction in the weight for non-commutative elementary walks that are on top of each other compared to walks that are separate.
The calculation is performed in App.~\ref{app:z3t} and yields the growth rate:
\begin{equation}
v_3 \simeq 1 - \frac{\ln \big( 2 + \frac{{3}}{\sqrt 2} \big) }{2\ln q} +  \frac{3\sqrt{2}}{4} \frac{1}{q^2 \ln q}. 
\end{equation}

For larger $n$, the combinatorial factors can no longer be absorbed into a simple effective attraction. It appears to be necessary to keep track  of the labelling of the walks explicitly. This is because different decompositions of $(12\ldots n)$ can be inequivalent. For example the decomposition ${(1234)= (14)(13)(12)}$  and the decomposition ${(1234) = (13)(12)(34)}$ are inequivalent: in the former case none of the adjacent domain wall pairs commute while in the latter case one adjacent pair commutes.

The above shows that different R\'enyi entropies grow at different rates following a quench (see also \cite{chan_solution_2017}).

 \section{The entanglement line tension $\lt_n$}
\label{subsec:line_tension}

Above, $v_2$ is  the coarse-grained line tension, or free energy per unit length of the elementary domain wall that appears in the calculation for $S_2$ (up to a factor of $\seq = \ln q$; `length' here is in the $t$ direction). To be more precise, this  is the line tension of a domain wall which is \textit{vertical} on large scales. As argued in \cite{jonay_coarse-grained_2018} (see also \cite{nahum_quantum_2017}) it is useful to define a more general line tension $\mathcal{E}_2(v)$  which is a function of the coarse-grained `velocity'  of the domain wall. The velocity $v(t)$ of the domain wall is its inverse slope, $\dd x(t)/\dd t$, at a given value of $t$. The free energy of the  domain wall scales as  ${\seq t \times \lt_2(v)}$ if its average velocity is fixed to be $v$, i.e. if its total displacement over time $t$ is $v t$.

Here we briefly review the role of the line tension $\lt_2$ and its generalization $\lt_n$. In Sec.~\ref{subsec:e3section} we discuss the meaning of $\lt_n$ for higher $n$ in more detail, and calculate $\lt_3$. This will introduce the concept of the `bound state' of domain walls, which will  be important to understand nonanlyticities in $\lt_{n>2}$, and later how Page's formula arises (Sec.~\ref{sec:saturation}) and the fluctuations of the higher R\'enyi entropies (Sec.~\ref{subsec:kpz_higher_n}).

It was conjectured that the line tension $\lt_n$ determines the time dependence of the entanglement entropy $S_n$, in an appropriate scaling limit,\footnote{This limit is where the length and timescales of interest are parametrically large and of the same order. Since $S$ is also of this  order, $\partial S/\partial x$ can be order 1 in this regime, but higher derivatives such as $\partial^2 S/\partial x^2$ are subleading.} for more general initial states that are not necessarily product states \cite{jonay_coarse-grained_2018}:
\begin{equation}
\label{eq:Sminimzation}
S_n(x,t) = \min_v \left[ S_n(x-vt,0) + \seq \lt_n(v)
\right].
\end{equation}
Consider first $S_2$. The above formula arises from the random walk picture when we consider only the leading behaviour at large time. In this scaling limit the walk's fluctuations are negligible, and it forms a straight line connecting $(x,t)$ to $(y,0)$. The position $y$ is determined by minimizing the free energy. This gives the above, if we assume that the initial state at $t=0$ simply contributes an `energy' equal to its entanglement across $y$. This will not be true for all possible initial states but may hold for states that are `typical' in some sense.

Ref.~\cite{jonay_coarse-grained_2018}  conjectured the general constraints
\begin{align}
\label{eq:generalconstraints}
\lt_n(v_B) &= v_B, &
\lt_n'(v_B) & = 1, & 
\lt_n(v)& \geq v.
\end{align}
In the next section we give a nontrivial check on these constraints. By definition we also have $\mathcal{E}_n(0) = v_n$.

Considering the free energy of a random walk with a fixed slope gives \cite{jonay_coarse-grained_2018}
\begin{align}
\label{linetensionformula}
\mathcal{E}_2 (v)  = &  \,
1+
\frac{\ln  \f{q^2+1}{q^2} 
 +  \f{1+v}{2}  \ln    \f{1+v}{2}   +    \f{1- v}{2}  \ln    \f{1-v}{2}}{\ln q}.
\end{align}
This function satisfies the relations (\ref{eq:generalconstraints}) above. If the `free energy' is defined using the replica limit, as is appropriate for calculating $\overline{S}_2$, then Eq.~\ref{linetensionformula} will be modified at order $1/(q^8\ln q)$.

The function $\lt_2(v)$ is analytic for all $|v|<1$, i.e. for all speeds up to the lightcone speed, including speeds greater than $v_B$. In fact, the minimum in Eq.~\ref{eq:Sminimzation} is always in the range $[-v_B, v_B]$. However, the mapping of \cite{nahum_operator_2017} shows that the function $\lt_2(v)$ is relevant to the scaling of the exponentially small tail of the out of time order correlator beyond the lightcone \cite{khemani_velocity-dependent_2018}.

 \subsection{Higher R\'enyi entropies:\\ The `bound state' phase transition}
\label{subsec:e3section}

As we saw above, the calculation of $S_3$ yields a pair of elementary domain walls with an attractive interaction. In 1+1D, two walks with an attractive interaction form a \textit{bound state}: the typical separation between the walks, in the $x$ direction, is of order one even in the limit ${t\rightarrow \infty}$. Therefore at large scales the two walks are paired and can be regarded as a single composite domain wall. (The `bound state' terminology is natural if we think of the walks as worldlines of fictitious particles.)

The line tension $\lt_3(v)$ is defined as $1/(2\seq)$ times the free energy per unit length of this composite domain wall, when its coarse-grained velocity is fixed to be $v$. The factor of $1/(2\seq)$ is to compensate the $2\seq$ in $Z_3=e^{-2\seq S_3}$.
For higher $n$ the combinatorial interactions between walks are much harder to treat, but we expect that the walks will again form a bound state with a spatial extent of order 1. Then $\mathcal{E}_n(v)$ is $1/[(n-1)\seq]$ times the free energy per unit length of this composite domain wall, when its coarse-grained velocity is $v$.

We find that the line tension for $n=3$ has interesting structure that is not present in $\lt_2(v)$. This is due to a \textit{phase transition} that is driven by varying $v$. As $v$ is increased towards a critical value $v_c$, the extent of the bound state (in the $x$ direction) diverges. For $v<v_c$, the `binding energy' of the bound state means that $\lt_3(v)$ is smaller than $\lt_2(v)$. But for $v\geq v_c$, the walks are unbound and their free energy is simply that of two independent walks: this means that in this regime $\lt_3(v) = \lt_2(v)$. 

We conjecture, and check explicitly to  leading nontrivial order, that the critical speed associated with this unbinding transition is precisely $v_B$:
\begin{equation}
v_c =v_B.
\end{equation}
This mechanism is how the conjectured constraint $\lt_3(v_B)=v_B$ in Eq.~\ref{eq:generalconstraints} is satisfied. We conjecture that a similar mechanism applies for higher $n$ also, with the $(n-1)$ walks becoming unbound at $v_B$.

We now give explicit formulas for $\lt_3$.
Firstly, we show in App.~\ref{app:e3v} that to order $1/\ln q$ the line tension for $S_3$ is 
\begin{equation}
 \lt_3(v) = 1- \f{\ln ( \phi^{-1} + \phi + \frac{3}{\sqrt{2}})- v \ln \phi}{2\ln q},
\end{equation}
with
\begin{equation}
\phi= \f{3v+\sqrt{8+v^2}}{\sqrt 8 (1-v)}.
\end{equation}
The functional form differs nontrivially from that for $\lt_2$. However, the bound state phase transition cannot be seen at this order in $q$. Therefore in App.~\ref{app:e3vlightcone} we perform a separate expansion for speeds close to the lightcone, writing 
\begin{equation}
\label{eq:speedscaling}
v = 1 - \f{\alpha}{q^2},\quad \text{with $\alpha$ of order 1}.
\end{equation}
$v$ close to $-1$ is of course equivalent.

First let us consider how the phase transition can occur in principle. Recall that each time the walks merge and split, they `gain' a weight $3/2$ for combinatorial reasons. This is an effective attraction that encourages them to bind together. 

However, examining the exact weights for the walks (App.~\ref{app:ex_s3}), we find that there is also a weak repulsion, of order $1/q^2$, for time steps in which the two walks are on top of each other (combined into a composite walk). For generic values of $v$, this weak repulsion is negligible compared with the $\mathcal{O}(1)$ attraction arising from the combinatorial effect. But for walks moving at speeds very close to unity, this repulsion is magnified as follows.
 
For a walk moving at the speed in Eq.~\ref{eq:speedscaling}, almost every step is to the right: only an $\mathcal{O}(1/q^2)$ fraction of steps are to the left. This means that when the two walks meet, they typically remain together for a long time, of order $q^2$ (both taking rightward steps) before one of them takes a leftward step and they split. Therefore the \textit{total} repulsion energy for the time interval between the merging and splitting is $\mathcal{O}(1)$, and can compete with the $\mathcal{O}(1)$ combinatorial attraction. For small enough $1-|v|$, the repulsion dominates and the bound state disappears.

In App.~\ref{app:e3vlightcone} we find that the critical speed for disappearance of the bound state  is 
\begin{equation}
\label{eq:unbindingvelocity}
v_c = 1- \f{2}{q^2} + \cdots.
\end{equation}
This is consistent with $v_c=v_B$. For $v>v_c$ we have $\lt_3(v)=\lt_2(v)$. For $v<v_c$ we find
\begin{equation}
\label{eq:e3nearlightcone}
\lt_3(v ) = 1- \f{\alpha}{q^2} + \f{A_3(\alpha)}{q^2 \ln q} + \ldots
\end{equation}
with
\begin{align}
\label{eq:A3formula}
&A_3(\alpha) = \\ 
&\,\,\,\f{
9
\hspace{-0.5mm}
- 
\hspace{-0.5mm}
\sqrt{
9\hspace{-0.5mm}+\hspace{-0.5mm} 4 \alpha^2  \left[  2 \hspace{-0.5mm} + \hspace{-0.5mm} \sqrt{4 \hspace{-0.5mm} +  \hspace{-0.5mm} \f{9}{\alpha^2} }  \right]
}
\hspace{-0.5mm}
 +
 \hspace{-0.5mm}
 2 \alpha \ln  \hspace{-0.5mm} \left[
\f{\alpha^2}{18} \left[ 2\hspace{-0.5mm} + \hspace{-0.7mm}\sqrt{4\hspace{-0.5mm}+\hspace{-0.5mm}\f{9}{\alpha^2} } \right]
\right]
 }{8}.
 \notag
\end{align}
In this regime $\lt_3(v)<\lt_2(v)$, which is necessary for Eq.~\ref{eq:Sminimzation} to be consistent with the general constraint $S_3\leq S_2$. 

Note that $A_3(2)=0$ and $A_3'(2)=0$, showing that the line tension $\lt_3(v)$ obeys the general constraints in Eq.~\ref{eq:generalconstraints} at least up to order $1/(q^2 \ln q)$.

\section{Fluctuations \& the replica limit}
\label{sec:kpz}

In this section we treat interactions \textit{between} replicas in the replicated partition function $\znk$. 
We study $S_2$ in detail, mapping it explicitly to the free energy of a directed polymer in a classical random medium 
and to the height field in a continuum KPZ equation.
The extension to $n>2$ is discussed more briefly in Sec.~\ref{subsec:kpz_higher_n}.

The Kardar-Parisi-Zhang equation was proposed to model  universal scaling in classical surface or interface growth \cite{kardar_dynamic_1986}. 
In that context there is a time-dependent function $h(x,t)$, representing the height of the interface, which obeys the nonlinear stochastic equation
\begin{equation}
\label{eq:kpz-height}
 \partial_t h = \nu  \partial_x^2 h + \frac{\lambda}{2}( \partial_x h )^2 + \eta( x, t ) + \text{const}.
\end{equation}
The first term represents diffusive relaxation. The non-linear term is a relevant perturbation of the linear theory that must be included in order to capture the generic universal behaviour of the interface growth problem.
The $\eta$ term is white noise (uncorrelated in space and time). 

There is a simple and important connection between this nonequilibrium growth problem in one spatial dimension plus time, and the \textit{equilibrium} statistical mechanics of a directed polymer (or a domain wall), subjected to a disordered potential energy, in two spatial dimensions. 
As a result there are  several equivalent ways to think about the universal properties of the KPZ universality class \cite{kardar_dynamic_1986,kardar_replica_1987,HuseHenleyFisherRespond}:  (i) in terms of the KPZ equation; 
(ii) in terms of the directed polymer in a disordered medium of spatial extent $t$ in the vertical dimension; and
(iii) in terms of a replicated system of $k\rightarrow 0$ polymers with attractive interactions arising from integrating out the disorder. These polymers can also be viewed as worldlines of  bosons in imaginary time, so (iii) is equivalent to a system of $k\rightarrow 0$ mutually attracting bosons in one spatial dimension.
We will use all of these languages. 

For a brief review: the relation between (i) and (ii) is given by the Cole-Hopf transformation \cite{kardar_dynamic_1986}. 
Defining  $Z(x,t) = e^{\f{\lambda}{2\nu} h(x,t)}$, this satisfies
  \begin{equation}\label{eq:partitionfneq}
    \partial_t  Z = \lf  \nu \partial_x^2 + \f{\lambda}{2\nu} \eta( x, t ) + \text{const.} \ri Z . 
    \end{equation}
This is a recursive equation for the partition function of a continuum polymer in a spatial sample of height $t$, in terms of the partition function for an infinitesimally shorter sample.
As a path integral, the partition function is ${Z(x,t) \propto \int \DD x'(t') Z\lf x'(0), 0 \ri e^{- \int_0^t  \f{\dd t'}{2\nu}\big[ \f{1}{2} \big( \f{\dd x'}{\dd t'}\big)^2 + \lambda \eta(x', t') \big]}}$, and the argument $x$ sets the position of the polymer's endpoint at the top of the sample: $x'(t) = x$. 

Eq.~\ref{eq:partitionfneq} is also the imaginary time Sch\"odinger equation for boson in a random potential. Using the replica trick and integrating out the disordered potential leads to a problem of $k\rightarrow 0$ bosons with pairwise attractive interactions (see \cite{kardar_replica_1987}). This is description (iii).

In the following we will work backwards, mapping the calculation of $S_2(t)$ to a replicated problem of type (iii) and using the above mappings to determine the coefficients in the KPZ equation (i) satisfied by $S_2(t)$.
Some of the universal consequences of the KPZ description for entanglement growth in noisy systems have been reviewed in Sec.~\ref{sec:overview}, see \cite{nahum_quantum_2017} for more details.

For $n = 2$,  each replica in $\overline{Z_2^k}$ gives only one domain wall, so that there are $k$ elementary walks in total. A  diagrammatic calculation shows that these $k$ walks have an effective pairwise attraction at order $\frac{1}{q^4}$.
This ultimately leads to KPZ  fluctuations of the entanglement of order $q^{-\frac{8}{3}} t^{\frac{1}{3}}$.

Because of the replica limit we have $k\rightarrow 0$ directed walks with a pairwise interaction, corresponding to (iii) above. 
Because the interaction is parametrically small at large $q$, we can make a \textit{controlled} continuum approximation.
Let us think of the coarse-grained walks as worldlines of bosons in 1+1D Euclidean spacetime, with attractive contact potentials between the bosons. 
In this language the partition function $\znk$ is the imaginary time path integral amplitude for the bosons
(and the entanglement growth rate is proportional to their ground state energy).
The resulting boson Hamiltonian is integrable \cite{kardar_replica_1987,calabrese_exact_2011} and this is one way to obtain the fluctuations of the entanglement.

But also, having applied the replica trick to $\Tr \rho_A^2$ in our original many-body quantum system, and mapped the resulting expression to an effective classical partition function for $k\rightarrow 0$ random walks, we can alternatively \textit{undo} the replica limit to obtain a \textit{classical} model with randomness. 
This can be done both in the continuum and on the lattice. We will discuss this in Sec.~\ref{subsec:polymermapping}.

In the continuum it is convenient to think in terms of the KPZ equation. 
Remarkably even the {nonuniversal} constants in this equation can be fixed. At large but finite $q$
 the second R\'enyi entropy $S\equiv S_2$ obeys\footnote{The sign of the nonlinear term in Eq.~\ref{eq:kpzeqconstsfixed} is opposite to that of Eq.~\ref{eq:kpz-height} because in the correspondence with the directed polymer, $S$ is proportional to the free energy, while $h$ is proportional to minus the free energy. The sign of the nonlinear term can be changed by a change of variable $h\rightarrow -h$ so does not affect the exponents.}
\begin{equation}
\label{eq:kpzeqconstsfixed}
  \partial_t S =  c+ \frac{1}{2} \partial_x^2 S - \frac{1}{2}( \partial_x S )^2 + \eta( x, t ) 
\end{equation} 
for a weak Gaussian noise
\begin{equation}
\label{eq:noisestrength}
\langle \eta( x, t )\eta( x', t' ) \rangle  = \frac{1}{4q^4} \delta( x - x' ) \delta( t- t' ). 
\end{equation}
Above, $c$ is a constant which contributes to the entanglement growth rate $v_2$ (given in Eq.~\ref{eq:v2}) which we fix  using the boson  mapping. The large-time scaling of $S_2$, for a cut across a given bond, may be written in terms of a fluctuating random variable $\chi(t)$ whose size is of order 1 at large times (below $\seq=\ln q$):
\begin{equation}
S_2(t) \simeq \seq\left[ v_2 t + B_2 t^{1/3} \chi(t)\right]
\end{equation}
Using the exact results in Ref.~\cite{calabrese_exact_2011}, the magnitude of the fluctuations are controlled by the constant
\begin{equation}
\label{eq:kpz_B2}
B_2 \simeq -\f{1}{4\times 2^{1/3}\times q^{8/3}}.
\end{equation}
The cumulative probability distribution of the random variable $\chi(t)$ is the Tracy-Widom distribution $F_1$. The mean and standard deviation are 
\begin{align}
\text{mean} (\chi)  & = -1.20653...,
&
\text{std}(\chi)   & = 1.26798...
\end{align}

On the lattice, undoing the replica limit on the classical side of the mapping gives a well-defined lattice directed polymer problem with a short-range correlated random potential. 
We make this mapping explicitly for large but finite $q$.

With the bound state concept introduced in Sec.~\ref{subsec:line_tension},
 we can generalize the above pictures to $S_n$ with ${n > 2}$. 
The composite domain walls in $S_n$ for $n > 2$ first form a bound state due to the leading order combinatorial interaction in Sec.~\ref{subsec:e3section}. 
Then by a similar mechanism as for $\overline{S}_2$, the weak pairwise interaction between the bound states from different replicas gives rise to the KPZ fluctuations, showing that such fluctuations are present in all R\'enyi entropies with $n\geq 2$.
We discuss this further in Sec.~\ref{subsec:kpz_higher_n}.

 \subsection{Interactions between replicas}
\label{subsec:int_bet_replica}

In this section, we focus on the $n = 2$ case where the leading order picture involves $k$ independent commutative elementary domain walls. 

First of all, the exact partition function for a single elementary domain wall ($k = 1$) is 
\begin{align}
\overline{Z_2}& = 2^t K^t, & K& =  \frac{q}{q^2 + 1},
\end{align}
where $2^t$ is the number of random walk configurations and $K^t$ is the product of weights of $t$ down-pointing triangles, {\it c.f.} Eq.~\eqref{eq:ex_weight}. Compared with the leading order result $1/q$, we see that the more accurate weight $K$ for each down-pointing triangle is
\begin{equation}
 \sdwfull =  K = \frac{1}{q} \left( 1- \frac{1}{q^2} + \frac{1}{q^4} + \cdots \right).
\end{equation}
These corrections determine the finite $q$ corrections to the energy per unit length, or the line tension, of a single walk. 

Similarly, we may consider the higher-order corrections to the weights of triangles that host $\ell >1$ walks. At leading order the weight of such a triangle is $q^{-\ell}$. At first sight we might expect corrections to this leading order result to be the dominant source of interactions between the replicas. However we find that to order $\frac{1}{q^4}$, we have 
\begin{equation}
\label{eq:ddw_decomp}
\begin{aligned}
\ddwfull &= \sdwfull  \times \sdwfull \times \left[ 1 + \mathcal{O}(\frac{1}{q^6}) \right]\\
\ddwlr  &= \sdwfull \times \sdwr \times \left[ 1 + \mathcal{O}(\frac{1}{q^6}) \right]
\end{aligned}
\end{equation}
This kind of decomposition also holds for all $\ell \ge 2$ up to  $\mathcal{O}(\frac{1}{q^4})$ order, see the perturbative calculation in App.~\ref{app:pert_calc}. Consequently, if there is an interaction at order $\frac{1}{q^4}$ it must come from \textit{additional} domain wall configurations which are \textit{absent} in the $q=\infty$ limit. This is indeed the case.

What are the lowest order (in $1/q$) modifications to the domain wall configurations described above? By Eq.~\ref{eq:ex_weight}, we cannot add isolated `bubbles', i.e. closed domain wall loops that are not attached to any of the $k$ walks: such configurations have weight zero. Similarly, the last formula in Eq.~\ref{eq:ex_weight} prevents us from modifying an isolated walk. However when two walks meet additional configurations are possible.

As mentioned in Sec.~\ref{subsec:map_leading_in_q}, the `naive' order in $1/q$ of a down-pointing triangle is equal to the number of elementary domain walls that pass through its lower edges. [The actual order may be higher, as a result of cancellations in the sum defining $J(\sigma_a;\sigma_b,\sigma_c)$.] There are two possibilities allowed by Eq.~\ref{eq:ex_weight} that are na\"ively of relative order $1/q^4$ compared to the leading order configurations. The first corresponds to adding a hexagonal `bubble' of the transposition $\alpha$ to a configuration of two walks, say $(12)$ and $(34)$:
\begin{equation}
\label{eq:regvert}
\regvert 
\end{equation}
However, the relative order of this configuration is in fact $1/q^6$, as the result of a cancellation between two values of the Weingarten function in Eq.~\eqref{eq:ex_weight}, see App.~\ref{app:pert_calc}.\footnote{The two values correspond to $\tau = \sigma_a$ and $\tau = \sigma_a \alpha^{-1}$ in Eq.~\eqref{eq:ex_weight}.}

The second possibility relies on the following decompositions of the product $(12)(34)$,
\begin{equation}
\label{eq:special_vertex}
(12)(34) =  (14) (23) \times (24)(13)  = (24)(13) \times  (14) (23).
\end{equation}
Each of these decompositions leads to a `Feynman diagram':
\begin{align}
\label{eq:ddw_bubble}
&\ddwbubbone  &  &\ddwbubbtwo.
\end{align}
Each such configuration has relative weight, compared to the dominant configurations, of  $1/q^4$ (plus higher order corrections).

These `special hexagons' are the only source of interactions in the bulk of the sample at order $1/q^4$. We may add the weight of these configurations to the weights of the leading-order configurations to obtain a `dressed' weight for a pair of walks which both visit a pair of triangles that are vertically adjacent as shown below:
\begin{equation}
\label{eq:ddwtwostep}
\begin{aligned}
\ddwtwostep =& 
\ddwtwostepll  + \ddwtwosteplr +  \ddwtwosteprl +  \ddwtwosteprr
\\
& +2\times \ddwbubble + \mathcal{O}(\frac{1}{q^6}),
\end{aligned}
\end{equation}
The factor of two indicates the two possibilities in Eq.~\ref{eq:ddw_bubble}. This gives the total weight (recall $K\equiv q/[q^2+1]$)
\begin{equation}
\begin{aligned}\label{eq:specialhex}
\ddwtwostep &= 
4 K^4 + 
\frac{1}{q^4} \frac{2}{q^4} + \mathcal{O}\lf\frac{1}{q^{10}}\ri \\
&= 4 K^4 
\left( 1 +  \frac{1}{2}\frac{1}{q^4} + \mathcal{O}\lf \frac{1}{q^{6}}\ri \right). \\
\end{aligned}
\end{equation}
This is an interaction of order $1/q^4$, and it is attractive, because it increases the Boltzmann weight for configurations in which two walks collide. 
Furthermore, it is a \textit{pairwise} interaction -- we can insert a `special hexagon' for any pair of (commutative) domain walls, and to leading order this insertion is not `seen' by any of the other $k-2$ walks. In Sec.~\ref{sec:numericalchecks} we also check these properties of the interaction numerically.

The fact that the attraction is small at large $q$ allows an analytical treatment which we discuss next (Secs.~\ref{subsec:polymermapping},~\ref{subsec:continuumdescription},~\ref{susec:bd_effect}). It should be noted however that for other random circuits, in which the microscopic probability distribution of gates is different, the interaction strength can remain of order one even in the limit $q\rightarrow \infty$. The simplest way to obtain an interaction of $\mathcal{O}(1)$ strength in the $q\rightarrow \infty$ limit is to allow the local unitaries to be equal to the identity with a nonzero probability $p$. 

These identity gates create `holes' in the circuit through which the domain walls can pass without costing any energy at all. Averaging over the locations of these holes gives an effective attractive interaction between replicas. This has similar effects to the attractive interaction described above, but the strength of the attraction remains finite at $q=\infty$ and can be controlled by varying $p$. This is essentially a model considered in \onlinecite{nahum_quantum_2017} where KPZ behaviour was obtained in the limit $q\rightarrow \infty$.

 \subsection{Mapping to polymer in random medium}
\label{subsec:polymermapping}

\begin{figure}[b]
\centering
\includegraphics[width=0.8\columnwidth]{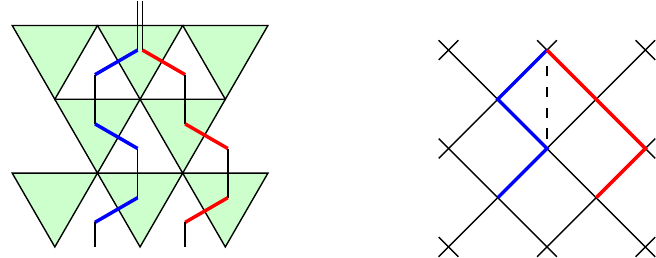}
\caption{Reduction from triangular lattice to square lattice.
Pairs of consecutive steps are combined into a single step on the square lattice.
Each blue (red) step on the left corresponds to a blue (red) step on the right. 
We also refer in the text to `visits' to  vertical bonds like the one indicated by the dashed line. 
According to our definition,  the vertical bond indicated here  is only visited by the blue walk and not the red (so there is no special hexagon interaction between these two walks).}
\label{fig:tri_squ_lattice}
\end{figure}

 For a circuit with regular structure the attractive interaction between replicas is small at large $q$. This allows both a controlled continuum description and an explicit mapping to a classical disordered model.

Let us simplify the lattice structure. Above, each random walk lives on the honeycomb lattice which is dual to the triangular lattice. Each honeycomb site corresponds to a triangle, either up- or down-pointing. However it is sufficient to draw only the sites corresponding to the down-triangles, as shown in Fig.~\ref{fig:tri_squ_lattice}.  That is, we can view the walks as living on a square lattice  (rotated by 45$^\circ$). Adjacent sites of this square lattice differ by $(\Delta x,\Delta t)  = (\pm 1, \pm 1)$. For an isolated walk, each step along a bond of this lattice is weighted by $K$.

It is useful to think of pairs of sites $(x,t)$ and $(x,t+2)$ as connected by vertical bonds, even though the walks cannot occupy such bonds. 
One such vertical bond is illustrated in Fig.~\ref{fig:tri_squ_lattice}, right.

If two walks both visit both of the sites $(x,t)$ and $(x,t+2)$, then the associated weight is not $4K^4$ but rather $ {4K^4\exp \lf 1/2q^4 \ri}$, by Eq.~\ref{eq:specialhex}. 
When any one  walk visits \textit{both} $(x,t)$ and $(x,t+2)$, we say that the vertical bond from $(x,t)$ to $(x,t+2)$ is visited by that walk. 
Two walks therefore interact if they both visit the same vertical bond.

For each vertical bound $b$, let the number of walks which visit it be $n_b$.
If $n_b\geq 2$ walks visit bond $b$, there is an interaction between each of the $n_b(n_b-1)/2$ pairs, as discussed below in Eq.~\ref{eq:specialhex}. The weight associated with the interactions is thus $\exp \lf 1/2q^4\times n_b (n_b-1)/2 \ri$. 

Using $D$ to denote the tilted square lattice  the effective partition function is:
\begin{equation}
\label{eq:Z2kinteractions}
\overline{Z_2^k} = K^{k t}  \sum_{\substack{
{\text{$k$ directed}}
\\
{\text{walks on $D$}}}}
\exp A,
\end{equation}
where,  neglecting boundary effects (which are discussed in Sec.~\ref{susec:bd_effect})
\begin{equation}
A = \sum_{\substack{
{\text{vertical}}
\\
{\text{bonds $b$}}}}
 \f{1}{2q^4} \f{n_b \lf n_b - 1 \ri}{2}.
\end{equation}
Remarkably, this form means that we can interpret $\overline{Z_2^k}$ as the average of a replicated \textit{classical} partition function for a single walk or polymer. Let us define the partition function for this fictitious classical polymer by:
\begin{equation}
\label{eq:explicitpolymerZ}
\mathcal{Z} = 
K^t
\sum_{\substack{
{\text{polymer}}
\\
{\text{on $D$}}}}
\exp 
\lf 
-
 \sum_{\substack{
{\text{vertical}}
\\
{\text{bonds $b$}}}}
\, n_b \times
V_b 
\ri.
\end{equation}
Since this partition function is for a \textit{single} polymer, $n_b$ is either $0$ or $1$. On each vertical bond $b$, the polymer experiences a Gaussian random potential $V_b$. 
We take these random potentials to be independent, with mean and variance 
\begin{align}
\label{eq:Vbmeanvariance}
\operatorname{mean} (V_b) &\simeq \f{1}{4 q^4}, &
\operatorname{var} (V_b) &\simeq \f{1}{2q^4}.
\end{align}
With these choices, averaging $\mathcal{Z}$ over the random potentials $V_b$ yields precisely the expression for $\overline{Z_2^k}$ in Eq.~\ref{eq:Z2kinteractions}. Writing the average over $V_b$ as $[\ldots]_V$:
\begin{equation}
\overline{Z_2^k} \sim \left[\mathcal{Z}^k\right]_V.
\end{equation}

The identity above implies that the statistics of $S_2$ in the quantum problem map onto the statistics of a classical polymer in a random potential that is  specified by Eqs.~\ref{eq:explicitpolymerZ},~\ref{eq:Vbmeanvariance}. 
Note that this makes the dynamics of the entropy efficiently simulable (modulo the large $q$ approximation used) for large $t$ that would be beyond the reach of a direct computation as in Sec.~\ref{sec:numericalchecks}.

 \subsection{Continuum description}
\label{subsec:continuumdescription}

Next we discuss the continuum limit.
Consider first $\overline{Z_2}$, i.e $k=1$. In the continuum the walk becomes a Brownian path characterized only by its free energy per unit time, ${f=-\ln 2 K}$, and its diffusion constant, which is easily seen to be $D=1/2$.\footnote{The mean squared displacement in the $x$ direction, for a section of duration $t$, is simply $2Dt = t$.} Viewing the walk as the Feynman path of a boson in Euclidean spacetime, with spatial coordinate $x$, the Hamiltonian for this boson is
\begin{equation}
  \mathcal{H} = -  \ln 2K    - \f{1}{2}   \f{\partial^2}{\partial x^2}.
\end{equation}
The scaling of the partition function is given by the ground state energy $E_0$ of this system of bosons:
 ${\overline{Z_2}\sim e^{-E_0 t}}$. For the above Hamiltonian $E_0$ is simply $-\ln 2K$.  
 
For $k\neq 1$ we must take into account the attractive interactions between bosons. The continuum Hamiltonian contains only a delta-function interaction and is solvable by Bethe ansatz in the $k\rightarrow 0$ limit \cite{kardar_replica_1987}:
\begin{equation}
\label{eq:ctmham}
\mathcal{H} = - k \ln 2K    - \f{1}{2}  \sum_{\alpha=1}^k \f{\partial^2}{\partial x_\alpha^2} - \lambda \sum_{\alpha<\beta} \delta(x_\alpha - x_\beta).
\end{equation}
Since $\lambda\ll 1$ at large $q$ we can fix it explicitly using the lattice results above (see  App.~\ref{app:interactionconstant}):
\begin{equation}
\lambda \simeq \f{1}{4 q^4}.
\end{equation}
Standard mappings \cite{kardar_replica_1987} relate the coefficients in Eq.~\ref{eq:ctmham} to those in the KPZ equation (\ref{eq:kpzeqconstsfixed}).

The energy of the system of bosons as $k\rightarrow 0$ gives the average free energy density $f$ of the polymer in the random medium, or equivalently the growth rate of the averaged entropy: $f= \seq v_2$. Using the result of \cite{kardar_replica_1987},
\begin{equation}
v_2 = \frac{1}{\ln q }\ln \lf  \f{q^2 + 1}{2q}  \ri  + \lf \f{1}{384 \, q^8 \ln q} + \ldots \ri.
\end{equation}
The first term is the `purity speed' $\widetilde v_2$ (Sec.~\ref{sec:overview}), and the second is a correction from replica interactions. The Bethe ansatz results in \cite{calabrese_exact_2011}  also fix the prefactor of the KPZ fluctuations, Eq.~\ref{eq:kpz_B2}. 

The above results apply in the  limit of large times.
Since at large $q$ the interaction is weak (but renormalization-group relevant) there is  a large crossover scale.
The crossover \textit{length} scale (in the spatial $x$ direction) is in the notation of Ref.~\cite{kardar_replica_1987}:
\begin{equation}
l_d = \f{2}{\lambda} \simeq 8 q^4.
\end{equation}
This corresponds to a timescale of order $l_d^2$.
For ${t\ll l_d^2}$ the polymer of the previous subsection resembles a random walk, with diffusive scaling between $x$ and $t$. For ${t\gg l_d^2}$ its conformation is strongly affected by the quenched randomness, and KPZ scaling exponents govern its statistics and the statistics of $S_2$.
 It is notable that in the present model the crossover timescale $l_d^2$ is large even for $q=2$. 
 We discuss crossovers in more detail in the next section.

 \subsection{Early-time crossovers for large $q$}
\label{susec:bd_effect}

So far we have considered KPZ scaling at asymptotically long times, which we expect to hold for any $q$.
However when $q$ is large there are interesting early and intermediate time regimes, while fluctuations in $S_2$ remain small, i.e. before the onset of KPZ scaling at times of order $l_d^2$.
In total there are three regimes, shown in Table~\ref{tab:3-regime}.

\begin{figure}[t]
\centering

\includegraphics[width=0.8\columnwidth]{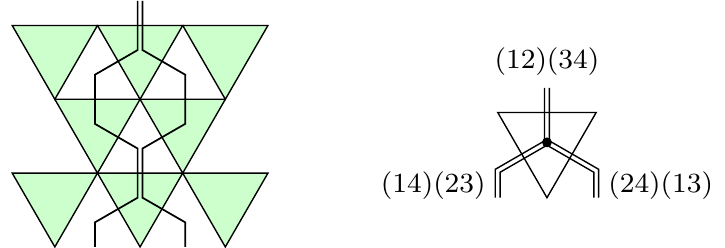}

 \caption{An example of the half special hexagon interaction at the boundary. (Left) The special hexagon in the bulk gives an interaction of order $\frac{1}{q^4}$, while the half hexagon at the bottom boundary gives an interaction of order $\frac{1}{q^2}$. Hence the boundary interaction will dominate the early time fluctuation. (Right) A domain wall configuration of the half special hexagon. Since there are only 2 extra legs, it is of order $\frac{1}{q^2}$.  }
\label{fig:bd_int}
\end{figure}

We first note that when two walks from different replicas meet at the $t=0$ boundary, there is an interaction that corresponds to `half' of the special hexagon in Sec.~\ref{subsec:int_bet_replica}. This is shown in Fig.~\ref{fig:bd_int}. This interaction is of order $\frac{1}{q^2}$. Since this is parametrically larger than the $\mathcal{O}(1/q^4)$ bulk interaction, it dominates at early times. In the polymer language it corresponds to a \textit{boundary} disorder potential of strength $\sim 1/q$. For times $1\ll t \ll q^2$, as we show below, this leads to fluctuations which \textit{decrease} with time as 
\begin{align}
\label{eq:earlytimedecrease}
\sqrt{\text{Var}(S_2 )} 
&\sim \frac{1}{( 4\pi t )^{\frac{1}{4}}}\frac{1}{q},
& 
&1\ll t \ll q^2.
\end{align}
The reason for the decrease of the fluctuations is that a polymer of length $t$ explores, through thermal fluctuations, a length of the boundary of size $\sim t^{1/2}$. It is therefore effectively subject to the disorder potential averaged over this region. The average of $\mathcal{O}(t^{1/2})$ local potentials with mean zero and typical magnitude $\sim 1/q$ gives the $1/(q t^{1/4})$  scaling above.

\begin{table}[b]
\centering
\begin{tabular}{ |c|c|c| } 
 \hline
  & $\sqrt{\text{Var}(S_2)} $  & $t$ \\\hline
 boundary-dominated & $\sim q^{-1} t^{-1/4}$ & $   t \lesssim q^{2}$ \\
 Edwards-Wilkinson & $ \sim q^{-2} t^{1/4} $ & $ q^{2} \lesssim t\lesssim q^8$ \\
 KPZ & $ \sim  q^{-8/3} t^{1/3}$ & $ t\gtrsim q^8$ \\
 \hline
\end{tabular}
\label{tab:3-regime}
\caption{When $q$ is large there are three time regimes for fluctuations in $S_2$. Here the initial state is a product state.}
\end{table}

\begin{figure}[h]
\centering
\includegraphics[width=.8\linewidth]{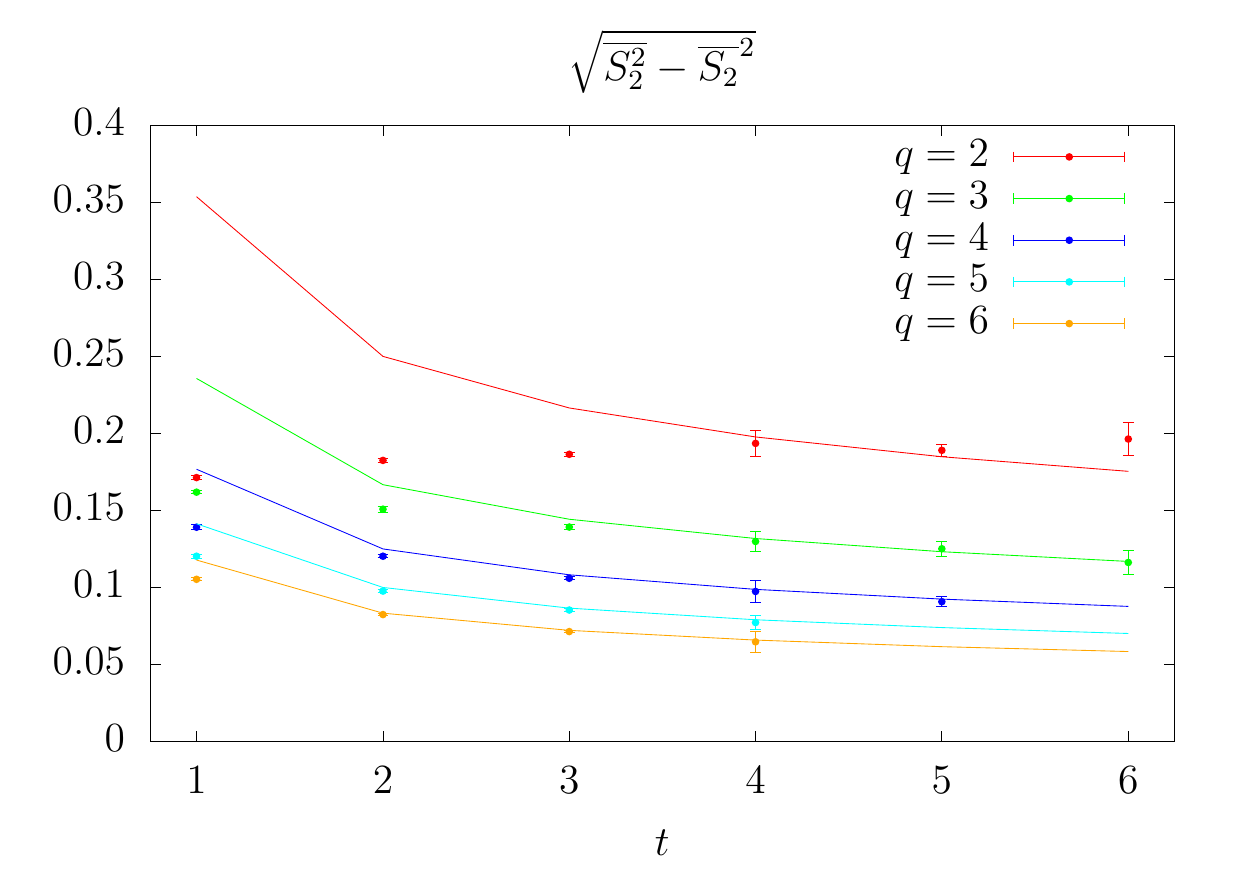}
\caption{Fluctuations of $S_2$ for different $q$ and number of layers of network.  Depending on the position of the entanglement cut the entanglement  increases  either in odd or even time steps. For each $t$ we have placed the cut so that the final layer of unitaries can create entanglement. The lines are the analytic lowest order result $ \frac{1}{q} \sqrt{\frac{2}{4^t} { 2( t- 1 ) \choose t- 1 }}$.}
\label{fig:S.pdf}
\end{figure}

More precisely, an exact combinatorial counting, involving pairs of walks from different replicas which meet at the $t=0$ boundary, gives
\begin{equation}
\label{eq:earlytime}
{\text{Var}(S_2 )} = {\frac{2}{q^2 4^t} { 2(t-1 ) \choose (t - 1 ) } } 
\end{equation}
and hence Eq.~\ref{eq:earlytimedecrease}.
We  calculate the early time fluctuations of $S_2$ numerically for various $q$, see Fig.~\ref{fig:S.pdf}. The largest $q$  values agree fairly well with the lowest order result in Eq.~\eqref{eq:earlytime} for larger $t$. 

When $t\gg q^2$, the bulk contribution to fluctuations dominates, and fluctuations grow with time. However they are not immediately governed by KPZ exponents.
From the continuum description in (\ref{eq:kpzeqconstsfixed},~\ref{eq:noisestrength})
the rescaled entropy $\widetilde S=\sqrt{4 q^4} {S}$ satisfies 
\begin{equation}\label{eq:rescaledkpzeq}
  \partial_t \widetilde{S} =  \f{1}{2} \partial_x^2 \widetilde{S}
  -  \f{1}{4q^2} (\partial_x \widetilde{S})^2  +  \widetilde \eta(x, t) + \widetilde c 
\end{equation}
subject to the normalized noise 
\begin{equation}
 \langle \widetilde \eta(x,t) \widetilde \eta(x',t' )  \rangle = \delta( x - x' ) \delta( t - t ' ).
\end{equation}
At early times the nonlinear term in (\ref{eq:rescaledkpzeq}) may be neglected due to its small coefficient. The resulting noisy linear equation (note that all coefficients are order 1)  is known as the Edwards-Wilkinson equation and gives fluctuations in $\widetilde S$ of order $t^{1/4}$  \cite{barabasi_fractal_1995}, or of order $q^{-2}t^{1/4}$ for $S$. These begin to dominate over the boundary fluctuations at time $t_\text{EW} \sim q^{2}$.

However the nonlinear term is RG relevant, and it can no longer be neglected at times $t\gtrsim t_{\rm KPZ}$, where ${t_{\rm KPZ}\sim q^8}$ \cite{kardar_dynamic_1986}. This is also the time at which fluctuations are of order one. Since $t_{\rm KPZ}\gg t_{\rm EW}\gg 1$ at large $q$, there are three regimes (Table~\ref{tab:3-regime}).

 \subsection{KPZ for R\'enyi entropies $S_n$ with $n\neq 2$}
\label{subsec:kpz_higher_n}

$S_2$ is the simplest entropy to calculate because each replica gives rise to only one elementary domain wall. However we can use the concept of the bound state (Sec.~\ref{subsec:map_leading_in_q}) to outline a generalization to larger $n$. For concreteness, consider the case $n=3$, with $q$ large. This limit simplifies the analysis by giving a clear separation of scales between two kinds of interactions.

First, within each replica there is a pair of walks, or equivalently quantum particles, with an attractive interaction between them of order 1 strength (Eq.~\eqref{eq:a3t-def}). Then at a parametrically smaller energy scale of order $q^{-4}$ there is the attractive interaction between walks in \textit{different} replicas which we have discussed in the previous section.

Therefore in the first step of RG --- at lengthscales of order one ---  the walks form independent bound states within each replica. On larger scales each bound state can be treated as a walk (or particle) with a single position coordinate $x_\alpha$ for $\alpha = 1,\ldots, k$. The bound states have a well-defined coarse-grained line tension and diffusion constant. Finally, there are weak attractive interactions \textit{between} bound states arising from the weak interactions between the microscopic walks. Therefore the next stage of the RG flow can again be described by a Hamiltonian like Eq.~\eqref{eq:ctmham}, but with different numerical constants. As a result we again expect KPZ scaling. We expect that a similar two-step RG picture applies for any $n>2$ when $q$ is large.

For each $n$, the continuum Hamiltonian for the bound states generalizing (\ref{eq:ctmham}) is characterised by three constants,
\begin{equation}
\label{eq:ctmham}
\mathcal{H} = - k \epsilon_n    - D_n  \sum_{\alpha=1}^k \f{\partial^2}{\partial x_\alpha^2} - \lambda_n \sum_{\alpha<\beta} \delta(x_\alpha - x_\beta),
\end{equation}
and the magnitude of the KPZ fluctuations in $S_n$ is proportional to $\lambda_n^{2/3}/[D_n^{1/3} (n-1)]$. 

If it was possible to compute these constants for arbitrary $n$, we could hope to analytically continue to $n=1$ to compute the 
fluctuations of the von Neumann entropy.\footnote{The scaling as $n\rightarrow \infty$  may be more easily tractable and is also interesting ($-S_\infty$  is the logarithm of the largest eigenvalue of the reduced density matrix).}

\section{Numerical checks using the operator entanglement}
\label{sec:numericalchecks}

In this section we perform numerical checks on some of the analytical arguments in Sec.~\ref{sec:kpz}. We argued that the dominant interaction between replicas, for large $q$, arose from a `special hexagon' diagram and that this is a pairwise interaction between replicas.

Here we check this result for the interactions by comparing numerics with the analytic form for
\begin{equation}
\overline{\exp\lf- k S_2[U(t)]\ri},
\end{equation}
where $S_2[U(t)]$ is the \textit{operator entanglement} of the time evolution operator. Recall that we may regard the tensor network defining $U(t)$ as a tensor network \textit{state} for $2L$ spins, $L$ at the top boundary and $L$ at the bottom boundary. $S_2[U(t)]$ is then the entanglement of a subsystem containing the $L/2$ spins on the left part of the bottom boundary together with the $L/2$ spins on the left part of the top boundary. This may be mapped to a lattice magnet by a simple extension of the above formulas.  The only change compared to the calculation of $\overline{Z_2^k}$ is the boundary condition at the bottom boundary. Since the top and bottom boundaries are treated on equal footing, the bottom boundary condition will be the same as the top one: there will be a composite domain wall $\tau_{n,k}$ at the bond of the entanglement cut. The configurations of incoming domain walls on the top and outgoing domain walls on the bottom are exactly the same as in Eq.~\eqref{eq:ddw_bubble}

\begin{figure}[h]
\centering
\includegraphics[width=.8\linewidth]{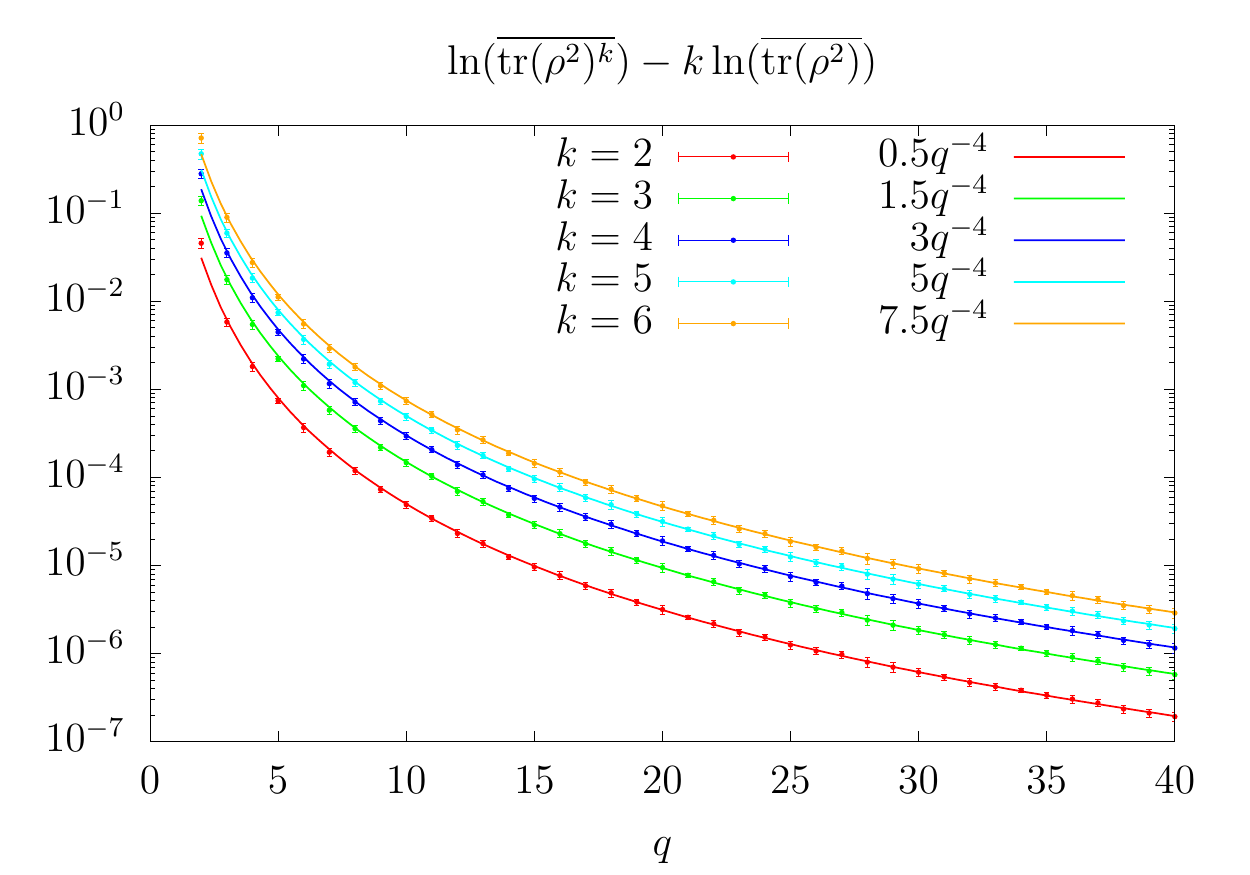}
\caption{Bipartite operator R\'enyi entanglement entropy for 2 site gate. This is the simplest one-layer random tensor network. The domain wall diagrams correspond exactly to those in Eq.~\eqref{eq:ddwtwostep}. We verified  the strength  $\frac{1}{2q^4}$ in Eq.~\eqref{eq:ddwtwostep} as well as the pairwise nature of the interaction. }
\label{fig:t1check}
\end{figure}

\begin{figure}[h]
\centering
\includegraphics[width=.8\linewidth]{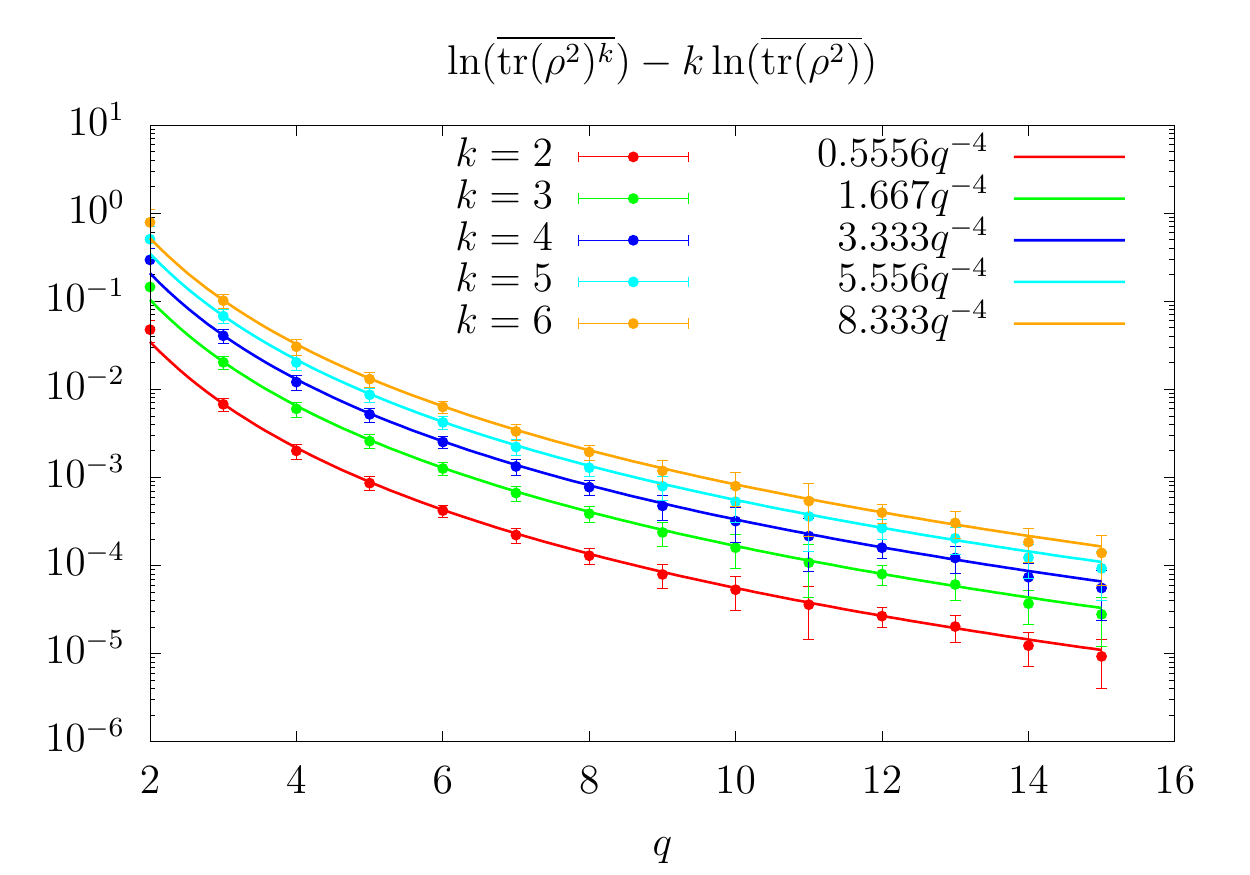}
\caption{Bipartite operator R\'enyi entanglement entropy of 4 sites. The tensor network has 3 layers: the first and last layers have one gate and middle layer has two gates. The interaction still fits $\frac{k(k-1)}{2} \frac{5}{9q^4}$, which is predicted by only considering the special hexagon interaction.}
\label{fig:t3check}
\end{figure}

We check the analytic result for $t=1$ and $t=3$. For $t=1$ we must consider a single unitary gate and the associated special hexagon interaction in Eq.~\ref{eq:ddw_bubble}. This gives
\begin{equation}
\ln \overline{\tr( \rho^2[U(1)])^k} - k \ln ( \overline{\tr( \rho^2[U(1)])} \simeq \frac{k(k-1)}{2} \frac{1}{2q^4}
\end{equation}
The subtraction on the LHS is to isolate the interaction contribution.
For $t=3$ we must sum over 6 configurations. At leading order in $1/q$ this gives
\begin{equation}
\ln \overline{\tr( \rho^2[U(3)])^k} - k \ln  \overline{\tr( \rho^2[U(3)])} \simeq \frac{k(k-1)}{2} \frac{5}{9q^4}
\end{equation}
In Figs.~\ref{fig:t1check} and \ref{fig:t3check} these results are compared against numerical results for $k=2,\ldots, 6$ and for various values of $q$. (We average over 4000 and 100 realizations  for $t=1$ and $t=3$ respectively.) In both cases the agreement is good at large $q$. 
This confirms that the special hexagon is indeed the interaction between replicas, at order $\frac{1}{q^4}$, in the bulk of the system. 

\section{Saturation at late time and Page's formula}
\label{sec:saturation}

For a finite system of even size $L$, when $t$ is far greater than the saturation time, we expect the half-chain entanglement to saturate to the value given by the generalized Page formula for R\'enyi entropy \cite{page_average_1993,nadal_statistical_2011}:
\begin{equation}
\label{eq:page_val} 
S_n^{\rm Page} = \frac{L}{2} \ln q  - \frac{\ln C_n}{ n -1}.
\end{equation}
In this formula $\frac{L}{2} \ln q$ is the maximal possible entanglement for the half-chain and $C_n$ is the $n$th Catalan number: $C_1 = 1$, $C_2 = 2$, $C_3 = 5$, $C_4 = 14$, etc. This formula is valid for all $q$ if $L$ is large, and the corrections are exponentially small in $L$. Fluctuations about the Page value are also exponentially small in $L$. 

We show that the constants $C_n$ have a simple and appealing explanation in terms of the domain walls. We first discuss the limit of large $q$, then give a sketch (which is partly conjectural) for how the domain wall picture allows the result to survive when $q$ is not large.

\begin{figure}[h]
\centering

\includegraphics[width=0.8\columnwidth]{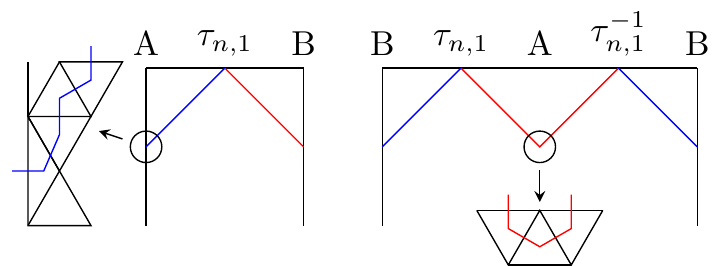}

 \caption{Domain wall paths for $t \gg L/2 v_2$. Left: entanglement of half of a chain with two boundaries. There are two possible paths at the leading order in $q$. They exit the system through a  tilted triangle on the boundary. Right: entanglement of half of a chain with periodic BCs.
 The two entanglement cuts give two domain walls, $\tau_{n,1}$ and its inverse. They meet at a down-pointing triangle either in region $A$ or in  $B$ again giving two possibilities.}
\label{fig:page_val}
\end{figure}

Consider the finite system with two spatial boundaries shown in Fig.~\ref{fig:page_val} (Left).
For a single elementary domain wall, as appears in the calculation for $\overline{Z_2}=\overline{e^{-S_2}}$, there are two possibilities at late time: it must exit the system via either the left or right spatial boundary. These possibilities are shown in red and blue respectively in the figure. At large $q$ the optimal slope for each domain wall (minimizing its total energy) is approximately unity, and the energy of such a domain wall is $(L/2) \ln q$.
The two possibilities lead to a factor of two: $\overline{Z_2}=2 q^{-L/2}$.

Now consider $\overline{Z_n^k}$. At leading order in $q$ the replicas decouple, so $\overline{Z_n^k}\simeq (\overline{Z_n})^k$. This is equivalent to the statement that at leading order in $q$ there are no fluctuations in the entanglement. The boundary condition for  $\overline{Z_n}$ introduces a domain wall of type $\tau_{n,1}=(12\ldots n)$ at the top boundary. This domain wall can split into two domain walls $\mu$ and $\nu$, satisfying 
\begin{equation}
\tau_{n,1}  = \mu \times \nu,
\end{equation}
with the $\mu$ domain wall exiting to the left and the $\nu$ wall to the right. 
The blue and red paths in Fig.~\ref{fig:page_val} both cross $\frac{L}{2}$ down-pointing triangles,  so $\overline{Z_n} \simeq c_n  q^{- (n-1) \frac{L}{2} } $. The number of configurations $c_n$ is the number of ways to \textit{factorize}  $\tau_{n,1}$ into a product $\mu \times \nu$. There are precisely $C_n$ such choices (see App.~B of \cite{zhou_operator_2017}), so $c_n = C_n$. This reproduces  Eq.~\ref{eq:page_val}.

The case of a finite interval (thus two cuts) of size $L/2$ in a chain with periodic boundary conditions is similar. Here we have boundary condition changes which insert domain walls $\tau_{n,1}$ and its inverse  $\tau_{n,1}^{-1}$ at the two cuts. The two optimal paths correspond to domain walls that meet either  inside $A$ or $B$ (Fig.~\ref{fig:page_val}, right). Otherwise the discussion is as above.
(Note that if the two subsystems contain different numbers of sites the energies of the different domain wall configurations will no longer be degenerate.)

This gives a combinatorial interpretation of the $\mathcal{O}(1)$ correction in the Page value as an entropy associated with the large-scale configurations of the random walks. 

For finite $q$ Eq.~\ref{eq:page_val} remains true so long as $L$ is large \cite{page_average_1993,nadal_statistical_2011}.
For this result to emerge from the random circuit, two things must happen. 
First, at finite $q$, the replicas must effectively decouple in the configurations that obtain at late time, to ensure that fluctuations about the Page value are parametrically small in $L$.
Second, all of the ways of splitting $\tau_{n,1}$ into $\mu\times \nu$ must have the same free energy.

This is closely connected to the conjectured constraints on $\lt_n(v)$ in Eq.~\ref{eq:generalconstraints} \cite{jonay_coarse-grained_2018}. 
Consider a domain wall that exits the boundary of the system (as in Fig.~\ref{fig:page_val}).
Approximately speaking,  Eq.~\ref{eq:generalconstraints} ensures that the preferred velocity of this domain wall, selected by free energy minimization, is $\pm v_B$ (rather than $\pm 1$, as at $q=\infty$),
and that the  line tension (free energy per unit time)  per elementary walk is independent of how the composite walk $\tau_{n,1}$ is split into smaller composites $\mu$ and $\nu$.
At a more microscopic level, what allows this to happen is the unbinding transition which we demonstrated for the case of two walks in Sec.~\ref{subsec:e3section}.
To be more accurate, we must also assume that different replicas, and more generally \textit{commuting} domain walls, aslo decouple by a similar mechanism: a vanishing of the effective attractive interaction when the coarse-grained speed is fixed to $v_B$.

 \subsection{The moment of saturation}

So far in this section we have discussed the asymptotic value of the entanglement at very late times. We may address the moment of saturation in a similar way. For definiteness, consider the entanglement of the first $\ell$ sites in a chain of size $L$, with $\ell \leq L/2$, so that the saturation time is approximately\footnote{The time at which the crossover happens will fluctuate by $\mathcal{O}(\ell^{1/3})$, due to KPZ fluctuations in the growth over this period. However, these fluctuations are \textit{between} realizations, and should not be confused with the rounding of the $S_n(t)$ profile \textit{within} a realization which we discuss here.} ${t_{\text{sat},n} \simeq \ell/v_n}$. Let's assume ${L/2-\ell \gg 1}$, so that we can neglect walks which travel to the right hand spatial boundary.

The leading order scaling picture for the moment of saturation is a sharp crossover in free energy, as a function of $t$, between `vertical' domain wall configurations which reach the $t=0$ boundary, and domain wall configurations which travel to the left spatial boundary.
Let us consider how this sharp transition is rounded out.

We must split $\tau_{n,1}$ into $\mu$, a composite walk which travels to the left boundary, and $\nu$, which travels to the $t=0$ boundary. 
 We will consider only the cases $S_2$ and $S_3$, i.e. $\tau_{n,1}=(12)$ and $\tau_{n,1}=(123)$. 
 In the first case we have, if we make the further simplification of neglecting fluctuations due to circuit randomness (this is controlled at large $q$):
\begin{equation}
\begin{aligned}
e^{-S_2(t)}  \simeq& e^{ - \seq \ell} + e^{- \seq v_2 t } \\
=& e^{-\seq \ell} \lf 1+e^{-\seq v_2 \left[ t - t_\text{sat} \right]} \ri. 
\end{aligned}
\end{equation}
where the first term represents the domain wall exiting at the left boundary and 2nd term is the domain wall going vertically as in the infinite system case.

For $S_3$ we at first sight have more terms, because we can choose $\nu=1$, $\nu=(123)$, or take $\nu$ to be an elementary domain wall. However, because $v_2> v_3$, the latter option is always exponentially subleading, so we have
\begin{equation}
e^{-2S_3(t)} \simeq e^{-2\seq \ell} \lf 
1+e^{-2\seq v_3 \left[ t - t_\text{sat} \right]}
\ri.
\end{equation}
Saturation is sharper for $S_3$ than for $S_2$. It is straightforward to extend these expressions to  similar situations, e.g. to the case $\ell \simeq L/2$ by including configurations with walks that travel to the right spatial boundary.

\section{Dynamics in realistic models}

The microscopic models we have studied here include randomness both in space and time.
The corresponding effective directed polymer partition function involves both thermal fluctuations and quenched disorder.
It is natural to expect that in realistic models without randomness, a mapping to a coarse-grained directed polymer problem will still be possible, and that this effective description will still include thermal fluctuations of the polymer. The quenched disorder will of course be absent in that case.
In fact this is very similar to what we have at large $q$, since as we have seen the effects of randomness are suppressed by a high order in $1/q$.
This picture is supported by the results of Ref.~\cite{chan_solution_2017}, which derived a domain wall picture for $\overline{Z_n}$ in a model with large $q$ unitaries that are random in space but not in time. At leading order in $q$ this picture coincides with that for the random circuits here.

The concept of local \textit{pairing of spacetime histories} that we find in random circuits is likely to be useful also in non-random models.
The basic point is that paired histories (one from a $U$ evolution and one from a $U^*$ evolution) contribute cancelling phases to the path integral, i.e. appear with positive weight.
This suggests that paired configurations, with domain walls between pairings that are enforced by boundary conditions, dominate the path integral for quantities such as $e^{-(n-1)S_n}$ even in the absence of any disorder average. 
We will discuss this further elsewhere.

It would be interesting to consider models in which the dynamics is time-independent but spatially random in more detail. A coarse-grained description for entanglement growth in such models was discussed in \cite{nahum_dynamics_2017}, in terms of a directed polymer subjected to randomness that depends on space but not on time.
In the replicated language, this corresponds to interactions between replicas that are local in space but nonlocal in time. 
In the case where the spatial randomness allows for `weak link' locations where the entanglement growth rate is arbitrarily small, a Gaussian average over disorder is not sufficient  since the weakest links, which are rare events, are important at late times. 
It would be interesting to search for these phenomena by applying the replica trick to the model of Ref.~\cite{chan_solution_2017} or extensions thereof.

 \section{Summary and outlook} 

We have shown that the minimal membrane picture for $\overline{S_n}$ makes sense beyond the $q=\infty$ limit, in a regime where it can no longer be identified with a `minimal cut' through the unitary circuit.
There is an emergent statistical mechanics governing the entanglement, in which the  R\'enyi  entropy is the free energy of an emergent domain wall. 
The interactions and `thermal' fluctuations of the domain walls play an important role in determining the entanglement velocities, and  entanglement line tensions $\mathcal{E}_n(v)$, which differ for different R\'enyi entropies. For example, a domain-wall unbinding phase transition (due to a delicate balance between interactions and `thermal' fluctuations) allows the general constraint $\mathcal{E}_n(v_B)=v_B$ to be satisfied for $n=3$.
Additional fluctuations associated with quenched disorder in the circuit are responsible for universal KPZ scaling at late times.
A different type of large scale fluctuation also governs the Page-like corrections to the entropies at late times.
For random circuits many properties are computable analytically in a power series expansion in $1/q$, 
where $q$ is the local Hilbert space dimension (and in some cases exactly).

The fact that entanglement entropies can be visualized in terms of domain walls can be understood heuristically as follows. The domain walls are between permutations that represent pairings of the forward and backward evolutions in the multi-layered path integral. When phase coherence is negligible, it is natural to guess that these paired configurations will dominate the path integral, since the pairing suppresses phase cancellation. This picture also applies to other quantities involving multiple-layer path integrals, for example the out-of-time-order correlator.

A lacuna in this work is an explicit treatment of the von Neumann entropy, as opposed to the higher R\'enyi entropies. This requires an additional  replica limit ($n\rightarrow 1$) which is likely to be more complicated than the  one we used here to average the R\'enyi entropies ($k\rightarrow 1$). Computing the von Neumann entropy for finite $q$ is a task for the future.  It is important to ask whether by focusing on $n>1$ we are missing important phenomena specific to $S_1$.

Another intriguing task is to obtain explicit numerical or analytical results for the entropies $S_n$ with $n>3$, extending the schematic picture above in terms of the bound state.
This would require us to understand the combinatorics associated with the labelling of the paths (by transpositions).\footnote{The `scattering matrix' for these domain walls does not appear to satisfy a Yang-Baxter equation for $n>3$, so perhaps the problem is non-integrable for $n>3$.}  This would shed light on the structure of the evolving entanglement spectrum. 
More detailed treatment of $S_{n>3}$ would also be interesting in the context of the bound state phase transition which we have argued occurs in $\lt(v)$ at $v=v_B$.

For $S_2$, our explicit mapping to a lattice directed polymer in a random medium problem means that the dynamics of the entanglement could be simulated, classically, over timescales which interpolate between the short times accessible in quantum simulations and the  large times required to see KPZ in the present model, at least for reasonably large $q$ (we have argued that a large crossover time is responsible for the apparent absence of KPZ scaling in short-time simulations).

In future work we will extend these mappings to related phenomena including light-cone effects in correlation functions, as well as entanglement growth for more general initial states.

\acknowledgements
We thank 
J. Haah,
D. Huse,
C. Jonay,
J. Ruhman,
and
S. Vijay
for discussions and collaboration on related work.
We thank P. Di Francesco, C. von Keyserlingk, T. Rakovszky,
and E. Vernier for discussions.
AN acknowledges EPSRC Grant No.~EP/N028678/1 and a Royal Society University Research Fellowship.

\appendix
\section{Combinatorial calculation of $\overline{Z}_3(t)$ at order $\frac{1}{q^2}$}
\label{app:z3t}

In this section, we use a combinatorial technique to calculate the partition function $\overline{Z}_3$ and the entanglement velocity $v_3$ to order $\frac{1}{q^2}$. In doing this, we define and calculate a slightly more general function $\Omega_3(t, q)$ which takes the $\frac{1}{q^2}$ correction into account.

According to the exact result in App.~\ref{app:ex_s3}, the two walks in $\overline{Z}_3$ have weight $K^2\simeq q^{-2} e^{-\frac{2}{q^2}}$ when they are separate. When they meet, there are two types of interactions:
\begin{enumerate}
\item The leading order attractive interaction resulting from the $3$ instead of $2$ ways to split in Eq.~\eqref{eq:123_split}. 
\item A weak repulsive interaction when the two walks are overlapping. Such a step has weight $q^{-2} e^{-\frac{3}{q^2}}$.
\end{enumerate}
Taking this into account, we can write the partition function as 
\begin{equation}
\overline{Z}_3(t) = q^{-2t} e^{-\frac{2t}{q^2}} \Omega_3( t, q )
\end{equation}
where $\Omega_3(t,q)$ is the following partition function for two walks
\begin{equation}
\Omega_3(t,q) = \sum_{\substack{\text{configs of} \\ \text{2 walks} }} \left( \frac{3}{2} \right)^{\text{$\#$ splitting}} \left( e^{-\frac{1}{q^2} } \right)^{\text{$\#$ overlapping steps}}.
\end{equation}
The entanglement velocity is 
\begin{equation}
v_3 = 1 + \frac{1}{q^2 \ln q} - \lim_{t\rightarrow \infty} \frac{\ln \Omega_3( t, q)}{2t \ln q }.
\end{equation}
Instead of counting $\Omega_3(t, q)$, we consider the partition function where each step is assigned a fugacity $\sqrt{x}$
\begin{equation}
\Omega_3(\sqrt{x}, q) = \sum_{t = 0}^{\infty} \Omega_3(t, q) x^{\frac{t}{2}}. 
\end{equation}
\subsection{$q = \infty$}
If we neglect the weak repulsive interaction at $\frac{1}{q^2}$ order, $\Omega_3(t, q)$ becomes $\Omega_3(t)$ defined in Eq.~\eqref{eq:a3t-def}. The corresponding partition function is 
\begin{equation}
\Omega_3(\sqrt{x},q = \infty) = \sum_t \sum_{\substack{\text{configs of} \\ \text{2 walks} }} \left( \frac{3}{2} \right)^{r} \sqrt{x}^{t},
\end{equation}
where $r$ is the number of splittings.

It is simpler to consider the relative motion of the two walks. There are 3 possible displacements in a single time step: $0, \pm 2$. A displacement of $\Delta = 0$ means that the two walks move in the same direction (left or right), while a displacement $\Delta = \pm 2$ means moving in  opposite directions. 

Let $t_0$ be the number of time steps in which $\Delta=0$ and $t'$ be the number of time steps in which $\Delta= \pm 2$. 
We can immediately perform the sum over $t_0$ by noting that, in between one $\Delta =\pm 2$ step and the next, 
there can be an arbitrary number of $\Delta=0$ steps.
The sub-partition function for these steps is
\begin{equation}
\sum_{n=0}^{\infty} (2\sqrt{x})^n = \frac{1}{1 - 2\sqrt{x}}
\end{equation}
where the $2$ represents the two choices of the center of mass going left or right. 
This leaves a partition function for a single random walk (with steps of $\pm 2$) representing the relative displacement, 
\begin{equation}\label{eq:omega3}
\Omega_3(\sqrt{x},q = \infty) = 
\sum_{t'} \sum_r 
 \left( \frac{3}{2} \right)^{r} \lf\f{\sqrt{x}}{1 - 2\sqrt{x}}\ri^{t'}
 {\sideset{}{'}\sum_{\text{one walk}}} 1.
\end{equation}
Where the prime indicates that the walks can now only take $\Delta=\pm 2$ steps, and where we sum over configurations with the specified $t'$, $r$.

To simplify, we assign $\frac{3}{2}$ to the meeting event (when the single walk returns to the origin) and assume the two walks meet at the end. This does not affect the asymptotic behavior. 
The final sum in the above equation is
the number of such single walks that return to the origin $r$ times, which we denote  $Z(t', r)$.
 This is \cite{spivey_enumerating_2012}
\begin{equation}
Z( t' = 2n , r ) = \frac{r 2^r}{2n - r }{ 2n - r  \choose n - r }.
\end{equation}
This has a generating function
\begin{equation}
\begin{aligned}
\sum_{n=r}^{\infty} Z(t'=2n, r) y^n  = [f(y)]^r
\end{aligned}
\end{equation}
where 
\begin{equation}
f(y) = 1 - \sqrt{ 1- 4y}.
\end{equation}
Therefore
\begin{align}
\Omega_3(\sqrt{x},q = \infty) & = 
\sum_r 
 \left( \frac{3}{2} \right)^{r} 
\left[ f \lf\f{{x}}{(1 - 2\sqrt{x})^2}\ri \right]^r \\
& = \f{1}{1-\f{3}{2} f \lf\f{{x}}{(1 - 2\sqrt{x})^2}\ri}.
\end{align}
We read off the smallest pole $\sqrt{x_*} = 3 \sqrt{2} - 4 $ which determines the asymptotic behavior
\begin{equation}
\Omega_3(t) \sim \left( \frac{1}{ \sqrt x_*} \right)^t = \left( 2 + \frac{3}{\sqrt{2}} \right)^t. 
\end{equation}

\subsection{$q$ large but finite}

The analysis is the same
except that the sub-partition function for a given string of consecutive $\Delta=0$ steps is different depending on whether the relative coordinate is zero or not. If it is, the sub-partition function is modified to
\begin{equation}
\sum_{n=0}^{\infty} (2e^{-\frac{1}{q^2}}\sqrt{x})^n = \frac{1}{1 - 2e^{-\frac{1}{q^2}}\sqrt{x}}.
\end{equation}
The modification to Eq.~\eqref{eq:omega3} is that
there are $r$ of these factors, and $t'-r$ of the factors we had before.
\begin{align}\notag
& \Omega_3(\sqrt{x},q = \infty) 
\\ \notag 
&= 
 \sum_{t'} \sum_r  
 \left( \frac{3}{2}\f{1-2\sqrt{x}}{1 - 2e^{-\frac{1}{q^2}}\sqrt{x}} \right)^{r} \lf\f{\sqrt{x}}{1 - 2\sqrt{x}}\ri^{t'}
 {\sideset{}{'}\sum_{\text{one walk}}} 1 \\
 &  = \f{1}{1-\left( \frac{3}{2}\f{1-2\sqrt{x}}{1 - 2e^{-\frac{1}{q^2}}\sqrt{x}} \right) f \lf\f{{x}}{(1 - 2\sqrt{x})^2}\ri}.
\end{align}
The pole satisfies
\begin{equation}
( 1 + \frac{4}{q^2}) x + (8 - \frac{2}{q^2})\sqrt{x} -2  = 0. 
\end{equation}
The smaller pole $\sqrt{x_*} = 3 \sqrt{2} - 4 + (3 - 2\sqrt{2} )^2 \frac{1}{q^2} $ determines the asymptotic behavior
\begin{equation}
\Omega_3(t, q ) \sim \left( \frac{1}{\sqrt x_*} \right)^t = \left( 2 + \frac{3}{\sqrt{2}} - \frac{1}{2q^2} \right)^t .
\end{equation}
This gives the partition function $\overline{Z}_3( t)$ and velocity to order $\mathcal{O}(\frac{1}{q^2 \ln q})$
\begin{equation}
v_3 = 1   - \frac{\ln \left( 2 + \frac{3}{\sqrt{2}} \right)  }{2  \ln q } + \frac{3\sqrt{2}}{4} \frac{1}{q^2 \ln q}.
\end{equation}

 \section{Slope-dependent Line Tension $\mathcal{E}_3( v ) $}
\label{app:e3v}

In this section, we derive the slope-dependent line tension $\lt_3(v)$ of the $n = 3$ bond state, which generalizes the partition function $\Omega_3(t)$ in App.~\ref{app:z3t}. To this end we must obtain the free energy of the walks as a function of their coarse-grained velocity.

Let $x$ be the total displacement of the bound state, i.e. the mean displacement of the two walks, and let $\Omega_3(x,t)$ be the partition function with a fixed displacement. We expect
\begin{equation}
\Omega_3(vt, t ) q^{-2 t} \sim e^{- 2 \ln q  \lt_3(v) t}.
\end{equation}
in other words
\begin{equation}
 \lt_3(v) \sim 2 \ln q - \ln \Omega_3(vt, t) .
\end{equation}

Consider the generating function of $\Omega_3(x,t)$,
\begin{align}
\tilde \Omega_3(\phi, t) & = \sum_x \Omega_3(x,t) \phi^{x}
\end{align}
where we assign weight $\phi$ for the mean displacement to go one step right and $\phi^{-1}$ for one step left. 

We break up the sum according to the number $t_0$ of time steps where the relative displacement is $0$. The steps with relative displacement 0 can change the mean displacement by $\phi^{\pm 1}$, while the steps with relative displacement $\pm 2$ do not change the mean displacement. We therefore have
\begin{align}
\tilde \Omega_3(\phi, t ) &= \sum_{t_0=0}^{t} { t \choose t_0} \lf \phi + \phi^{-1}\ri^{t_0} Z( t - t_0 ) \\
& \sim  \sum_{t_0=0}^{t} { t \choose t_0} \lf \phi + \phi^{-1}\ri^{t_0} 
\lf\f{9}{2}\ri^{\f{t-t_0}{2}} \\
&\sim \lf \phi + \phi^{-1} + \f{3}{\sqrt 2} \ri^t.
\end{align}
If we regard $\tilde \Omega_3(\phi, t)$ as the partition function for a modified ensemble, the total displacement of the bound state, i.e. the mean displacement of the two walks, is
\begin{equation}
 \f{\partial}{\partial \ln \phi} \ln \tilde \Omega_3(\phi, t) = {t}\times \f{\phi-\phi^{-1}}{\phi + \phi^{-1} + \f{3}{\sqrt 2 } }.
\end{equation}
Then the mean velocity is 
\begin{equation}
v(\phi) =   \f{\phi-\phi^{-1}}{\phi + \phi^{-1} + \f{3}{\sqrt 2 } }
\end{equation}
which we can solve for the fugacity
\begin{equation}
\phi= \f{3v+\sqrt{8+v^2}}{\sqrt 8 (1-v)}.
\end{equation}
By saddle-point reasoning,
\begin{align}
\tilde \Omega_3(\phi, t) \sim \Omega_3(v(\phi) t, t ) \phi^{v(\phi) t}.
\end{align}
Therefore
\begin{equation}
\begin{aligned}
  \lt_3(v) =& 1 - \frac{1}{2 \ln q} \left(  \ln (\frac{3}{\sqrt{2}} + \phi^{-1} + \phi) - v \ln \phi\right)  \\
  =& 1- \frac{1}{2 \ln q} \Bigg(  \frac{3v^2 + \sqrt{v^2 + 8}}{\sqrt{2} ( 1 - v^2 ) } + \frac{3}{\sqrt{2}} \\
&- v \ln \frac{3v + \sqrt{ 8 + v^2 }}{\sqrt{8}( 1 - v ) }  \Bigg) .
\end{aligned}
\end{equation}

 \section{Line tension $\lt_3(v)$ close to lightcone}
\label{app:e3vlightcone}

In this section we calculate $\lt_3(v)$ for $v$ very close to the lightcone $v=1$. Writing  
$v = 1 - \f{\alpha}{q^2}$, with $\alpha$ of order 1 and $q$ large, we obtain ${\lt_3(1 - \f{\alpha}{q^2})}$ up to terms of order $1/q^2 \ln q$:  see Eqs.~\eqref{eq:unbindingvelocity},~\eqref{eq:e3nearlightcone} and~\eqref{eq:A3formula} in the main text. This allows a nontrivial check on the relation $\lt_3(v_B)=v_B$, and reveals an unbinding transition for the two walks appearing in the partition function $Z_3$ when the boundary conditions are modified so that their coarse-grained speed exceeds a critical value ${v_c \simeq  1-2/q^2}$. This is consistent at this order with ${v_c = v_B}$, which we conjecture is true to all orders. 

To the order at which we are working we can neglect interactions between replicas. Then
\begin{equation}
\lt_3(v) = \lim_{t\rightarrow\infty}  \frac{-\ln Z(vt;t)}{2 \ln q \times t},
\end{equation}
where $Z(vt;t)$ is a partition function for two walks with the constraint that the displacement of their centre of mass is $v t$. For definiteness we can take their relative coordinate $\Delta$ (the difference in the coordinates of the two walks) to be zero at the initial and final time.

From the exact triangle weights in Appendix.~\ref{app:S3weights}, to relative order $1/q^2$, the weight is 
\begin{equation}
q^{-2}e^{-2/q^2}
\end{equation}
for a time step in which the walks are separate, and
\begin{equation}
q^{-2}e^{-3/q^2}
\end{equation}
for a time step in which the walks are in a composite walk. Finally, for a time step in which the walks either split or merge we may take the weight to be
\begin{equation}
\sqrt{\frac{3}{2}}\times q^{-2}e^{-2/q^2}.
\end{equation}
Here we have shared the statistical weight $3/2$ for each merge-split event (see Sec.~\ref{subsec:weight_down_tri}) equally between the splitting event and the merging event. We neglect boundary terms, which are unimportant in the $t\rightarrow \infty$ limit.

In order to fix the velocity $v$, we introduce a `fugacity' $\phi$ for leftward steps.
Since $v$ is close to one almost all steps are rightward, and the fugacity $\phi$ will be small. If $Z(\phi;t)$ is the partition function with fugacity $\phi$ but with no constraint on the total displacement of the centre of mass, and if ${v(\phi)=1-\alpha(\phi)/q^2}$ is the average speed in the ensemble with fixed $\phi$, then
\begin{equation}
\label{eq:Zphidefn}
Z(\phi; t) 
\sim 
Z(v(\phi);t) \phi^{\alpha(\phi) t/q^2}.
\end{equation}
In the present regime only an $\mathcal{O}(1/q^2)$  fraction  of the time steps involve a walk taking a step to the left. We can neglect configurations in which \textit{both} walks take a step to the left in the same time step, since such an event occurs only once in every $\mathcal{O}(1/q^4)$ time steps.

The configuration is then determined entirely by the relative displacement $\Delta$ as a function of time, and we can write $Z(\phi,t)$ in terms of a transfer matrix $T_{\Delta, \Delta'}$. This transfer matrix contains a factor of $\phi$ for each time step in which $\Delta\neq \Delta'$, since in such a time step one of the walks takes a step to the left:
\begin{equation}
Z(\phi;t) = q^{-2t} e^{-2t/q^2} \lf T^t \ri_{0,0}
\end{equation}
with (expanding in $1/q^2$)
\begin{equation}
T = 
\begin{bmatrix}
 \ldots &   & & &  & &  \\
   &  1 & \phi & &  & &  \\
     &  \phi & 1 & \phi \sqrt{3/2}  & &  &  \\
     &   & \phi \sqrt{3/2} & 1-q^{-2}  &  \phi \sqrt{3/2} &  &  \\
          &   & &  \phi \sqrt{3/2}  &  1 & \phi &  \\
                  &   & &  &  \phi & 1 &  \\
                             &   & &  &   &  &  \ldots
\end{bmatrix}.
\end{equation}
Let us define the $\mathcal{O}(1)$ quantities
\begin{equation}
\Phi= q^2 \phi.
\end{equation} 
For $\Phi>1$ the largest eigenvalue of the transfer matrix, determining the scaling of the partition function, is 
\begin{equation}
\label{eq:largesteigenvaluelambda}
\lambda = 1 + \f{3 \sqrt{1 + 8 \Phi^2}-1}{4q^2},
\end{equation}
corresponding to the bound state `wavefunction' $\psi_\Delta$ with 
\begin{align}
\label{eq:boundstatepsi}
\psi_0 & =1, & 
\psi_{\Delta\neq 0} & = \sqrt{\f{3}{2}} \mu^{|\Delta|},
&
&\text{with} & \mu &= \f{1+\sqrt{1 + 8 \Phi^2}}{4 \Phi}.
\end{align}
The bound state exists for $\mu< 1$, i.e. for $\Phi>1$. At $\Phi=1$ the bound state disappears. For the range of velocities where $\Phi<1$, when the walks are unbound, their typical separation is $\sqrt t$ at large $t$. They are therefore effectively independent and their free energy is  twice that of a single walk, leading to $\lt_3(v)=\lt_2(v)$. It is straightforward to check that $\Phi=1$ corresponds to $\alpha_c=2$: for $\Phi\leq 1$ the walks can be treated as independent, and $v$ is simply related to the weight ${\phi = \Phi/q^2}$ for a left step by ${v=1-2\phi}$. 

For the range of velocities where $\Phi>1$, Eqs.~\eqref{eq:Zphidefn},~\eqref{eq:largesteigenvaluelambda} together with ${\ln (T^t)_{00}\sim t\times \ln \lambda}$ give
\begin{equation}
\label{eq:e3fmlaintermediate}
\lt_3(v) \simeq 1  - \f{\alpha}{q^2} + \f{1}{q^2 \ln q} \lf \f{9}{8}- \f{3\sqrt{1 + 8 \Phi^2}}{8}+\f{\alpha}{4} \ln \Phi^2 \ri.
\end{equation}
We still need to relate $\Phi$ and $v$. 

In the bound region, we note that $v$ is equal to the probability that in a given time step the change in $\Delta$ is zero. The sum over such configurations is obtained by replacing $T$ with $T_\text{diag}$ for the given time step, where $T_\text{diag}$ is the diagonal part of $T$. This leads to
\begin{equation}
v= \f{1}{\lambda} 
\frac{
 \bra{\psi}  T_\text{diag} \ket{\psi}
 }{
 \bra{\psi}\psi\rangle
 }.
\end{equation}
Using Eq.~\eqref{eq:boundstatepsi} gives 
\begin{align}
\notag
v =& 1- \f{1}{q^2} \f{6 \Phi^2 (1+\sqrt{1 + 8 \Phi^2})}{1+8\Phi^2+\sqrt{1 + 8 \Phi^2}},
&
\Phi^2 =& \f{\alpha^2}{18} \lf 2 +\sqrt{4+\f{9}{\alpha^2}} \ri.
\end{align}
Together with Eq.~\eqref{eq:e3fmlaintermediate} this gives Eqs.~\eqref{eq:e3nearlightcone},~\eqref{eq:A3formula} in the main text.

 \section{The Weingarten function}
\label{app:wein}

In this section, we introduce the properties of the Weingarten function used in the main text. 

We begin with the general formula for the average of the tensor product of a Haar random unitary\cite{weingarten1978asymptotic,collins_integration_2006}
\begin{equation}
\begin{aligned}
&\int [dU_{d \times d} ] U_{i_1,j_1} U_{i_2,j_2}  \dots  U_{i_n, j_n}  U^*_{i'_1,j'_1} U^*_{i'_2,j'_2}  \dots  U^*_{i'_n, j'_n}  \\
=& \sum_{ \sigma, \delta \in S_n} \delta_{i_1 i'_{\tau(1)}} \dots \delta_{i_n i'_{\tau(n)}} {\rm Wg}(d, \sigma \tau ^{-1} ) 
\delta_{j_1 j'_{\sigma(1)}} \dots \delta_{j_n j'_{\sigma(n)}}.   \\
\end{aligned}
\end{equation}
In the main text, $d = q^2$ and we pack formula compactly in the bracket notation (Eq.~\eqref{eq:fourU}), where the products of delta functions are identified as components of the permutation states $| \sigma \rangle$ and $|\tau \rangle$. 

The Weingarten function ${\rm Wg}( \sigma ) \equiv {\rm Wg}( q^2, \sigma )$ is a function of the conjugacy class of the permutation. Its defining property can be obtained from the left/right invariance of the Haar ensemble, 
\begin{equation}
\stackP = \oneP. 
\end{equation}
Translating this into algebra, we have 
\begin{equation}
\sum_{\tau_1 \sigma_1} {\rm Wg}(\sigma \tau_1^{-1} ) d^{N - | \tau_1^{-1} \sigma_1 | } {\rm Wg}(\sigma_1 \tau^{-1} ) = {\rm Wg}(\sigma \tau^{-1} ).
\end{equation}
If we regard ${\rm Wg}( \sigma \tau^{-1} )$ as an invertable matrix with $\sigma$ and $\tau$ as its row and column indices, then
\begin{equation}
\label{eq:wg_inv}
\sum_{ \tau} {\rm Wg}(\sigma_a \tau^{-1} ) d^{N - ( \tau^{-1} \sigma_b ) } =  \delta_{\sigma_a \sigma_b} .
\end{equation}
Therefore ${\rm Wg( \sigma_a \sigma_b^{-1})}$ is the inverse of $d^{N - ( \sigma_a^{-1} \sigma_b ) }$. This is the key to all the exact weights, see App.~\ref{app:ex_weight}. 

The Weingarten function can be expanded perturbatively, and the leading order term for each permutation is \cite{collins_integration_2006}
\begin{equation}
{\rm Wg}( \sigma ) = \frac{1}{d^{N}}\left[ \frac{{\rm Moeb} ( \sigma ) }{d^{|\sigma|}} + \mathcal{O}\lf\frac{1}{d^{|\sigma| + 2}}\ri \right]
\end{equation}
where the M\"obius function for a permutation with cycle decomposition $\sigma = c_1 c_2 \cdots c_k$ is defined as 
\begin{equation}
{\rm Moeb}( \sigma ) = \prod_{i=1}^k {\rm Catalan}_{|c_i| - 1 } (-1)^{|c_i| - 1 }.
\end{equation}

Some elementary examples are
\begin{align}
  {\rm Moeb} ( \I ) &= 1,  &{\rm Moeb} ( ( 12 ) ) &= -1,\\
  {\rm Moeb} ( (12)(34)) &= 1, & {\rm Moeb} ( (123) ) &= 2.
\end{align}

For the convenience of the perturbative calculation, we define
\begin{equation}
{\rm wg} ( \sigma ) = d^N { \rm Wg}( \sigma ) . 
\end{equation}
Up to $\frac{1}{d^2}$ ($\frac{1}{q^4}$) order, the only non-vanishing wg functions are
\begin{equation}
\label{eq:wg_exp}
\begin{aligned}
{\rm wg}( \I ) &= 1 + \frac{{ N \choose 2}}{d^2} + \mathcal{O}\lf \frac{1}{d^3}\ri, \\
{\rm wg}( (12) ) &= -\frac{1}{d} + \mathcal{O}\lf\frac{1}{d^3}\ri, \\
{\rm wg}( (123) ) &= \frac{2}{d^2} + \mathcal{O}\lf\frac{1}{d^3}\ri, \\
{\rm wg}( (12)(34) ) &= \frac{1}{d^2} + \mathcal{O}\lf\frac{1}{d^4}\ri,
\end{aligned}
\end{equation}
where the particular permutations inside, like $(12)$, are representatives of their conjugacy classes. 
The last three relations come from the leading term expansion of ${\rm Wg}$. The first one can be worked out by subtracting all the other $\frac{1}{d^2}$ terms from the sum \cite{harrow_church_2013}
\begin{equation}
\sum_\sigma {\rm wg} ( \sigma ) = \frac{d^N ( d - 1)!}{( d+ N - 1 )!}.
\end{equation}

 \section{Exact weights with $\leq 1$ incoming domain wall}
\label{app:ex_weight}

In this section, we derive the exact wight of some down-pointing triangles by using the orthogonality relation in Eq.~\eqref{eq:wg_inv}. We will denote the number of cycles in a permutation by ${\chi( \sigma) = N - | \sigma|}$. 

First, according to the definition in  Eq.~\eqref{eq:Jabc}
\begin{equation}
J( \sigma_b, \sigma_b; \sigma_a ) = \sum_{\tau} \text{Wg}( \sigma_a \tau^{-1} ) q^{2N - 2 | \sigma_b^{-1} \tau| } . 
\end{equation} 
Comparing this with the orthogonality relation in Eq.~\eqref{eq:wg_inv} and setting $q^2 = d$, we obtain
\begin{equation}
\label{eq:J_delta_ab}
J(\sigma_b, \sigma_b ; \sigma_a) = \delta_{ \sigma_a, \sigma_b}.
\end{equation}

Next, we consider the weight of a single domain wall
\begin{equation}
\sdwfull = \sum_{\tau_a} {\rm Wg} (\tau_a  ) q^{\chi( \tau_a ) + \chi(\tau_a(12))}.
\end{equation}
We define
\begin{equation}
\Sigma^{\pm} = \sum_{\chi(\tau_a(12))= \chi(\tau_a ) \pm 1 } {\rm Wg} (\tau_a  ) q^{\chi( \tau_a ) + \chi(\tau_a)}
\end{equation}
then
\begin{equation}
K = \sdwfull = q \Sigma^+ + \frac{1}{q}\Sigma^- .
\end{equation}
By taking $\sigma = \I$ and $(12)$ in the variant of the orthogonality relation
\begin{equation}
\sum_{\delta }  {\rm Wg} (\delta  ) d^{\chi( \delta \sigma  )}
 = \delta_{\I, \sigma} ,
\end{equation}
we have 
\begin{equation}
\begin{aligned}
\Sigma^+ + \Sigma^- &= 1\\
d \Sigma^+ + \frac{1}{d} \Sigma^- &= 0 .
\end{aligned}
\end{equation}
The solution is
\begin{equation}
\Sigma^+ = \frac{-1}{d^2 - 1}, \quad \Sigma^- = \frac{d^2}{d^2 - 1}.
\end{equation}
Therefore
\begin{equation}
\label{eq:K_ex}
K =  \frac{-q}{q^4 - 1} + \frac{q^3}{q^4 - 1} = \frac{q}{1+ q^2}.
\end{equation}

Further, we consider a single domain wall that creates a pair of new (possibly composite) domain walls
\begin{equation}
K_{\wedge}=\sdwbubb = \sum_{\tau_a } {\rm Wg}( \tau_a\mu ) q^{\chi( \tau_a ) + \chi( \tau_a (12)) } .
\end{equation}
We define 
\begin{equation}
\Sigma^{\pm} = \sum_{\chi(\tau_a(12))= \chi(\tau_a ) \pm 1 } {\rm Wg} (\tau_a\mu  ) q^{\chi( \tau_a ) + \chi(\tau_a)},
\end{equation}
then
\begin{equation}
 K_{\wedge} = q \Sigma^+ + \frac{1}{q} \Sigma^-.
\end{equation}
On the other hand, from the orthogonality relation
\begin{equation}
\begin{aligned}
  &\sum_{\tau_a } {\rm Wg} (\tau_a\mu  ) q^{\chi( \tau_a ) + \chi(\tau_a)} \\
 =& \sum_{\tau_a } {\rm Wg} (\tau_a\mu  ) q^{\chi( \tau_a(12) ) + \chi(\tau_a(12))} = 0 
\end{aligned}
\end{equation}
we have
\begin{equation}
\Sigma^+ + \Sigma^- = 0 \quad  d \Sigma^+ + \frac{1}{d} \Sigma^- = 0  \implies \Sigma^{\pm} = 0 .
\end{equation}
Hence
\begin{equation}
\label{eq:k-wedge}
K_{\wedge} = 0.
\end{equation}
We conclude that
\begin{equation}
J(\I,  (12); \sigma_a ) = \frac{q}{q^2 +1} \left( \delta_{ \I, \sigma_a} + \delta_{(12) ,\sigma_a}  \right).
\end{equation}

 \section{Exact weights for $N=3$}
\label{app:ex_s3}
\label{app:S3weights}

This section presents the exact weights $J( \sigma_b, \sigma_c; \sigma_a )$ for $\overline{Z_1^3}$ (the single replica of $S_3$). 

There are $6$ elements in the order $3$ permutation group: $\I, (1,2), (1,3),(2,3),(1,2,3),(1,3,2)$. The relevant Weingarten functions are (see for example \cite{gu_moments_2013}; $d = q^2$)
\begin{equation}
\begin{aligned}
  \text{Wg}( d, [1,1,1] ) &= \frac{q^4-2}{q^2 ( q^4-1) (q^4-4)} \\
  \text{Wg}( d, [1,2] ) &= \frac{-q^2}{q^2 ( q^4-1) (q^4-4)} \\
  \text{Wg}( d, [3] ) &= \frac{2}{q^2 ( q^4-1) (q^4-4)} \\
\end{aligned}
\end{equation}
where the numbers inside the square brackets are the cycle sizes of the permutation. Then $J( \sigma_b, \sigma_c; \sigma_a )$ becomes
\begin{equation}
\label{eq:JabcS3}
\begin{aligned}
J( \sigma_b,& \sigma_c; \sigma_a )=   {\rm Wg}( d, I ) \langle \sigma_a | \sigma_b \rangle  \langle \sigma_a | \sigma_c\rangle \\
  +&{\rm Wg}( d, [1,2]  )  \Big\{\langle (1,2)^{-1} \sigma_a | \sigma_b \rangle  \langle  (1,2)^{-1} \sigma_a | \sigma_c\rangle  \\
  & + \langle (1,3)^{-1} \sigma_a | \sigma_b \rangle  \langle  (1,3)^{-1} \sigma_a | \sigma_c\rangle  \\
  & + \langle (2,3)^{-1} \sigma_a | \sigma_b \rangle  \langle  (2,3)^{-1} \sigma_a | \sigma_c\rangle  \Big\}\\
  +&{\rm Wg}( d, [3] ) \Big\{ \langle (1,2,3)^{-1} \sigma_a | \sigma_b \rangle  \langle (1,2,3)^{-1} \sigma_a | \sigma_c\rangle  \\
  +&  \langle (1,3,2)^{-1} \sigma_a | \sigma_b \rangle  \langle  (1,3,2)^{-1} \sigma_a | \sigma_c\rangle \Big\}. \\ 
\end{aligned}
\end{equation}

The computation in Eq.~\eqref{eq:JabcS3} gives the same results for the exact weights in App.~\ref{app:ex_weight}, and also additional non-trivial weights
\begin{equation}
\begin{aligned}
\dptri{\I}{\I}{(123)} &= \frac{q^4 - 2q^2 -2}{(q^2 +1 ) (q^4 - 4)} \\
\dptri{\I}{(12)}{(23)} &= \frac{q^2( q^2 - 1 )}{(q^2 + 1 )( q^4 - 4 )} \\
\dptri{\I}{(123)}{(132)} &= \frac{-2 (q^2 - 1 )}{(q^4 - 4) (q^2 + 1 )} .\\
\end{aligned}
\end{equation}

 \section{Perturbative calculation of the triangle weights}
\label{app:pert_calc}

In this section, we present the perturbative calculation of the weight of a down-pointing triangle.

The weight of the down-pointing triangle is obtained by integrating out the $\tau$ spin. Formally
\begin{equation}
J( \sigma_b, \sigma_c; \sigma_a) = \sum_{\tau \in S_N} {\rm wg}( \tau^{-1} \sigma_a )  q^{ - | \sigma_b^{-1} \tau| - |\tau^{-1} \sigma_c |}.
\end{equation}
To represent this in diagrams, we put the $\tau$ spin in the center of the triangle and use dashed lines to connect the $\tau$ spin and the three neighboring $\sigma$s. 
The links between $\tau$ and $\sigma_{c}$ or $\sigma_b$ give an exact factor  $\frac{1}{q}$ for each elementary domain wall, and 
 the link between $\tau$ and $\sigma_a$ gives ${ \rm wg}( \tau^{-1} \sigma_a)$. 
 
 First consider $K$ (which we know exactly).
 The leading order diagram is the one where $\tau = \sigma_a$, and we have
\begin{equation}
 K = \sdwfull = \sdwi + \mathcal{O}(\frac{1}{q^2}) = \frac{1}{q} +  \mathcal{O}(\frac{1}{q^2}).
\end{equation}
Now using the higher order series expansion  of ${\rm wg}$ in Eq.~\eqref{eq:wg_exp}, we can obtain a more accurate value of $K$
\begin{equation}
\begin{aligned}
K&= \sdwfull = \sdwi + \sdwturn + \sdwtri  + \mathcal{O}(\frac{1}{q^6})\\
&= \frac{1}{q} {\rm wg}( d, \I ) + \frac{1}{q} {\rm wg}( d, (12) ) + \frac{ { N \choose 2 } - 1 }{q^3} { \rm wg} ( d, ( 13) )  + \mathcal{O}(\frac{1}{q^7})\\
&= \frac{1}{q} \left[ \left(  1 + \frac{{N \choose 2}}{q^4}  \right)- \frac{1}{q^2} - \frac{1}{q^2} \frac{ \left( { N \choose 2 } - 1 \right)}{q^2}  \right] + \mathcal{O}(\frac{1}{q^7})\\
&= \frac{1}{q} \left( 1 - \frac{1}{q^2} + \frac{1}{q^4}+ \mathcal{O}(\frac{1}{q^6})  \right)
\end{aligned}
\end{equation}
Here we see that the number of choices for the 
elementary domain wall on the vertical link in the last diagram cancels the $N$ dependence in the expansion of ${\rm wg}( \I)$ from the first diagram, generating an $N$ independent weight, which is consistent with the exact result in Eq.~\eqref{eq:K_ex}. 

Now we consider two commutative incoming and outgoing domain walls $(12)$ and $(34)$, which is relevant to evaluating $\zk$,
\begin{equation}
\begin{aligned}
&\ddwfull = \ddwi + 2 \ddwiturn  + \ddwtt + \ddwtri \\
=& \frac{1}{q^2} {\rm wg}( d, \I ) + 2 \frac{1}{q^2} { \rm wg}( d, (12) )  + \frac{1}{q^2} {\rm wg} ( d, (12)(34) ) \\
& \qquad \qquad  + \frac{1}{q^4} { \rm wg} ( d, (12) ) \left( { N\choose 2 } - 2 \right)  + \mathcal{O}(\frac{1}{q^8})\\
=& \frac{1}{q^2} \left[  \left( 1 + \frac{{ N\choose 2 }}{q^4}-\frac{2}{q^2}\right)  + \frac{1}{q^4}- \frac{1}{q^2}\frac{{ N\choose 2 } - 2}{q^2} +  \mathcal{O}(\frac{1}{q^6}) \right] \\
=& \frac{1}{q^2} \left[ 1 - \frac{2}{q^2}  + \frac{3}{q^4} +  \mathcal{O}(\frac{1}{q^6})\right] = \frac{1}{q^2} \left[ 1- \frac{1}{q^2} + \frac{1}{q^4} + \mathcal{O}(\frac{1}{q^6})\right]^2 \\
=& \sdwfull \times \sdwfull \times \left( 1 + \mathcal{O}(\frac{1}{q^6})\right) .
\end{aligned}
\end{equation}
The calculation for the outgoing domain walls exiting in the opposite directions is similar. We thus obtain the factorization condition in Eq.~\eqref{eq:ddw_decomp}.

The factorization fails if the incoming domain wall is a product of non-commutative transpositions. 
We take it to be $(123)$, which is relevant to $S_3$. There are now 3 ways to assign one elementary domain wall to the vertical link, and the weight ${\rm wg}(d, 123))$ is $2/q^4$. Taking these into account, we have
\begin{equation}
\label{eq:comp_ddw_int}
\begin{aligned}
&\ddwfull = \ddwi + 3\ddwiturn + \ddwtt + \ddwtri \\
&= \frac{1}{q^2} {\rm wg}( d, \I ) + 3\frac{1}{q^2} { \rm wg}( d, (12) ) + \frac{1}{q^2} {\rm wg} ( d, (123) ) \\
& \qquad \qquad  + \frac{1}{q^4} { \rm wg} ( d, (12) ) \left( { n\choose 2 } - {3 \choose 2} \right)  + \mathcal{O}(\frac{1}{q^8})\\
&= \frac{1}{q^2} \Big[  \left( 1 + \frac{{ n\choose 2 }}{q^4}\right)  + \left( -\frac{1}{q^2}\right) \times 3 + \frac{2}{q^4} \\
&+ \frac{1}{q^2} \left( - \frac{1}{q^2}\right)\left( { n\choose 2 } - 3 \right) +  \mathcal{O}(\frac{1}{q^6}) \Big] \\
&= \frac{1}{q^2} \left[ 1- \frac{1}{q^2} + \frac{1}{q^4} + \mathcal{O}(\frac{1}{q^6})\right]^2 \left[ 1 - \frac{1}{q^2} + \mathcal{O}(\frac{1}{q^6}) \right]\\
&= \sdwfull \times \sdwfull \times \left[ 1 - \frac{1}{q^2} + \mathcal{O}(\frac{1}{q^6}) \right]
\end{aligned}
\end{equation}
We see that the factor of $\left[ 1 - \frac{1}{q^2} + \mathcal{O}(\frac{1}{q^6}) \right]$ gives rise to a repulsive interaction between the domain walls. 

Next we turn to corrections from adding `bubbles' to the domain wall configurations as in Fig.~\ref{fig:bubb_diagram}. 
{ 
Many such bubble configurations vanish due to the exact results in Eq.~\eqref{eq:J_delta_ab} and Eq.~\eqref{eq:k-wedge}. The leading non-trivial diagrams corresponds to configurations (e) and (f) in Fig.~\ref{fig:bubb_diagram}. Na\"ive domain wall number counting suggests a bubble is a $\frac{1}{q^4}$ correction to the diagram without the bubble. It is however at most a $\frac{1}{q^6}$ correction if the bubble is created simply by adding a closed loop of a given domain wall type. 
\begin{figure}[h]
\centering

\includegraphics[width=0.8\columnwidth]{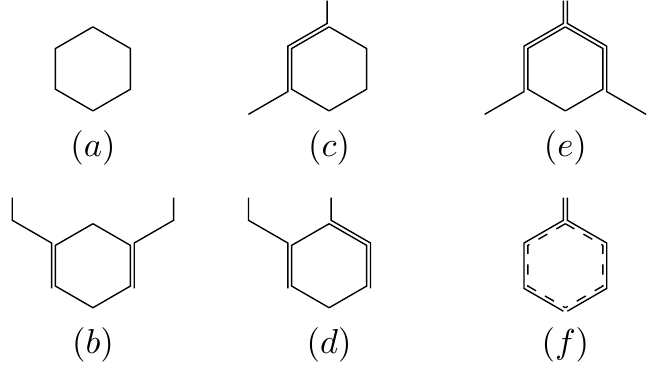}

 \caption{Possible bubble diagrams. Each line represents an elementary domain wall. (a) (b) (c) (d) have $0$ weight: (a) and (b) vanish because the tip of the hexagon is $J( \I, \I ; (12)) =0 $; (c) and (d) vanish because on the top $K_{\wedge} = 0 $. (e) is an order $\frac{1}{q^6}$ correction. (f) can be an order $\frac{1}{q^6}$ correction if the dashed loop is an elementary domain wall. It is an order $\frac{1}{q^4}$ correction if it is a special hexagon as in Eq.~\eqref{eq:ddw_bubble}. }
\label{fig:bubb_diagram}
\end{figure}
For example, consider the bubble corrections to the leading diagram for $K_{\wedge}$. Let the incoming domain wall be $(12)$ and the outgoing ones be $(12)(34)$ and $(34)$. We can always choose a vertical link carrying $(34)$ to cancel the leading order diagram
\begin{equation}
\sdwpfull = \sdwpi + \sdwpw + \mathcal{O}(\frac{1}{q^5}) =  \mathcal{O}(\frac{1}{q^5})
\end{equation}
so that it is consistent with the exact result $K_{\wedge} = 0$ in Eq.~\eqref{eq:k-wedge}. 
The cancellation mechanism also exists for two commutative incoming domain walls in Eq.~\eqref{eq:regvert}:
\begin{equation}
\ddwpfull = \ddwpi + \ddwpw + \mathcal{O}(\frac{1}{q^6}) = \mathcal{O}(\frac{1}{q^6}).
\end{equation}
}
When the newly generated domain wall pair  annihilates in the time step immediately below, this gives  a $\frac{1}{q^6}$ correction in the bulk (Fig.~\ref{fig:bubb_diagram} (f)). In contrast, the special hexagons in Eq.~\eqref{eq:ddw_bubble} do not suffer from the cancellation mechanism. As a result they are $\frac{1}{q^4}$ corrections in the bulk and lead to the dominant pairwise attraction in $\overline{Z_2^k}$.

 \section{Continuum interaction constant}
\label{app:interactionconstant}

Continuing  from the discussion in Sec.~\ref{subsec:continuumdescription}, we use $Z^{(k)}$ to denote the partition function for $k$  bosons on a ring, or equivalently $k$ walks on  a torus. (Note that these BCs are not related to the entanglement calculation.) To fix $\lambda$ we take $L$ and $t$ large enough that the continuum approximation is valid but small enough that the interaction may be treated as as a perturbation: this is possible when $\lambda \ll 1$. If $\Delta E$ is the change in the ground state energy of a pair of bosons when the small interaction is switched on then $Z^{(2)} / [Z^{(1)}]^2 = e^{-t \Delta E}$. Since in the noninteracting problem the ground state wavefunction is spatially constant, 
\begin{equation}
\label{eq:deltaE}
\Delta E = - \f{\lambda}{L^2} \int \dd x_1 \dd x_2 \delta (x_1 - x_2) = - \f{\lambda}{L}.
\end{equation}
On the other hand, on the lattice
\begin{equation}
\f{Z^{(2)}}{ [Z^{(1)}]^2}=
\<
\exp 
\sum_{\substack{
{\text{vertical}}
\\
{\text{bonds $b$}}}}
\lf
\f{1}{2q^4} \f{n_b \lf n_b - 1 \ri}{2}
\ri
\>,
\end{equation}
where the expectation value is taken for a pair of \textit{noninteracting} walks on $D$. Expanding the exponential, and using translational invariance in both dimensions,
\begin{equation}
\label{eq:pmeet}
\f{Z^{(2)}}{ [Z^{(1)}]^2} \simeq 1 + \f{t L}{2} \f{1}{2q^4}  P_\text{meet}.
\end{equation}
Here $tL/2$ is the number of vertical bonds on the square lattice $D$, and $P_\text{meet}$ is the probability that a given bond is visited by both walks. Using the independence of the walks, this is  $P_\text{meet} = 1/L^2$. Matching Eqs.~\eqref{eq:pmeet},~\eqref{eq:deltaE} gives $\lambda = 1/(4 q^4)$, as stated in the main text.

\bibliographystyle{unsrt}
\bibliography{TZ_ref}

\end{document}